\documentclass[iop]{emulateapj}
\usepackage[utf8]{inputenc}
\usepackage{amssymb}
\usepackage{graphicx}
\usepackage{natbib}
\usepackage{amsmath}

\def \ms{m\,s$^{-1}$\,}

\def \msun{M$_{\odot}$}
\def \rsun{R$_{\odot}$}

\def \mjup{M$_{\rm JUP}~$}

\def \mearth{M$_{\oplus}$}

\newcommand{\rhk}{\ensuremath{\log R'_{hk}}}

\begin{document}





\slugcomment{paper\_rev1}
\shorttitle{HD~34445 Six-planet system}

\shortauthors{Vogt et al.}

\title{A Six-planet system around the star HD~34445}

\author{Steven S. Vogt\altaffilmark{1}, 
R. Paul Butler\altaffilmark{2},  
Jennifer Burt\altaffilmark{1}, 
Mikko Tuomi\altaffilmark{3}, 
Gregory Laughlin\altaffilmark{4}, 
Brad Holden\altaffilmark{1}, 
Johanna K. Teske\altaffilmark{5},
Stephen A. Shectman\altaffilmark{5},
Jeffrey D. Crane\altaffilmark{5},
Mat\'{\i}as D\'{\i}az\altaffilmark{6},
Ian B. Thompson\altaffilmark{5},
Pamela Arriagada\altaffilmark{2},
 and Sandy Keiser\altaffilmark{2}}
\altaffiltext{1}{UCO/Lick Observatory, 
Department of Astronomy and Astrophysics, 
University of California at Santa Cruz,
Santa Cruz, CA 95064}
\altaffiltext{2}{Carnegie Institution for Science, 5241 Broad Branch Road NW, Washington, DC, USA 20015-1305}
\altaffiltext{3}{University of Hertfordshire, Centre for Astrophysics Research, Science and Technology Research Institute, College Lane, AL10 9AB, Hatfield, UK}
\altaffiltext{4}{Department of Astronomy, Yale University, New Haven CT 06511}
\altaffiltext{5}{The Observatories, Carnegie Institution for Science, 813 Santa Barbara Street, Pasadena, CA 91101-1292}
\altaffiltext{6}{Departamento de Astronomia, Universidad de Chile, Camino
El Observatorio 1515, Las Condes, Santiago, Chile}

\begin{abstract}
We present a new precision radial velocity (RV) dataset that reveals a multi-planet system orbiting the G0V star HD~34445. Our 18-year span consists of 333 precision radial velocity observations, 56 of which were previously published, and 277 which are new data from Keck Observatory, Magellan at Las Campanas Observatory, and the Automated Planet Finder at Lick Observatory. These data indicate the presence of six planet candidates in Keplerian motion about the host star with periods of 1057, 215, 118, 49, 677, and 5700 days, and minimum masses of 0.63, 0.17, 0.1, 0.05, 0.12 and 0.38 $M_{\rm J}$ respectively. The HD~34445 planetary system, with its high degree of multiplicity, its long orbital periods, and its induced stellar radial velocity half-amplitudes in the range $2 \,{\rm m\, s^{-1}} \lesssim K \lesssim 5\,{\rm m\, s^{-1}}$ is fundamentally unlike either our own solar system (in which only Jupiter and Saturn induce significant reflex velocities for the Sun), or the \textit{Kepler} multiple-transiting systems (which tend to have much more compact orbital configurations).
\end{abstract}


\section{Introduction}

HD~34445 was first reported to host a planet by \cite{Howard2010}. Their combined data set of 68 precision radial velocities (RV's) measured over 12 years with the Keck Observatory's HIRES spectrometer \citep{Vogt94} and 50 RV's \citep{Howard2010} measured over 6 years with the HRS spectrometer \citep{Tull1998} on the McDonald Observatory Hobby-Eberly Telescope (HET), yielded a 1049-day planet of minimum mass 0.79 \mjup in a mildly eccentric (e=0.27) orbit. \cite{Howard2010} also reported indications of a second possible planet candidate signal at 117 days with minimum mass of 52 \mearth, but which needed a significant number of additional measurements to confirm, and to rule out other possible periods.

By the time of the \cite{Howard2010} publication, it was clear to us that this star hosted a quite complex multi-planet system, with periods long enough that it would take years of further observations to adequately characterize. Accordingly, HD~34445 remained a high-priority target on our Keck observing list and we obtained an additional 160 unbinned HIRES velocities up until the end of our final Keck observing run on Jan. 17, 2014. In addition, we obtained a single HIRES velocity harvested from the Keck archive from UT date August 26, 2014 (P.I. Leslie Rogers). We also included HD~34445 as a high-priority target on our target lists for the Planet Finding Spectrometer (PFS) spectrometer \citep{Crane2010} on Carnegie's Magellan telescope, and the Levy Spectrometer of the Automated Planet Finder (APF) at Lick Observatory \citep{Vogt2014a}. Over the past 18 years we have accumulated a total of 333 precision RV's that indicate a system of at least five additional planets orbiting HD~34445. In this paper, we present the full RV data set and discuss the planetary configuration it implies.

\section{Basic properties of HD~34445}
HD~34445 (HIP~24681) is a bright (V = 7.31) G0V star at a distance of 45.4 pc \citep{Gaia2016}. The basic properties of this star were given in detail by \cite{Howard2010} and are reproduced here for the reader's convenience in Table \ref{tab:stellar}. As detailed by \cite{Howard2010}, they indicate an old, relatively inactive star of age 8.5 $\pm$ 2 Gyr, that lies slightly above the main sequence, is slightly metal-rich with respect to the Sun, and has an expected rotation period (from $\rhk$) of $\sim$22 days. \citet{Howard2010}'s estimates of the age and the rotation rate were obtained using the correlations with $\log\,R\prime_{\rm HK}$ presented by \citet{Noyes1984}. In re-reporting these values in Table \ref{tab:stellar}, we note that the values should be taken with special caution. HD~34445 is slightly evolved, and the \citet{Noyes1984} relations were derived for main sequence stars. Finally, we note that  directed surveys of the star by \citet{Mason2011} using speckle interferometry and by \citet{Ginski2012} using lucky imaging found no evidence for stellar companions to HD~34445.

\begin{deluxetable}{rll}
\tablecaption{\label{tab:stellar} Stellar parameters for HD~34445}
\tablehead{{Parameter}&{Value}}
\tablecolumns{2}
\startdata
Spectral type & G0V \\
$M_V$ & 4.04 $\pm$ 0.10 \\
$V$ & 7.31 $\pm$ 0.03 \\
$B-V$ & 0.661 $\pm$ 0.015 \\
Mass (\msun) & 1.07 $\pm$ 0.02 \\
Radius (\rsun) & 1.38 $\pm$ 0.08 \\
Luminosity ($L_\sun$) & 2.01 $\pm$ 0.2 \\
$S_{hk}$ & 0.148 \\
$\rhk$ & -5.07 \\
Age (Gyr) & 8.5 $\pm$ 2.0 \\
$[\mathrm{Fe/H}]$ & +0.14 $\pm$ 0.04 \\
T$_{eff}$ ($K$) & 5836 $\pm$ 44 \\
$\log(g)$ (cm s$^{-2}$) & 4.21 $\pm$ 0.08 \\
P$_{rot}$ (days) & $\sim$22 \\
\enddata
\end{deluxetable}

\section{Radial velocities}
The data consist of 333 precision radial velocities obtained with a variety of telescopes. \cite{Howard2010} published a 12-year set of HIRES velocities, along with a 6-year set of 50 HET velocities. To that, we added 200 RV's obtained with the HIRES spectrometer from  Jan 24, 1998 to Aug 26, 2014. The HIRES spectra used by \cite{Howard2010} were shared by the California Planet Search team (CPS) and Lick-Carnegie Exoplanet Survey (LCES) teams before the two teams split in late-2007. The CPS team's HIRES data reduction pipeline is slightly different than that of the LCES team, in that it uses a somewhat different deconvolution routine. In August 2004, we performed a major upgrade of the HIRES CCD and optical train. Unlike the CPS data reduction pipeline of \cite{Howard2010}, we find no velocity offset between the pre-fix and post-fix epochs of HIRES data. So in this paper, for velocities obtained from shared HIRES spectra between the CPS and LCES teams, we used our own data reduction pipeline, and refer to these velocities as ``LCES/HIRES" velocities (see Table 2). The six velocities named ``CPS/HIRES" listed in Table 3 were as published in \cite{Howard2010} and were obtained from the CPS data reduction pipeline on their HIRES spectra taken after the team split. We also obtained a set of 54 radial velocities with the APF and a set of 23 radial velocities with the PFS. We traditionally bin our velocities over 2-hour time periods to supress stellar jitter. However, in the present paper, we show all of our HIRES, APF, and PFS velocities as unbinned. To summarize, our analysis is based on 54 APF velocities, 50 HET velocities, 200 Keck velocities reduced with our pipeline, 6 Keck velocities reduced by Howard et al. (2010), and 23 PFS velocities.

The HIRES, PFS, and APF RVs were all obtained by placing a gaseous Iodine absorption cell in the converging beam of the telescope, just ahead of the spectrometer slit \citep{butler96}. The absorption cell superimposes a rich forest of Iodine lines on the stellar spectrum over the 5000-6200~\AA\ region, thereby providing a wavelength calibration and proxy for the point spread function (PSF) of the spectrometer. The Iodine cell is sealed and maintained at a constant temperature of 50.0 $\pm$0.1$^{\circ}$ C such that the Iodine gas column density remains constant over decades. Our typical spectral resolution was set at $\sim$60,000 with HIRES, and $\sim$80,000 with PFS and APF. The wavelength range covered 3700-8000~\AA\ with HIRES, 3900-6700~\AA\ with PFS, and 3700-9000~\AA\ with APF, though only the 5000-6200~\AA\ Iodine region was used in obtaining Doppler velocities.

\begin{figure}
\center
\includegraphics[angle=0, width=0.5\textwidth,clip]{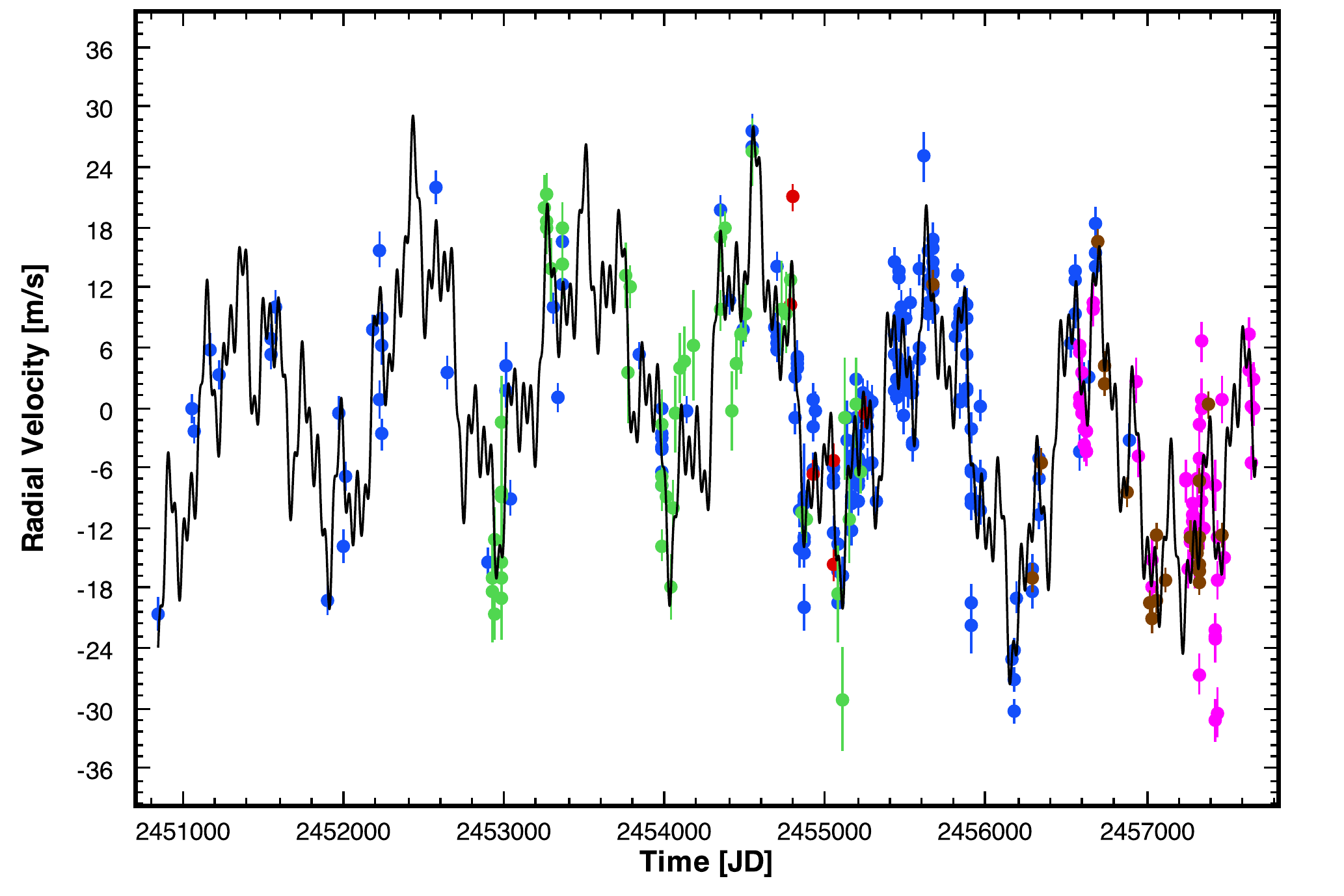}
\caption{ Relative radial velocities of HD~34445 from Keck/HIRES (blue), HET10 (green), Keck10 (red), Magellan/PFS (brown) and APF/Levy (magenta). The black solid curve denotes the final 6-planet MAP Keplerian model.}\label{fig:rvs}
\end{figure}

That iodine spectral region was divided into $\sim$700 chunks of about 2 \AA\ each for HIRES and $\sim$818 for PFS. Each separate chunk produced an independent measure of the wavelength, PSF, and Doppler shift. The Doppler shifts were determined using the spectral synthesis technique described by \cite{butler96}. The final reported Doppler velocity is the weighted mean of the velocities of all the individual chunks. The final uncertainty of each velocity is the standard deviation of all 700 chunk velocities about that mean. A plot of the entire set of 333 precision relative RV's together with our final 6-planet Keplerian model (solid line) is shown in Figure \ref{fig:rvs}.

We derived Mt. Wilson S-index \citep{Duncan1991} measurements from each of our HIRES, PFS, and APF spectra to serve as proxies for chromospheric activity in the visible stellar hemisphere at the moments when the spectra were obtained. The S-index is obtained from measurement of the emission reversal at the cores of the Fraunhofer H and K lines of Ca II at 3968 {\AA} and 3934 {\AA} respectively. In addition, we also report H-index measurements for our post-fix (JD 2453303.12) HIRES as well as PFS spectra. Similarly to the S-index, the H-index quantifies the amount of flux within the H$\alpha$ line core compared to the local continuum. We use the \citet{Gomes2011} prescription, which defines the H-index as the ratio of the flux within $\pm$0.8{\AA} of the H$\alpha$ line at 6562.808 {\AA} to the combined flux of two broader flanking wavelength regions: 6550.87 $\pm$ 5.375 {\AA} and 6580.31 $\pm$ 4.375 {\AA}.

\begin{deluxetable}{ccccc}
\tablecaption{LCES/HIRES radial velocities for HD~34445 (\textit{Sample: full table in electronic version})
\label{tab:rvdata_LCES/HIRES}}
\tablecolumns{5}
\tablehead{{JD}&{RV [m\,s$^{-1}$]}&{Unc. [m\,s$^{-1}$]}&{S-index}&{H-index}}
\startdata
2450838.76 & -21.60 & 1.65 & 0.1595 & -1.00 \\ 
2451051.11 & -1.06 & 1.34 & 0.1505 & -1.00\\ 
2451073.04 & -3.24 & 1.20 & 0.1581 & -1.00\\ 
2451171.85 & 4.81 & 1.51 & 0.1476 & -1.00\\ 
2451228.81 & 2.35 & 1.37 & 0.1566 & -1.00\\ 
2451550.88 & 5.90 & 1.38 & 0.1436 & -1.00\\ 
2451551.88 & 4.35 & 1.48 & 0.1454 & -1.00\\ 
2451581.87 & 9.07 & 1.55 & 0.1452 & -1.00\\ 
2451898.03 & -20.13 & 1.40 & 0.1508 & -1.00\\ 
2451974.76 & -1.57 & 1.52 & 0.1307 & -1.00\\ 
2452003.74 & -14.86 & 1.57 & 0.1336 & -1.00\\ 
2452007.72 & -7.84 & 1.41 & 0.1288 & -1.00\\ 
2452188.14 & 6.77 & 1.45 & 0.1360 & -1.00\\ 
2452219.15 & 14.76 & 1.73 & 0.1440 & -1.00\\ 
2452220.08 & -0.22 & 1.89 & 0.1386 & -1.00\\ 
2452235.86 & -3.68 & 1.38 & 0.1356 & -1.00\\ 
2452238.89 & 5.10 & 1.64 & 0.1452 & -1.00\\ 
2452242.93 & 7.91 & 1.43 & 0.1424 & -1.00\\ 
2452573.00 & 21.00 & 1.51 & 0.1422 & -1.00\\ 
2452651.94 & 2.51 & 1.51 & 0.1400 & -1.00\\ 
2452899.10 & -16.48 & 1.50 & 0.1310 & -1.00\\ 
2453016.87 & 3.16 & 2.20 & 0.1362 & -1.00\\ 
2453016.90 & 0.69 & 1.54 & 0.1423 & -1.00\\ 
2453044.79 & -9.99 & 1.74 & 0.1355 & -1.00\\ 
2453303.12 & 8.96 & 1.48 & 0.1243 & 0.03046\\ 
2453338.90 & 0.06 & 1.23 & 0.1462 & 0.03046\\ 
2453340.02 & 0.00 & 1.30 & 0.1306 & 0.03036\\ 
2453368.95 & 15.54 & 1.22 & 0.1428 & 0.03021\\ 
2453369.80 & 11.23 & 1.24 & 0.1492 & 0.03023\\ 
2453841.76 & 4.22 & 1.31 & 0.1445 & 0.03017\\ 
\enddata
\end{deluxetable}

\begin{deluxetable}{ccc}
\tablecaption{CPS/HIRES radial velocities for HD~34445
\label{tab:rvdata_CPS/HIRES}}
\tablecolumns{3}
\tablehead{{JD}&{RV [m\,s$^{-1}$]}&{Uncertainty [m\,s$^{-1}$]}}
\startdata
2454777.95 & 4.68 & 1.02\\ 
2454791.01 & 15.34 & 1.26\\ 
2454927.76 & -12.33 & 1.36\\ 
2455045.13 & -11.01 & 1.59\\ 
2455049.13 & -21.37 & 1.43\\ 
2455251.86 & -6.17 & 1.43\\ 
\enddata
\end{deluxetable}

\begin{deluxetable}{ccc}
\tablecaption{HET radial velocities for HD~34445 (\textit{Sample: full table in electronic version})
\label{tab:rvdata_HET}}
\tablecolumns{3}
\tablehead{{JD}&{RV [m\,s$^{-1}$]}&{Uncertainty [m\,s$^{-1}$]}}
\startdata
2452926.87 & -23.82 & 4.99\\ 
2452932.00 & -22.46 & 2.09\\ 
2452940.99 & -18.74 & 3.23\\ 
2452942.98 & -26.21 & 2.37\\ 
2452978.74 & -24.64 & 4.03\\ 
2452979.74 & -22.54 & 4.01\\ 
2452983.88 & -13.97 & 2.21\\ 
2452984.89 & -20.95 & 3.41\\ 
2452986.73 & -6.99 & 4.35\\ 
2452986.86 & -14.40 & 2.99\\ 
2453255.96 & 14.40 & 3.08\\ 
2453258.96 & 12.95 & 3.01\\ 
2453260.96 & 15.60 & 2.06\\ 
2453262.95 & 12.35 & 1.76\\ 
2453286.89 & 8.19 & 3.05\\ 
2453359.69 & 8.66 & 3.28\\ 
2453365.68 & 12.38 & 2.36\\ 
2453756.76 & 7.60 & 3.08\\ 
2453775.70 & -2.11 & 4.85\\ 
2453780.69 & 6.51 & 2.08\\ 
2453978.97 & -13.27 & 2.42\\ 
2453979.98 & -19.30 & 1.56\\ 
2453985.99 & -7.28 & 3.29\\ 
2453988.96 & -12.49 & 3.31\\ 
2454003.93 & -14.39 & 1.91\\ 
2454043.81 & -23.50 & 3.20\\ 
2454055.77 & -15.55 & 2.73\\ 
2454071.90 & -6.16 & 3.70\\ 
2454096.67 & -1.56 & 3.40\\ 
2454121.60 & -0.97 & 3.34\\ 
\enddata
\end{deluxetable}

\begin{deluxetable}{cccc}
\tablecaption{APF radial velocities for HD~34445 (\textit{Sample: full table in electronic version})
\label{tab:rvdata_APF}}
\tablecolumns{4}
\tablehead{{JD}&{RV [m\,s$^{-1}$]}&{Unc. [m\,s$^{-1}$]}&{S-index}}
\startdata
2456585.98 & 13.30 & 1.47 & 0.1476\\ 
2456585.99 & 12.54 & 1.30 & 0.1468 \\ 
2456587.97 & 7.32 & 1.36 & 0.1492 \\ 
2456587.98 & 8.00 & 1.52 & 0.1486 \\ 
2456602.94 & 10.49 & 1.52 & 0.1513\\ 
2456602.95 & 6.56 & 1.64 & 0.1507\\ 
2456605.93 & 3.26 & 1.61 & 0.1480 \\ 
2456605.94 & 4.93 & 1.58 & 0.1459 \\ 
2456620.89 & 4.74 & 1.72 & 0.1547 \\ 
2456620.90 & 2.74 & 1.50 & 0.1544 \\ 
2456672.75 & 17.56 & 1.72 & 0.1567\\ 
2456672.76 & 16.85 & 1.69 & 0.1604\\ 
2456938.94 & 9.67 & 2.12 & 0.1720\\ 
2456947.96 & 2.26 & 2.00 &0.1699 \\ 
2457027.89 & -10.81 & 2.15 & 0.1775\\ 
2457029.89 & -8.19 & 2.20 & 0.1846\\ 
2457240.01 & -0.32 & 1.93 & 0.1677\\ 
2457240.02 & -0.05 & 1.78 & 0.1717\\ 
2457255.00 & -9.05 & 1.87 & 0.1593\\ 
2457266.93 & -5.53 & 1.73 & 0.1636\\ 
2457272.91 & -6.32 & 2.49 & 0.2727\\ 
2457285.88 & -3.54 & 2.97 & 0.2643\\ 
2457285.89 & -2.44 & 2.44 & 0.2525\\ 
2457291.89 & -4.24 & 1.78 & 0.1597\\ 
2457307.85 & -8.80 & 1.98 & 0.1671 \\ 
2457309.83 & -5.43 & 1.88 & 0.1603\\ 
2457316.81 & -0.07 & 2.51 & 0.1950\\ 
2457316.82 & -4.15 & 2.46 & 0.2324\\ 
2457322.98 & -19.58 & 1.96 & 0.1612\\ 
2457331.76 & 5.45 & 2.23 & -1.000\\ 
\enddata
\end{deluxetable}

\begin{deluxetable}{ccccc}
\tablecaption{PFS radial velocities for HD~34445
\label{tab:rvdata_PFS}}
\tablecolumns{5}
\tablehead{{JD}&{RV [m\,s$^{-1}$]}&{Unc. [m\,s$^{-1}$]}&{S-index}&{H-index}}
\startdata
2455663.49 & 25.21 & 1.46 & 0.2040 & 0.03212 \\ 
2456281.70 & -4.01 & 1.40 & 0.2380 & -1.000\\ 
2456345.56 & 7.57 & 1.27 & 0.1513 & 0.02941\\ 
2456701.57 & 29.58 & 1.12 & 0.1345 & 0.02959\\ 
2456734.57 & 15.39 & 1.15 & 0.4032 & 0.02916 \\ 
2456734.57 & 17.25 & 1.22 & 0.3406 & 0.02930\\ 
2456879.93 & 4.54 & 1.42 & 0.1376 & 0.02954\\ 
2457023.68 & -6.52 & 1.22 & 0.1392 & 0.03001\\ 
2457029.68 & -7.99 & 1.32 & 0.1346 & 0.02974\\ 
2457055.58 & -6.19 & 1.27 & 0.1340 & 0.02993\\ 
2457067.54 & 0.17 & 1.30 & 0.1453 & 0.02999\\ 
2457117.48 & -4.19 & 1.19 & 0.1340 & 0.02913\\ 
2457267.91 & 0.14 & 1.43 & 0.1410 & 0.02930\\ 
2457319.84 & -1.05 & 1.21 & 0.1481 & 0.03004 \\ 
2457320.83 & -1.57 & 1.08 & 0.1367 & 0.02921\\ 
2457321.82 & -0.76 & 1.11 & 0.1384 & 0.02935\\ 
2457322.80 & 0.00 & 1.01 & 0.1347 & 0.02907\\ 
2457323.85 & -4.45 & 1.07 & 0.1344 & 0.02919\\ 
2457324.80 & -3.26 & 1.03 & 0.1355 & -1.000\\ 
2457325.79 & -2.68 & 1.01 & 0.1384 & 0.02917\\ 
2457327.83 & 5.73 & 1.17 & 0.1398 & 0.02929\\ 
2457389.64 & 13.29 & 1.22 & 0.1383 & 0.02909 \\ 
2457471.50 & 0.28 & 1.18 & 0.1502 & 0.02935 \\ 
\enddata
\end{deluxetable}

\section{Keplerian solution}

We followed the approach of \citet{tuomi2014} and \citet{butler2017} when analyzing the combined radial velocities of HD~34445. That is, we applied a statistical model accounting for Keplerian signals and red noise, as well as correlations between radial velocities and activity data. Moreover, we searched for the signals in the data with posterior samplings by applying the delayed-rejection adaptive Metropolis Markov chain Monte Carlo algorithm \citep{haario2006,butler2017}. With this technique, we were able to search the whole period space for unique probability maxima corresponding to significant periodic signals. When prominent probability maxima were identified, we estimated the parameters of the corresponding models with $k=0, 1, 2, ...$ Keplerian signals with the common adaptive Metropolis algorithm \citep{haario2001}. When assessing the significances of the Keplerian signals, we applied the Bayesian information criterion \citep[see][]{liddle2007,feng2016} but also tested the significances with a likelihood-ratio test \citep[e.g.][]{butler2017}.

The statistical model \citep[see][]{tuomi2014,butler2017} for the $i$th radial velocity measurement obtained at telescope, $l$, is
\begin{equation}
\begin{split}
    m_{i,l}=\, & f_k(t_i) + \gamma_{0,l} + \gamma_1\,t_i + \gamma_2\,t_{i}^{2} + \epsilon_{i,l} \\
    & + \phi_l \exp\{-\alpha(t_{i-1}-t_{i})\}\epsilon_{i-1,l} +c_l\, \xi_{i,l}\, . 
\end{split}
\end{equation}
This model contains the standard Keplerian parameters ($P$, $K$, $e$, $\omega$, and $M_{o}$) that comprise $f_k(t_i)$, along with a number of nuisance parameters. These nuisance parameters include the individual telescope velocity zero-point offsets, $\gamma_{0,l}$, the second-order polynomial acceleration, with parameters $\gamma_{1}$ (first-order term) and $\gamma_{2}$ (second-order term), parameters quantifying the linear dependence of radial velocities on S-indices, $c_{l}$ (where $\xi_{i,l}$ is the $i$th S-index measurement obtained at telescope, $l$), and moving average parameters $\phi_{l}$. The quantity $\epsilon_{i,l}$ is a Gaussian random variable with zero mean and variance $\sigma_{i,l}^2+\sigma_l^2$, where $\sigma_{i,l}$ is the internal uncertainty estimate on the velocity (as derived from the $i^{\rm th}$ spectra at telescope $l$), and $\sigma_l$ is an additional white noise component that is a free parameter and specific to telescope $l$. The parameters $\epsilon$, $\phi$, and $c$ are assumed to be independent for each instrument $l$. The decay constant $\alpha$ is the inverse of the time-scale $\tau$. As discussed in \citet{butler2017}, we adopt $\tau = 4\,$d. The fit is referenced to epoch $t_0=\rm{JD}~2450000.0$.

The analyses were complemented by calculating likelihood-ratio periodograms for spectroscopic activity indicators and photometric data. These periodograms were calculated as in \citet{butler2017}, i.e. by calculating the likelihood ratio of models with and without a sinusoidal periodicity and by investigating this ratio as a function of the period of the signal. This is essentially a generalized likelihood-ratio version of the Lomb-Scargle periodogram \citep[see e.g.][]{lomb1976,scargle1982,cumming2004,anglada2012} and we could incorporate a linear trend in the baseline model without signals to enable searching for strictly periodic features in the data. The assessment of signal significances was based on $\chi^{2}$ distributed likelihood ratios.

\begin{figure}
\includegraphics[angle=270, width=0.23\textwidth,clip,trim=3cm 0 0 3cm]{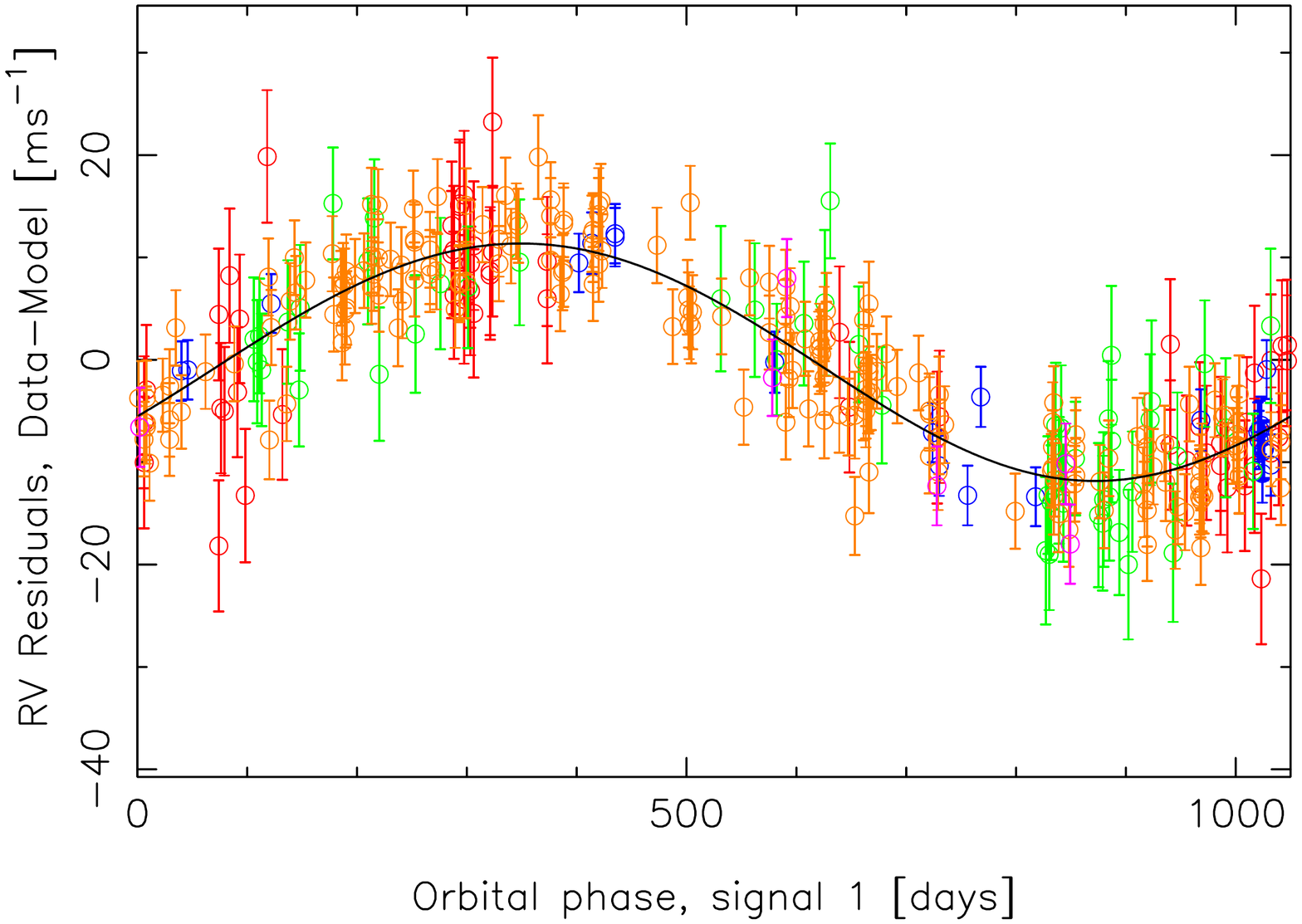}
\includegraphics[angle=270, width=0.23\textwidth,clip,trim=3cm 0 0 3cm]{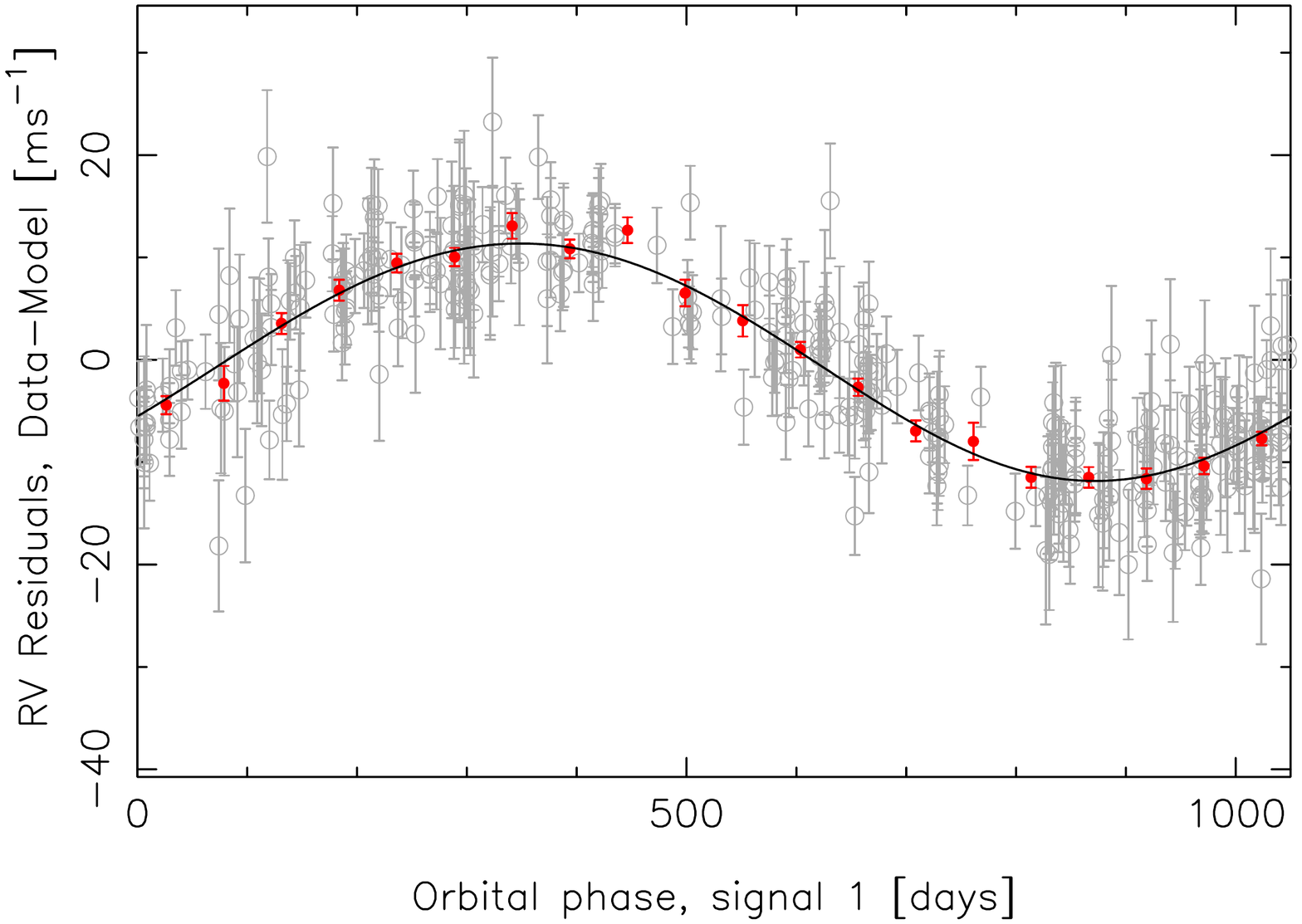}

\includegraphics[angle=270, width=0.23\textwidth,clip,trim=3cm 0 0 3cm]{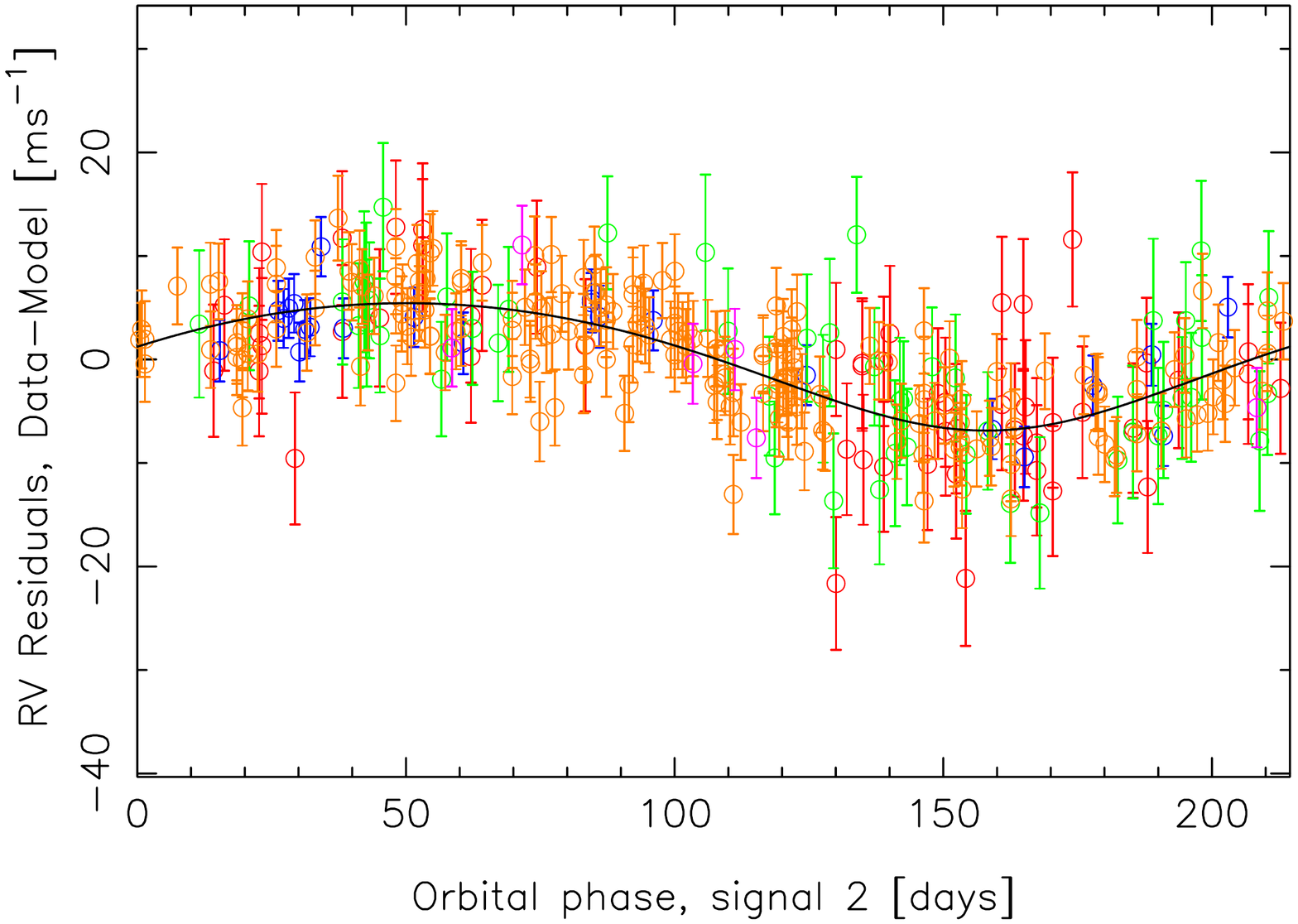}
\includegraphics[angle=270, width=0.23\textwidth,clip,trim=3cm 0 0 3cm]{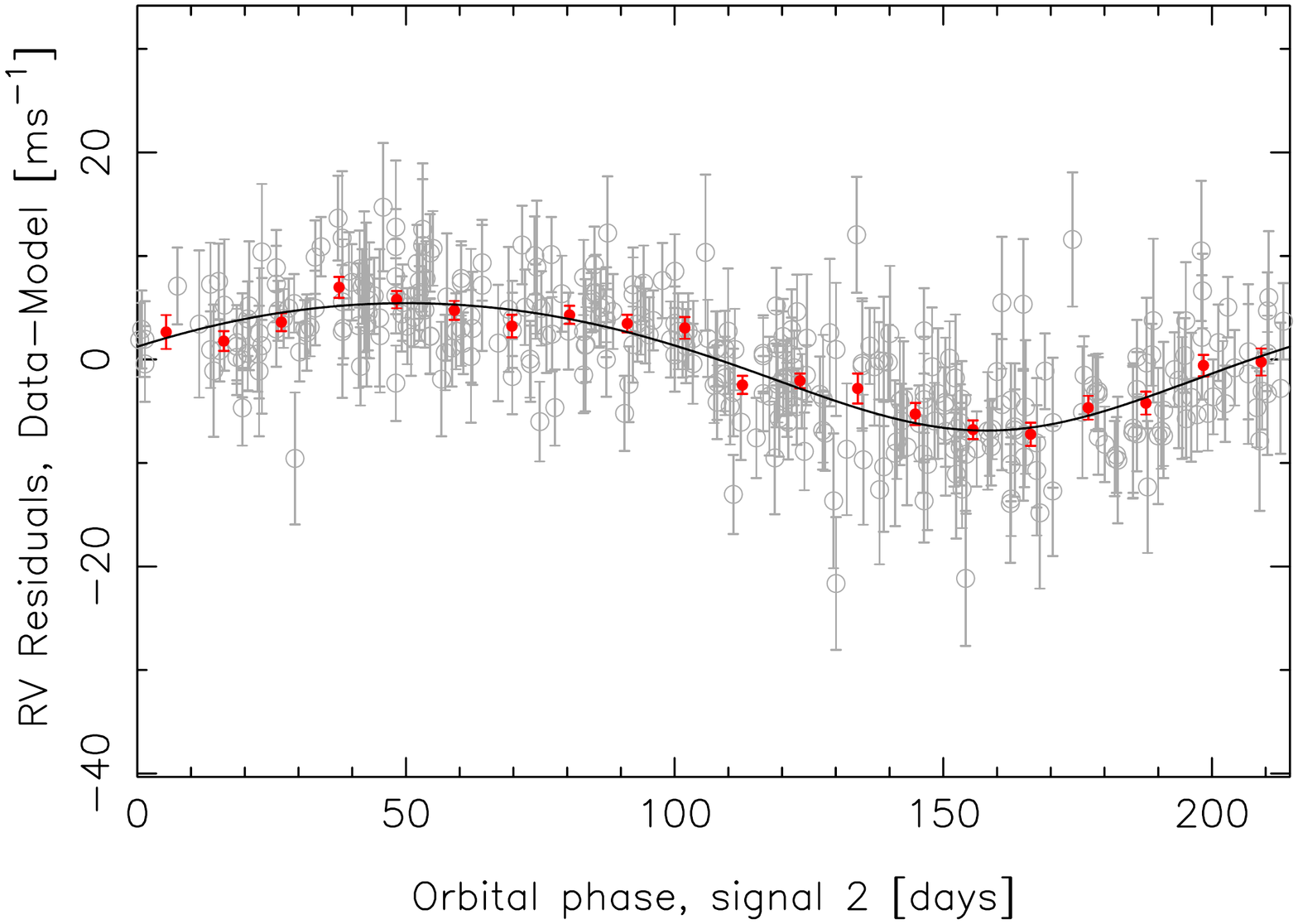}

\includegraphics[angle=270, width=0.23\textwidth,clip,trim=3cm 0 0 3cm]{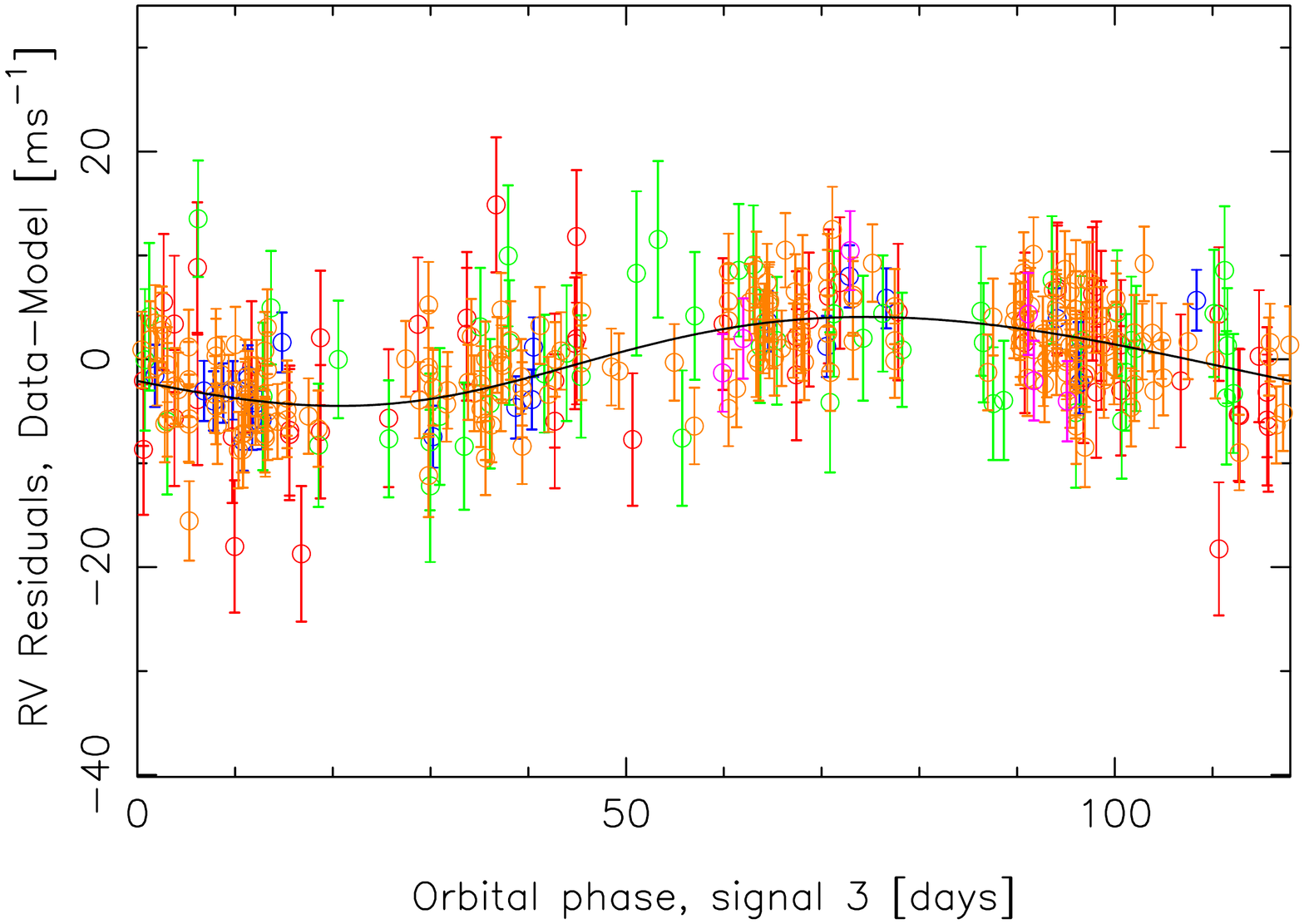}
\includegraphics[angle=270, width=0.23\textwidth,clip,trim=3cm 0 0 3cm]{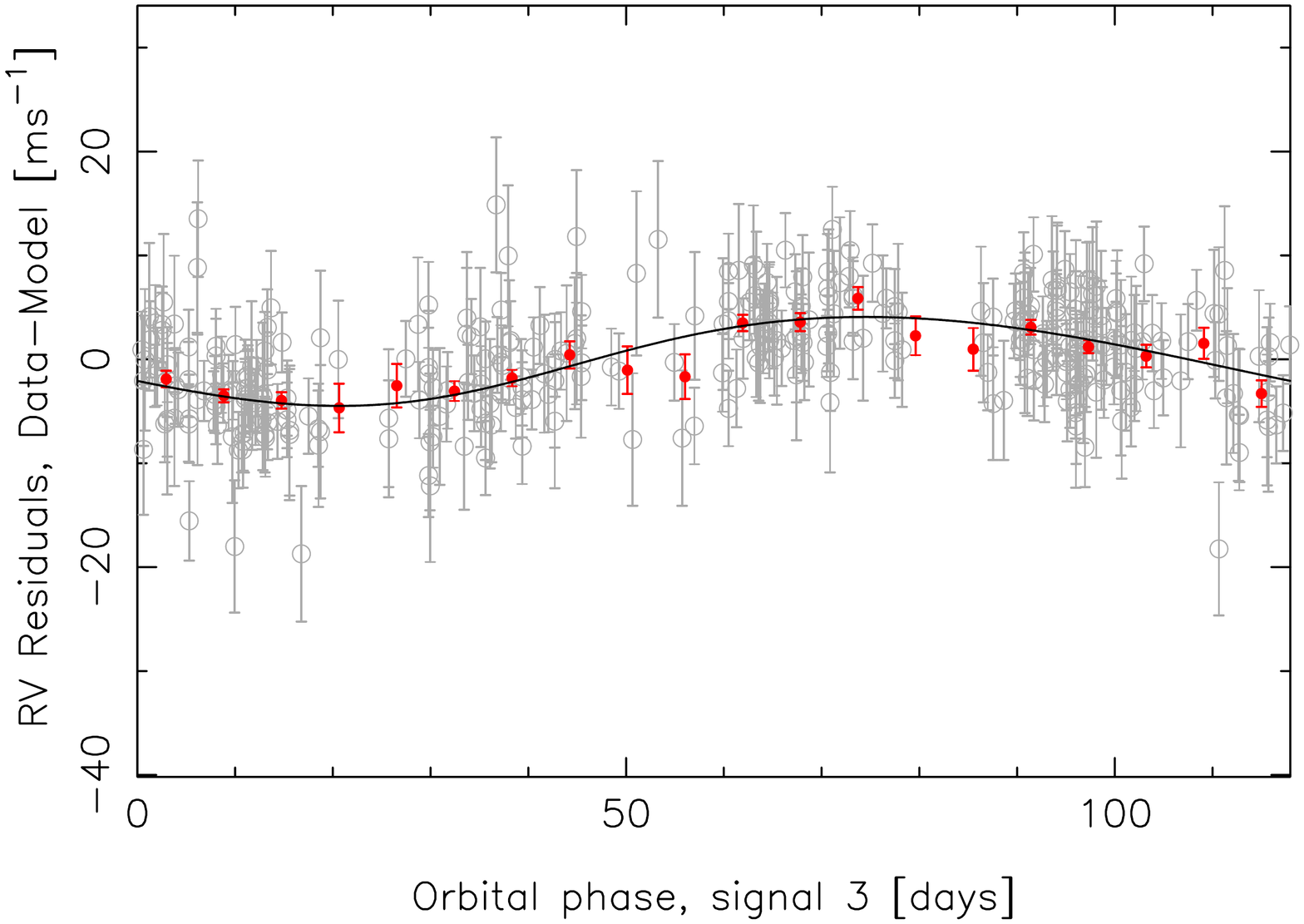}

\includegraphics[angle=270, width=0.23\textwidth,clip,trim=3cm 0 0 3cm]{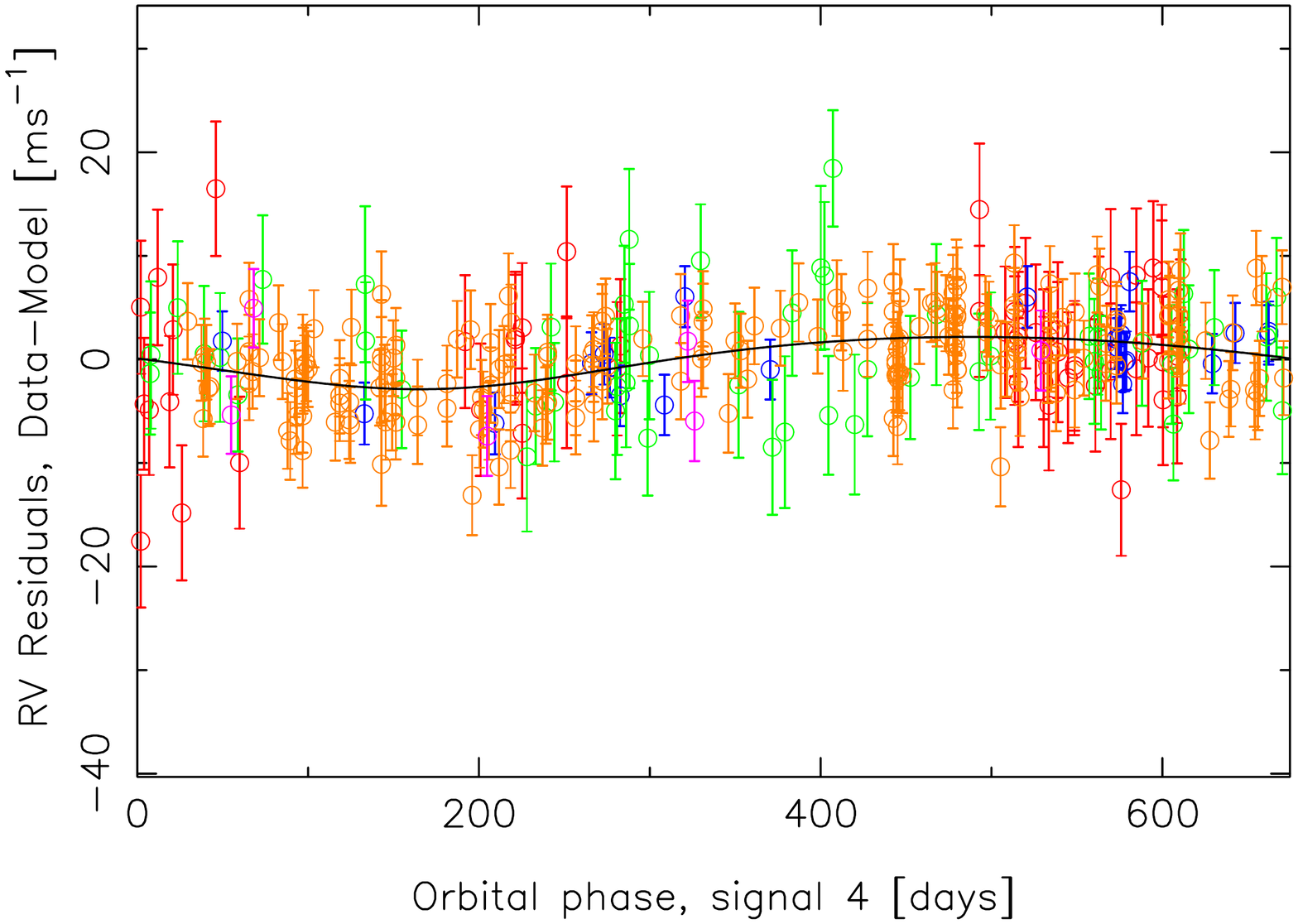}
\includegraphics[angle=270, width=0.23\textwidth,clip,trim=3cm 0 0 3cm]{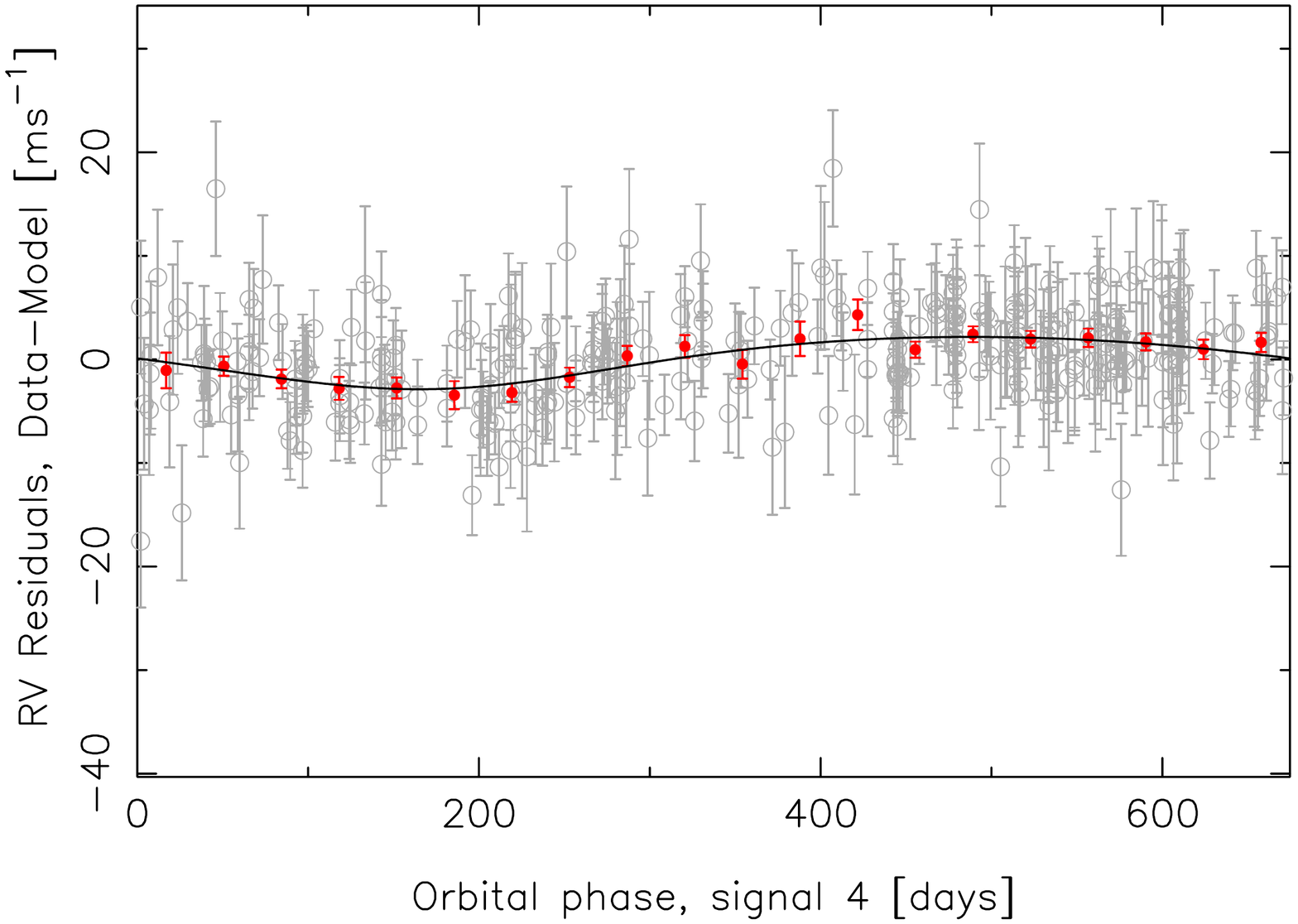}

\includegraphics[angle=270, width=0.23\textwidth,clip,trim=3cm 0 0 3cm]{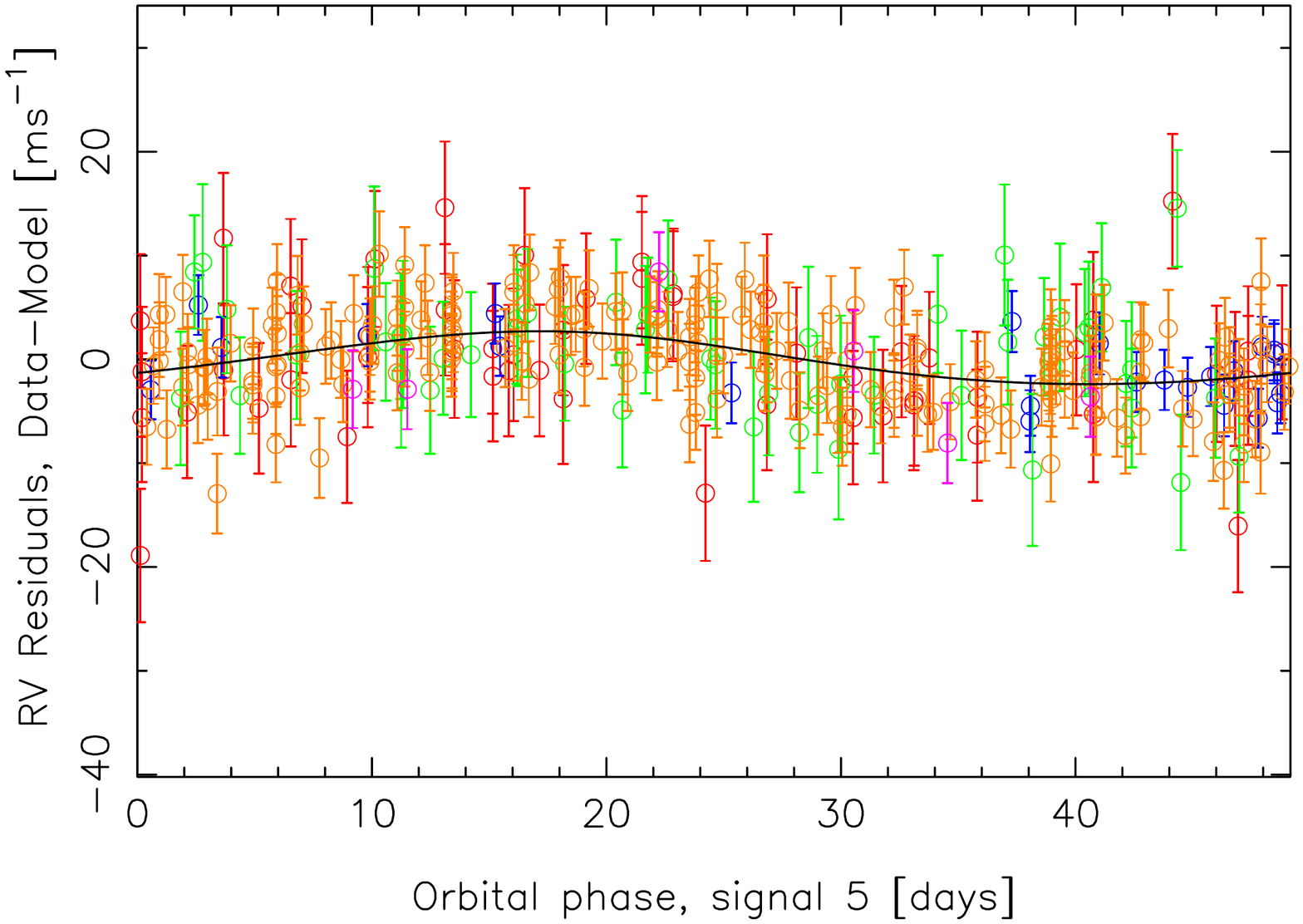}
\includegraphics[angle=270, width=0.23\textwidth,clip,trim=3cm 0 0 3cm]{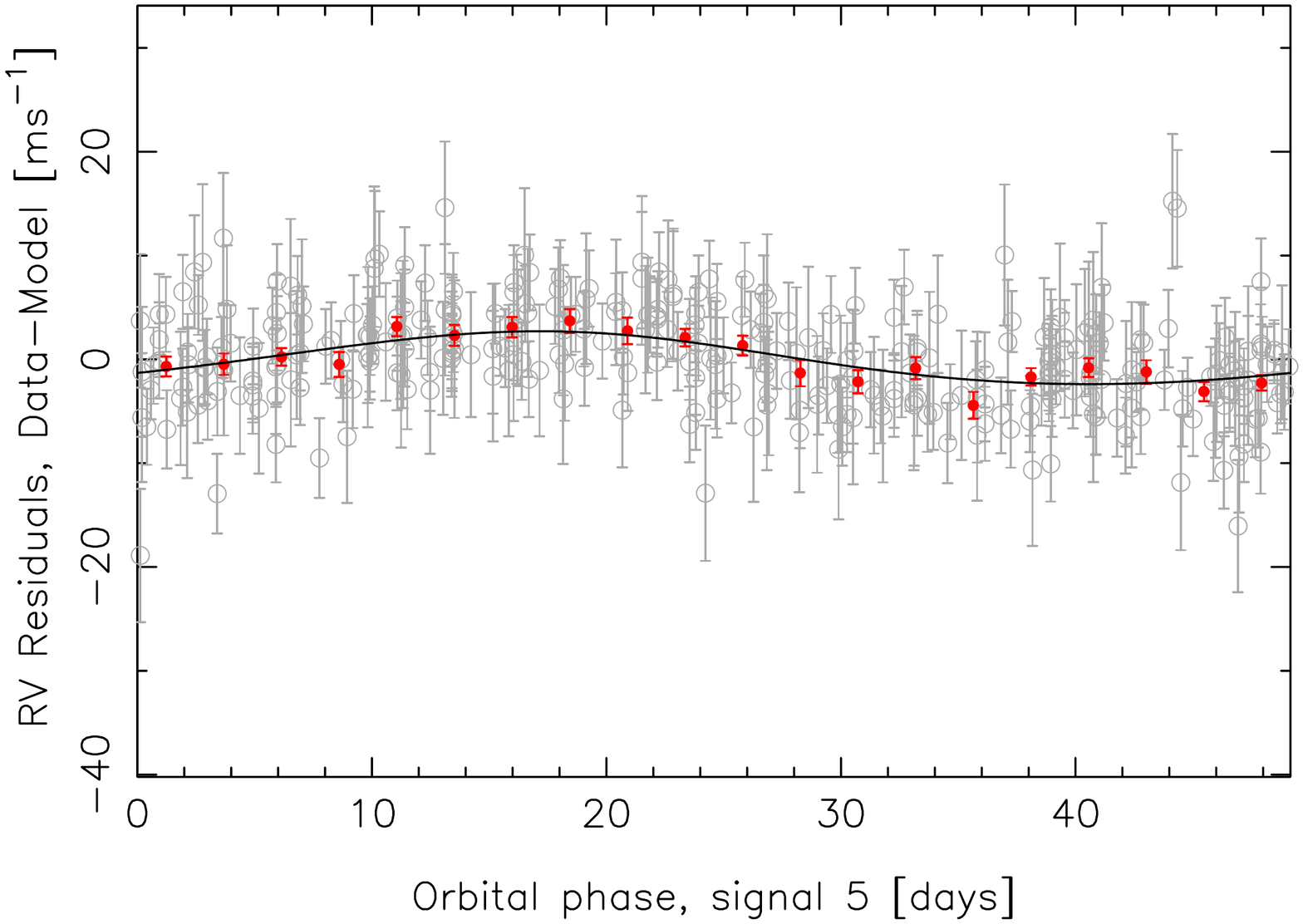}

\caption{Combined APF (red), PFS (blue), HET10 (green), KECK (orange), and KECK10 (purple) radial velocities folded on the phases of the five signals with the other four signals subtracted from each panel. Left panels show all the data demonstrating different instruments with different colour codes whereas right panels show the velocities binned into 20 bins denoted by red filled circles. The black solid curves denote the maximum \emph{a posteriori} Keplerian models.}\label{fig:phased}
\end{figure}

\begin{figure}
\center
\includegraphics[angle=270, width=0.45\textwidth,clip,trim=4cm 0 1cm 0]{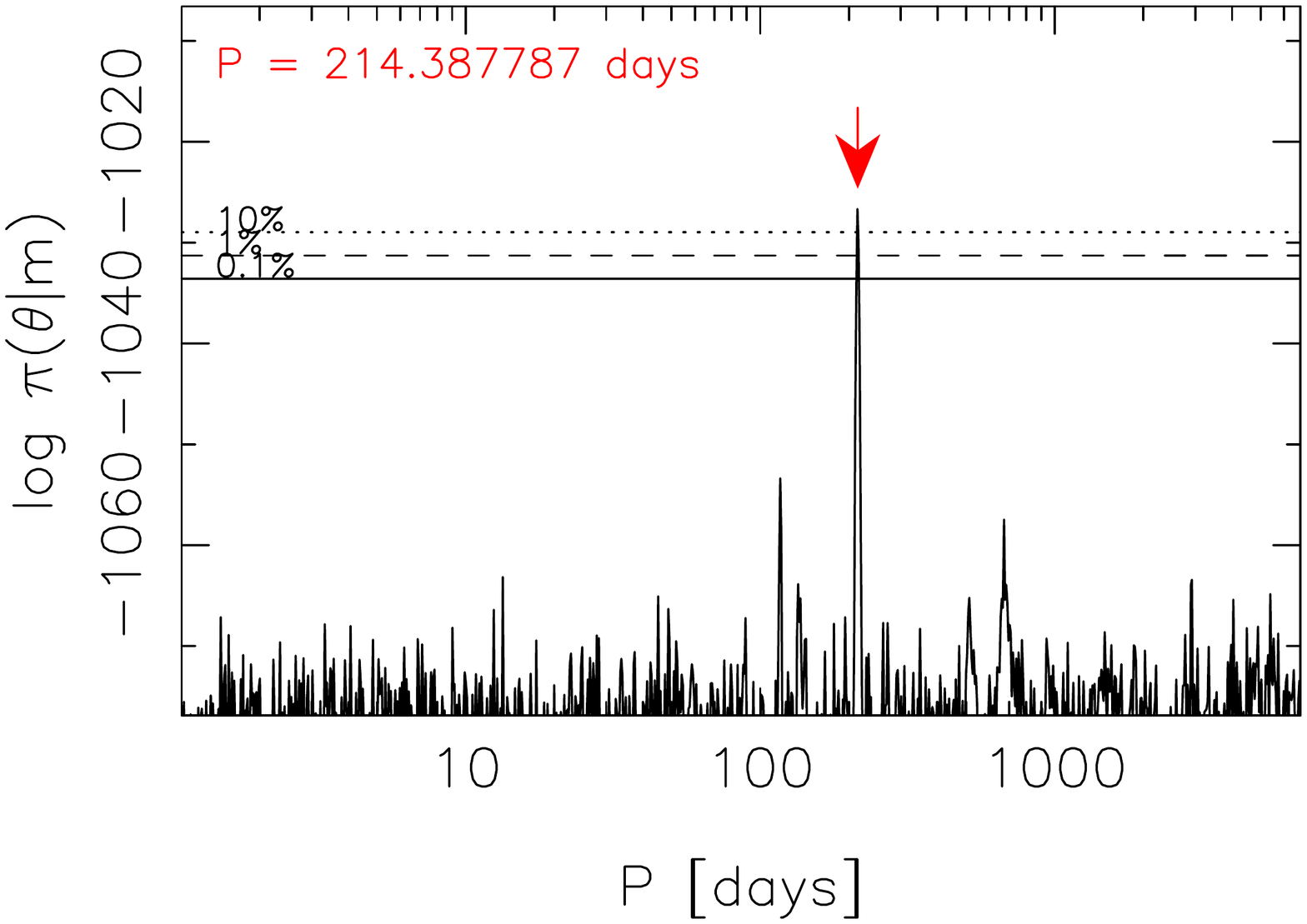}

\includegraphics[angle=270, width=0.45\textwidth,clip,trim=4cm 0 1cm 0]{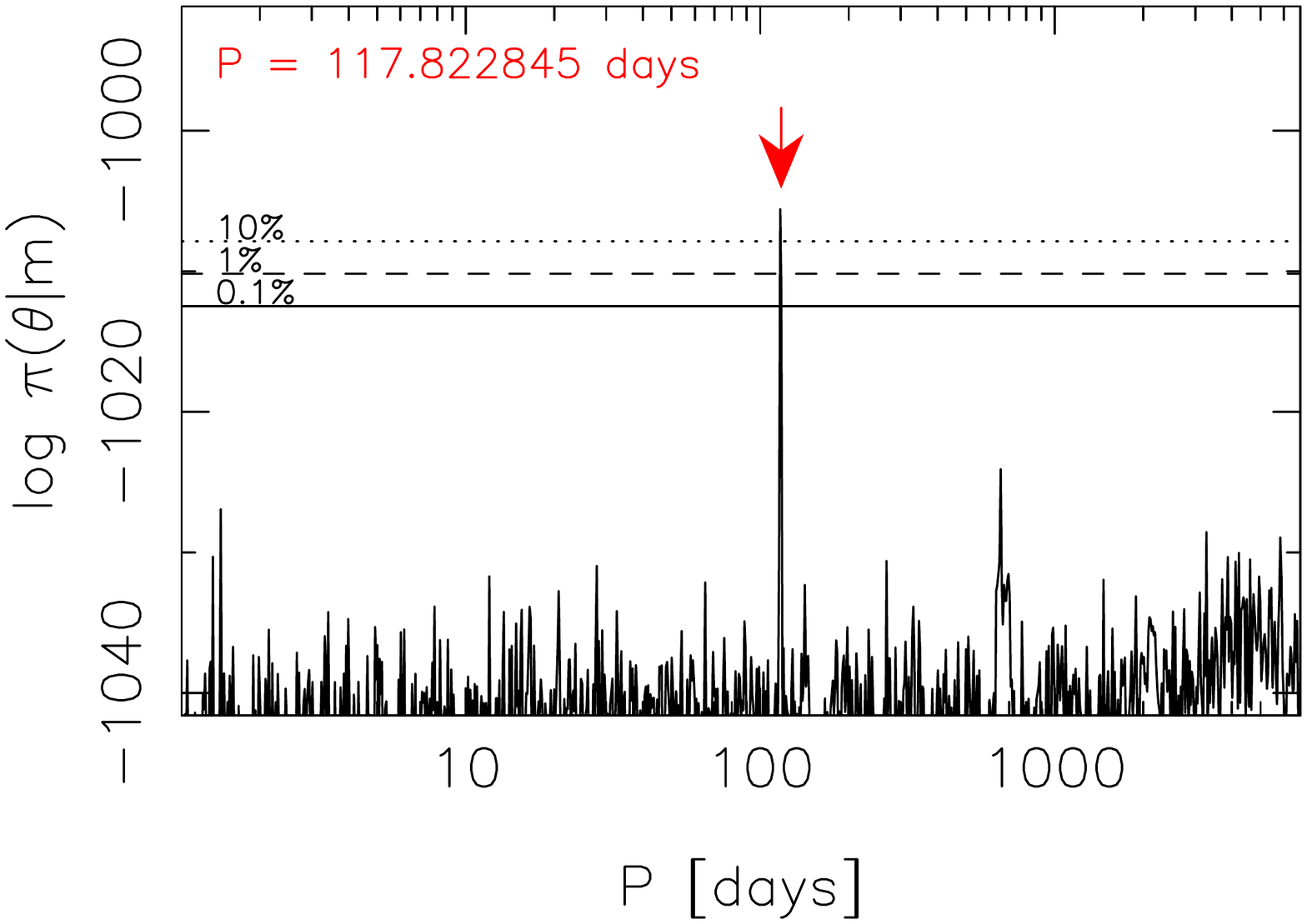}

\includegraphics[angle=270, width=0.45\textwidth,clip,trim=4cm 0 1cm 0]{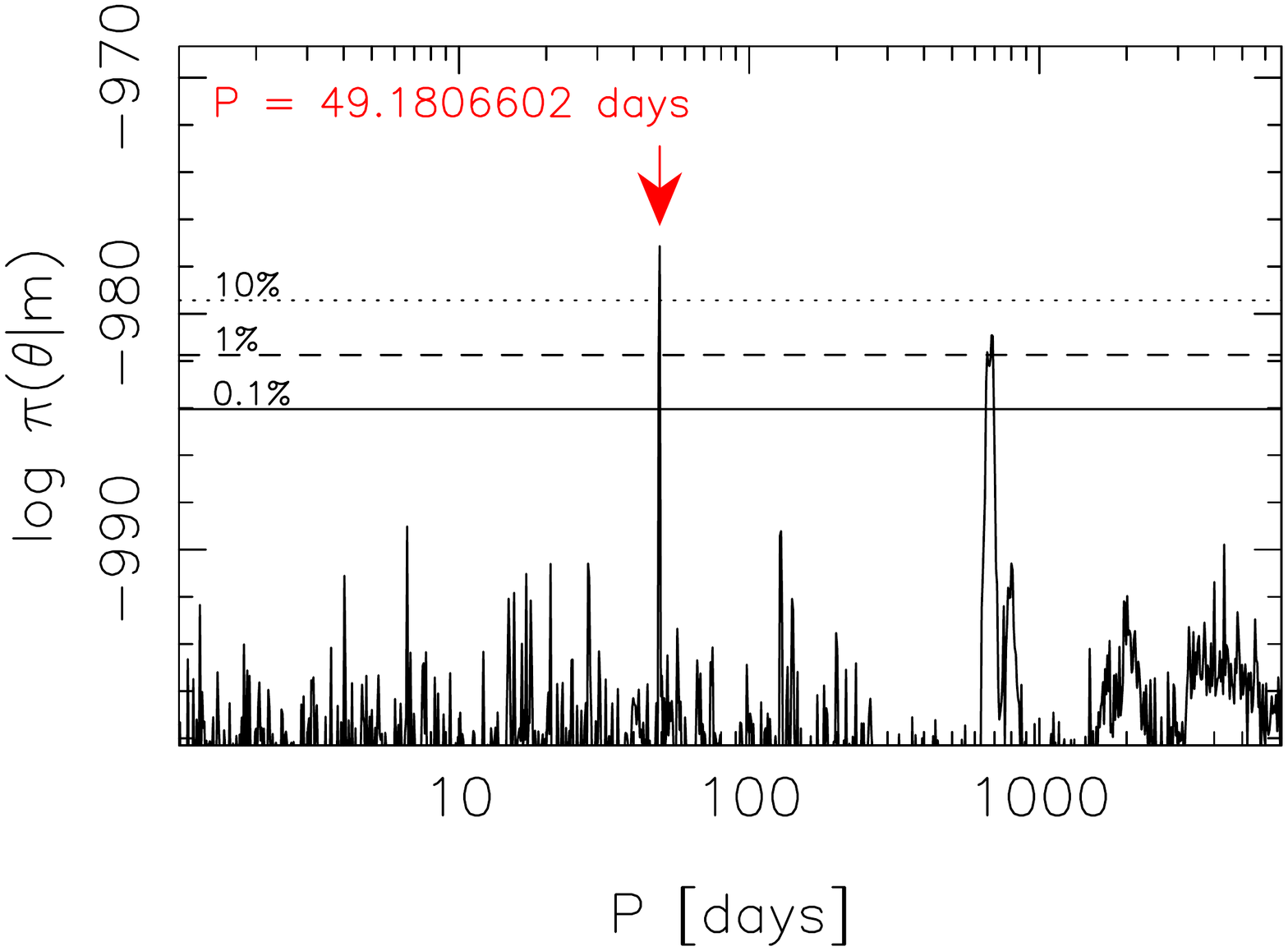}

\includegraphics[angle=270, width=0.45\textwidth,clip,trim=4cm 0 1cm 0]{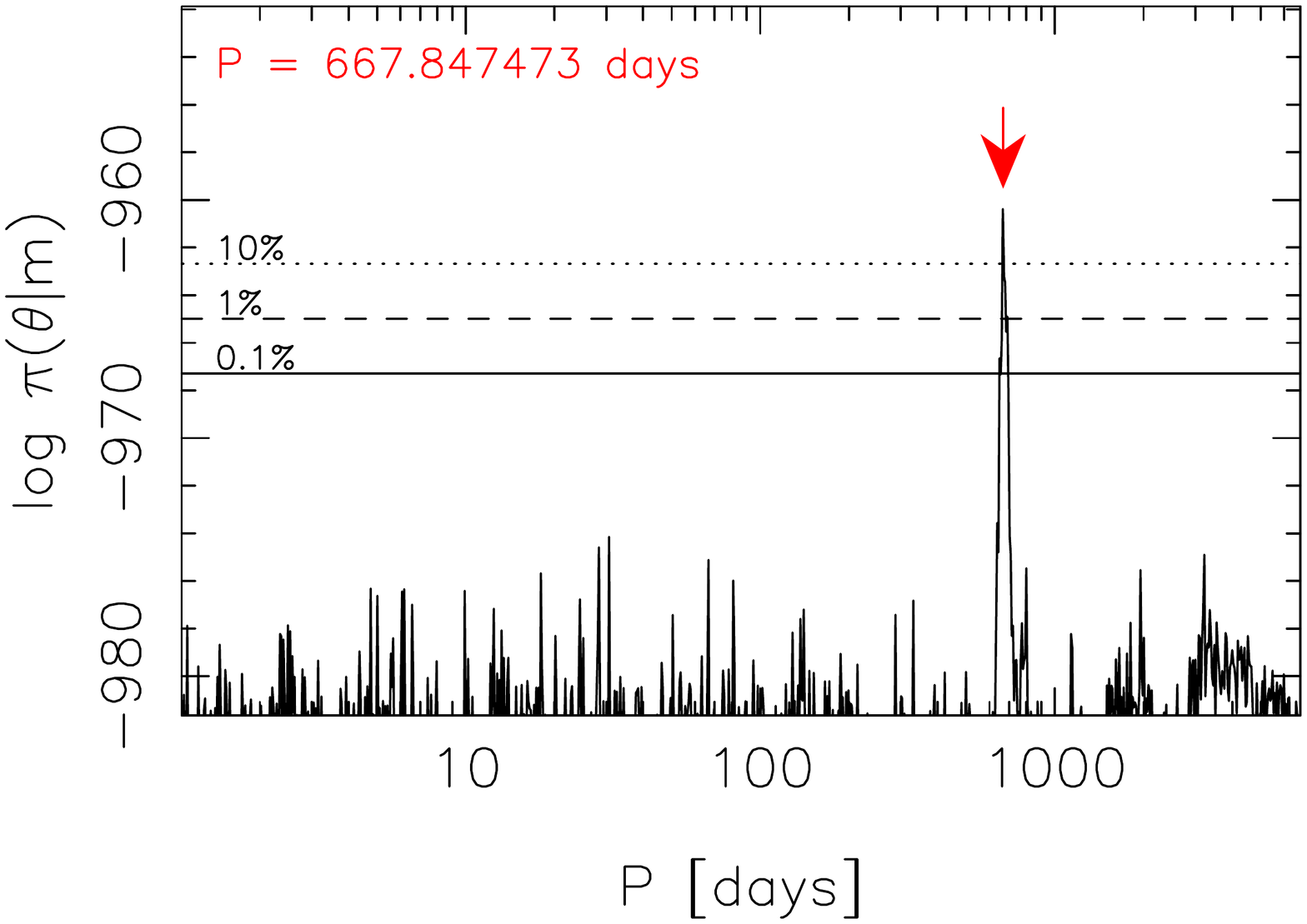}
\caption{Estimated posterior probability densities of models with $k$ Keplerian signals as functions of the period parameter of the $k$th signal for $k = 2, ..., 5$. The red arrows denote the global probability maxima in the period space whereas the horizontal lines indicate the 10\% (dotted), 1\% (dashed), and 0.1\% (solid) equiprobability thresholds with respect to the maxima indicating that all the detected signals correspond to unique probability maxima in the period space.}\label{fig:searches}
\end{figure}

Our analysis reveals evidence for at least five Keplerian signals in the combined APF, PFS, LCES/HIRES, HET, and CPS/HIRES data (Table \ref{tab:likelihoods} and Figs. \ref{fig:phased} and \ref{fig:searches}). The most obvious periodic signal, already reported by \citet{Howard2010}, was easy to find in all periodograms and posterior sampling analyses of the data. The other four signals are presented as unique posterior probability maxima in Fig. \ref{fig:searches}. We present the phase-folded radial velocities from all the instruments in Fig.~\ref{fig:phased} for all the signals, illustrating visually that, although some of the signals are weak, they are well-supported by all instruments. In addition to the signals, we observed polynomial acceleration in the data possibly corresponding to a long-period companion to the star. All the model parameters for the 5-planet model are presented in Table \ref{tab:parameters}.

\begin{table*}
\caption{Parameter estimates of the statistical model with five Keplerian signals. The maximum \emph{a posteriori} point estimates and 99\% Bayesian credibility intervals.}\label{tab:parameters}
\begin{minipage}{\textwidth}
\begin{center}
\begin{tabular}{lcccccc}
\hline \hline
Parameter & HD 34445 b & HD 34445 c & HD 34445 d & HD 34445 e & HD34445 f \\
\hline
$P$ (days) & 1054.1 [1041.4, 1068.1] & 214.66 [213.35, 215.84] & 117.76 [117.27, 118.31] & 680.1 [653.5, 698.2] & 49.192 [49.064, 49.319] \\
$K$ (ms$^{-1}$) & 11.92 [10.56, 13.40] & 5.75 [4.28, 7.21] & 3.82 [2.49, 5.16] & 2.77 [1.46, 3.94] & 2.76 [1.33, 3.96] \\
$e$ & 0.014 [0, 0.116] & 0.041 [0, 0.223] & 0.016 [0, 0.237] & 0.030 [0, 0.263] & 0.018 [0, 0.245] \\
$\omega$ (rad) & 2.8 [0, 2$\pi$] & 3.0 [0, 2$\pi$] & 4.1 [0, 2$\pi$] & 3.9 [0, 2$\pi$] & 0.5 [0, 2$\pi$] \\
$M_{0}$ (rad) & 4.9 [0, 2$\pi$] & 4.5 [0, 2$\pi$] & 2.0 [0, 2$\pi$] & 2.0 [0, 2$\pi$] & 2.2 [0, 2$\pi$] \\
\hline
$\gamma_{1}$ (ms$^{-1}$year$^{-1}$) & 2.21 [0.92, 3.63] \\
$\gamma_{2}$ (ms$^{-1}$year$^{-2}$) & -0.122 [-0.178, -0.060] \\
$\sigma_{\rm APF}$ (ms$^{-1}$) & 5.77 [4.31, 8.27] \\
$\sigma_{\rm PFS}$ (ms$^{-1}$) & 3.60 [1.94, 6.23] \\
$\sigma_{\rm HET10}$ (ms$^{-1}$) & 4.65 [3.30, 6.90] \\
$\sigma_{\rm KECK}$ (ms$^{-1}$) & 3.43 [2.88, 4.09] \\
$\sigma_{\rm KECK10}$ (ms$^{-1}$) & 4.98 [1.85, 11.46] \\
$\phi_{\rm APF}$ & 0.83 [0.25, 1] \\
$\phi_{\rm PFS}$ & 0.70 [-1, 1] \\
$\phi_{\rm HET10}$ & 0.30 [-0.66, 1] \\
$\phi_{\rm KECK}$ & 0.61 [0.27, 0.85] \\
$\phi_{\rm KECK10}$ & 0.70 [-1, 1] \\
$c_{\rm S,APF}$ (ms$^{-1}$) & -8.94 [-22.48, 6.03] \\
$c_{\rm S,PFS}$ (ms$^{-1}$) & -13.8 [-49.9, 16.6] \\
$c_{\rm S,KECK}$ (ms$^{-1}$) & 195 [21, 369] \\
$\gamma_{0,\rm APF}$ (ms$^{-1}$) & 5.77 [4.31, 8.27] \\
$\gamma_{0,\rm PFS}$ (ms$^{-1}$) & 3.60 [1.94, 6.22] \\
$\gamma_{0,\rm HET10}$ (ms$^{-1}$) & 4.65 [3.30, 6.90] \\
$\gamma_{0,\rm KECK}$ (ms$^{-1}$) & 3.43 [2.88, 4.09] \\
$\gamma_{0,\rm KECK10}$ (ms$^{-1}$) & 4.98 [1.85, 11.46]\\
\hline \hline
\end{tabular}
\end{center}
\end{minipage}
\end{table*}

\begin{figure*}
\center
\includegraphics[angle=270, width=0.23\textwidth,clip,trim=4cm 0 0 0]{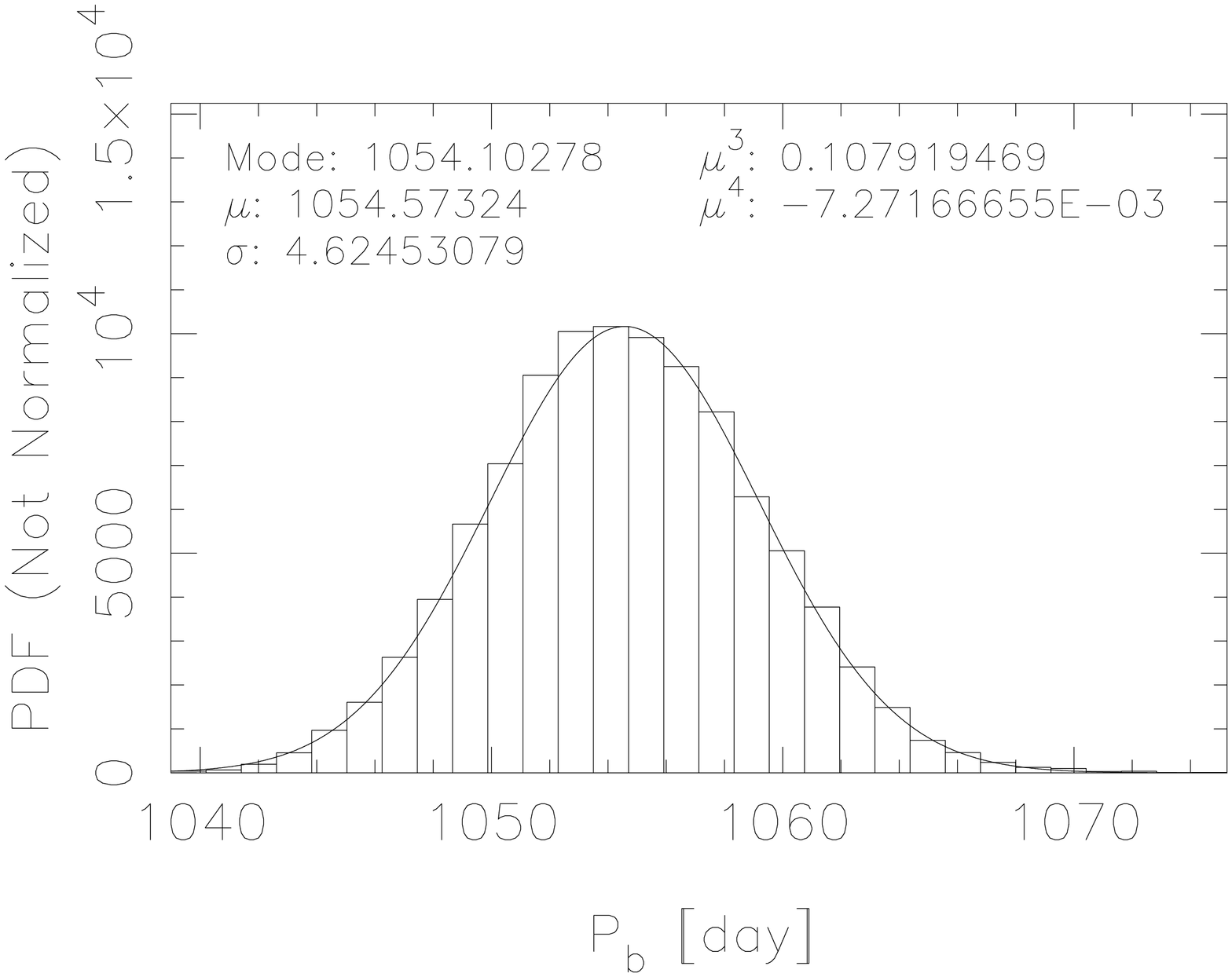}
\includegraphics[angle=270, width=0.23\textwidth,clip,trim=4cm 0 0 0]{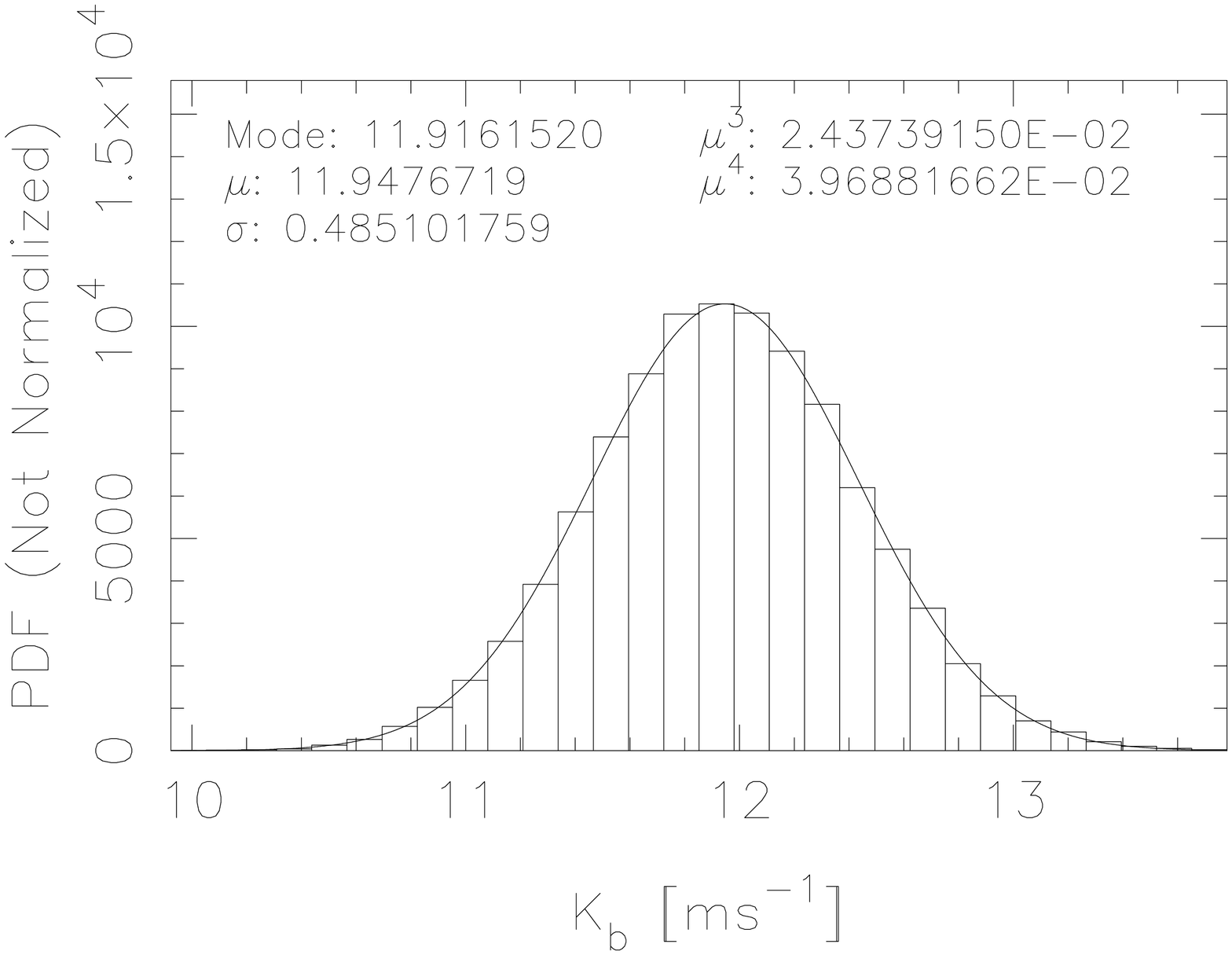}
\includegraphics[angle=270, width=0.23\textwidth,clip,trim=4cm 0 0 0]{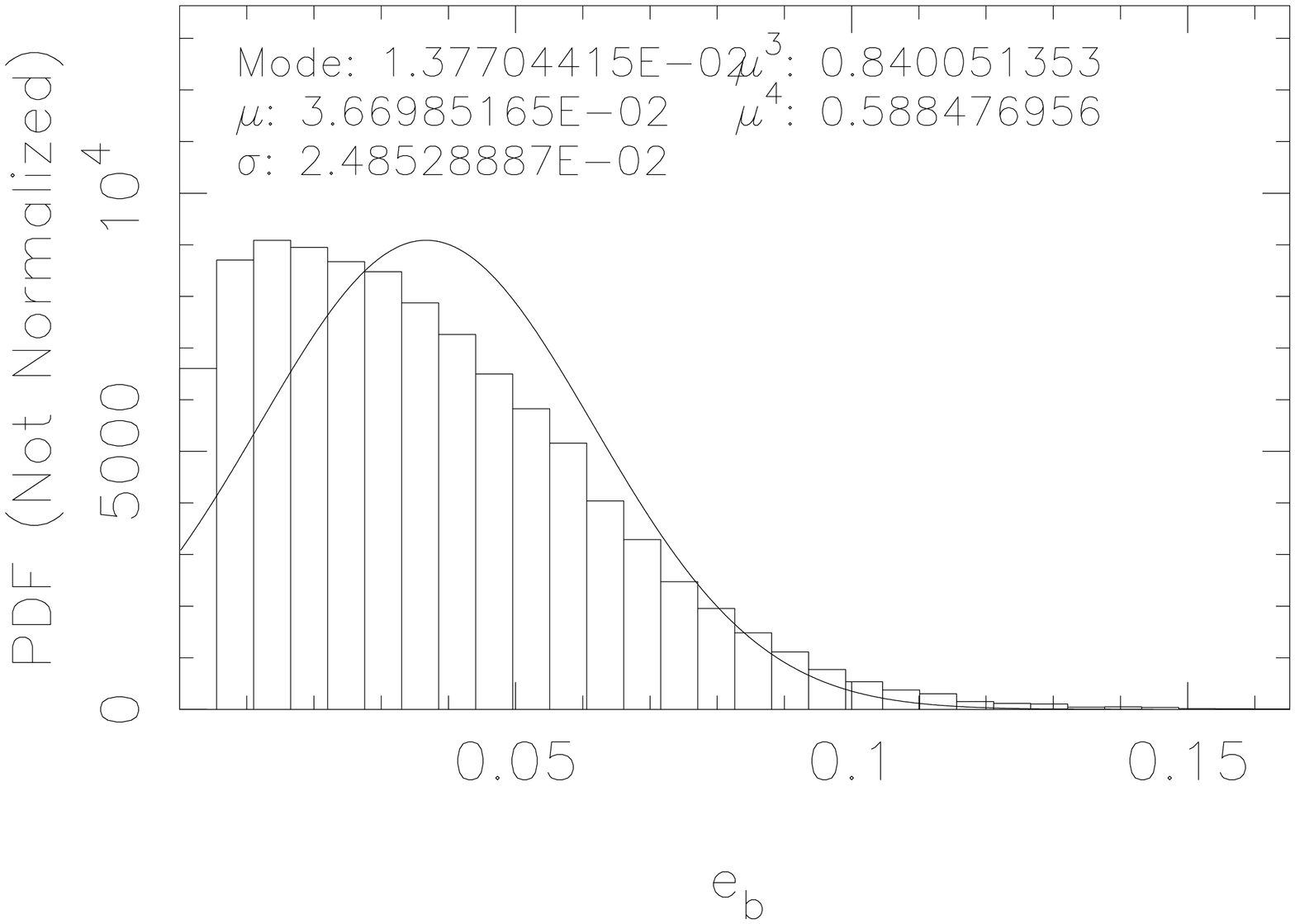}
\includegraphics[angle=270, width=0.23\textwidth,clip,trim=4cm 0 0 0]{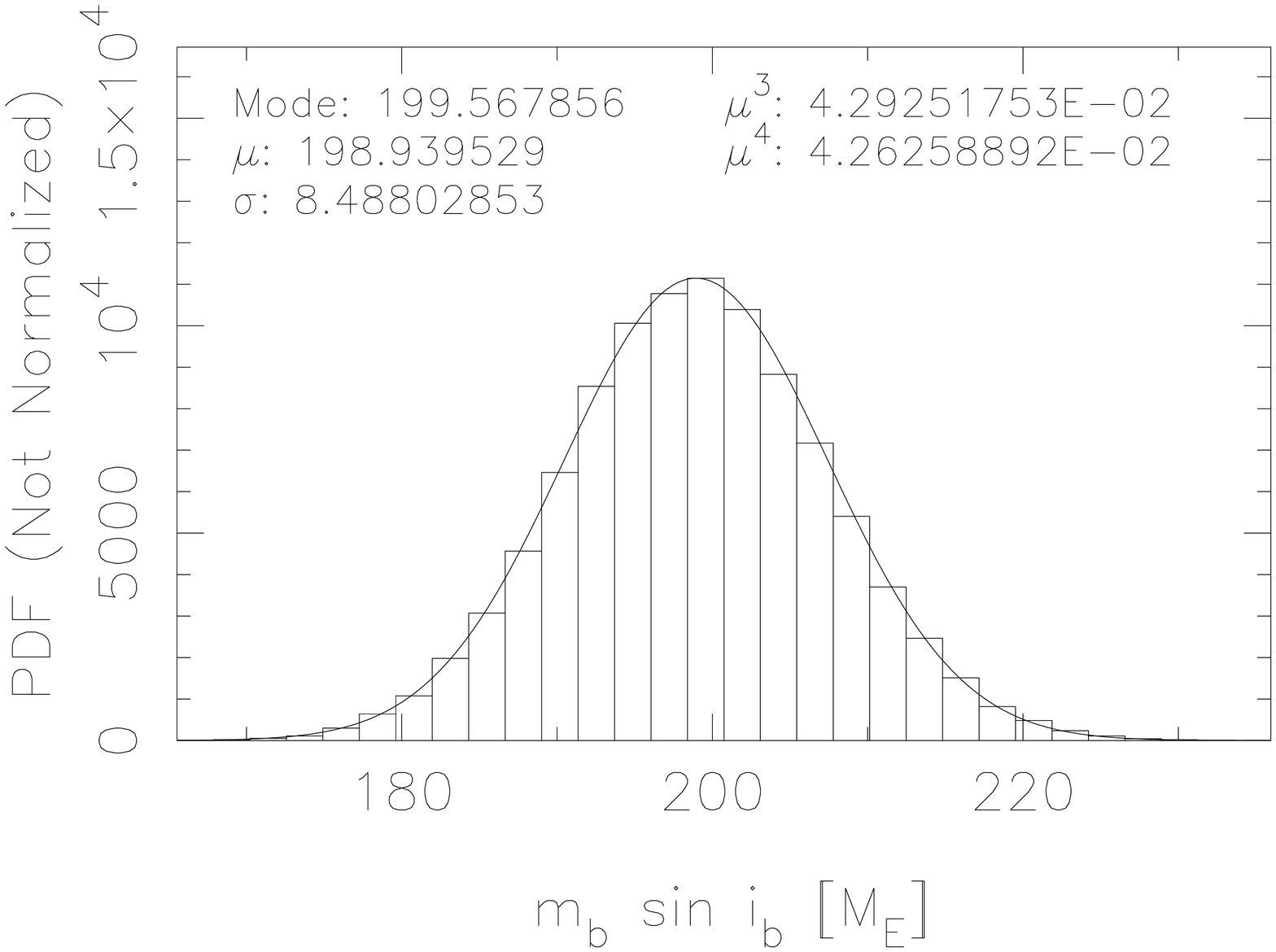}

\includegraphics[angle=270, width=0.23\textwidth,clip,trim=4cm 0 0 0]{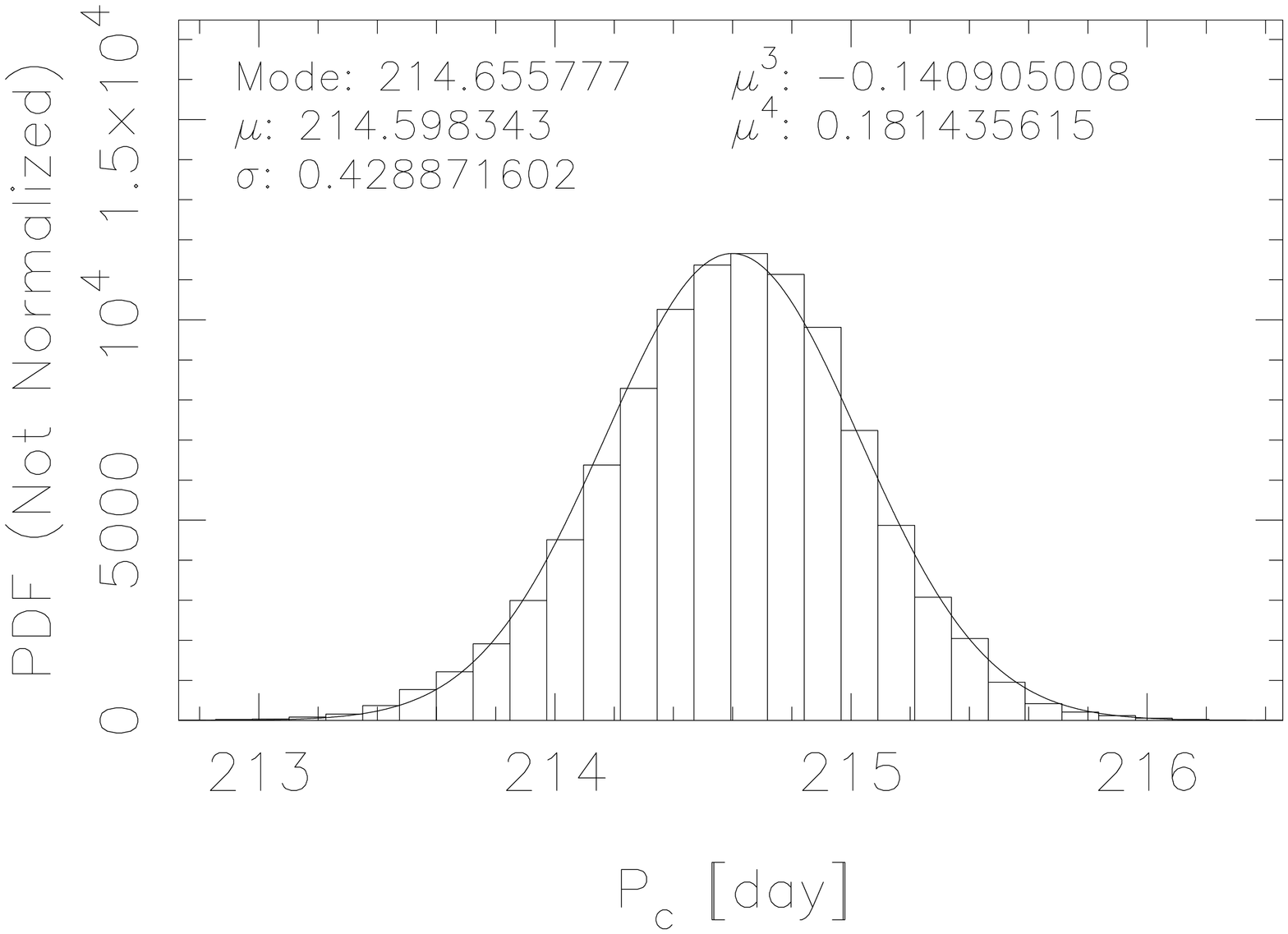}
\includegraphics[angle=270, width=0.23\textwidth,clip,trim=4cm 0 0 0]{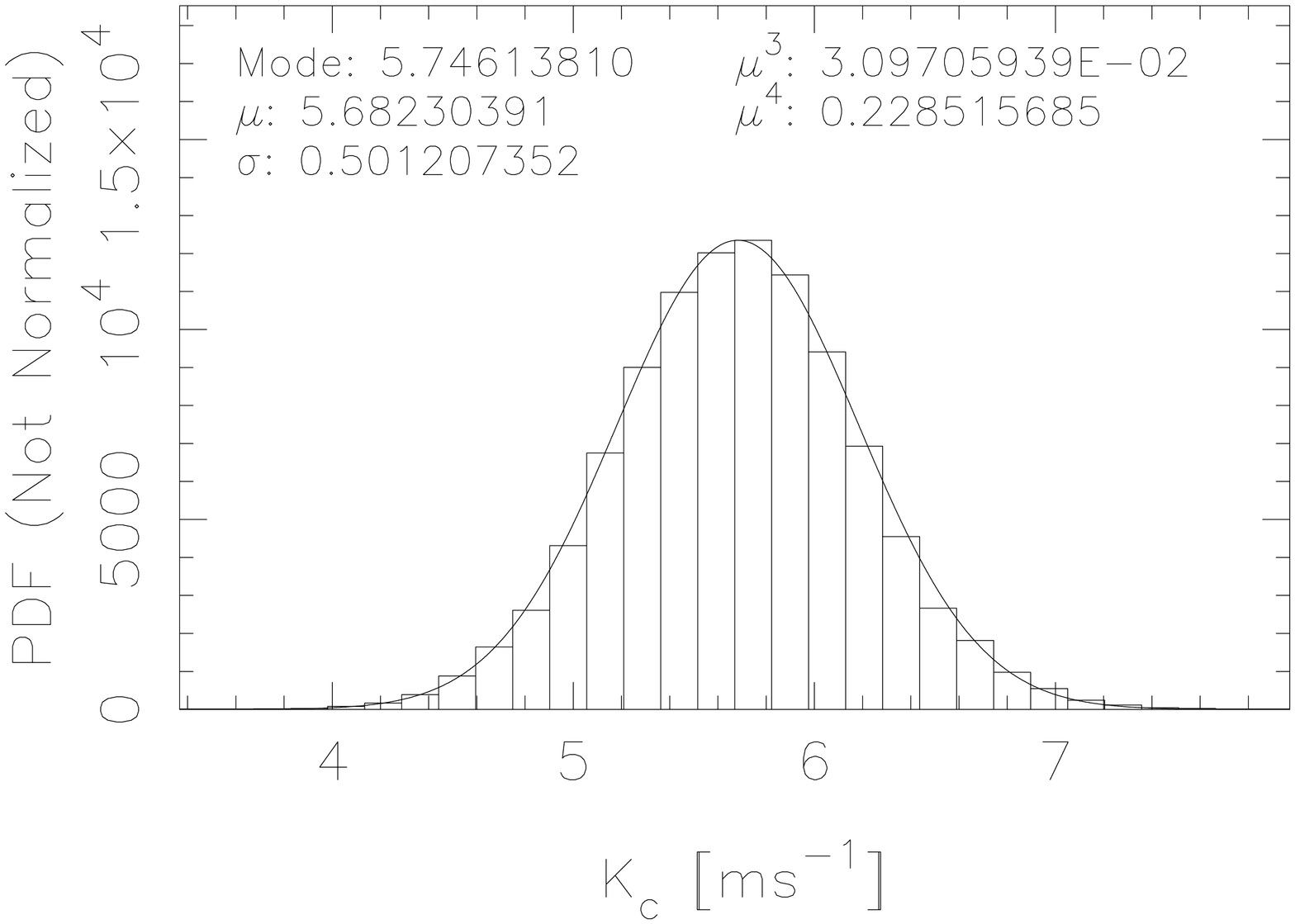}
\includegraphics[angle=270, width=0.23\textwidth,clip,trim=4cm 0 0 0]{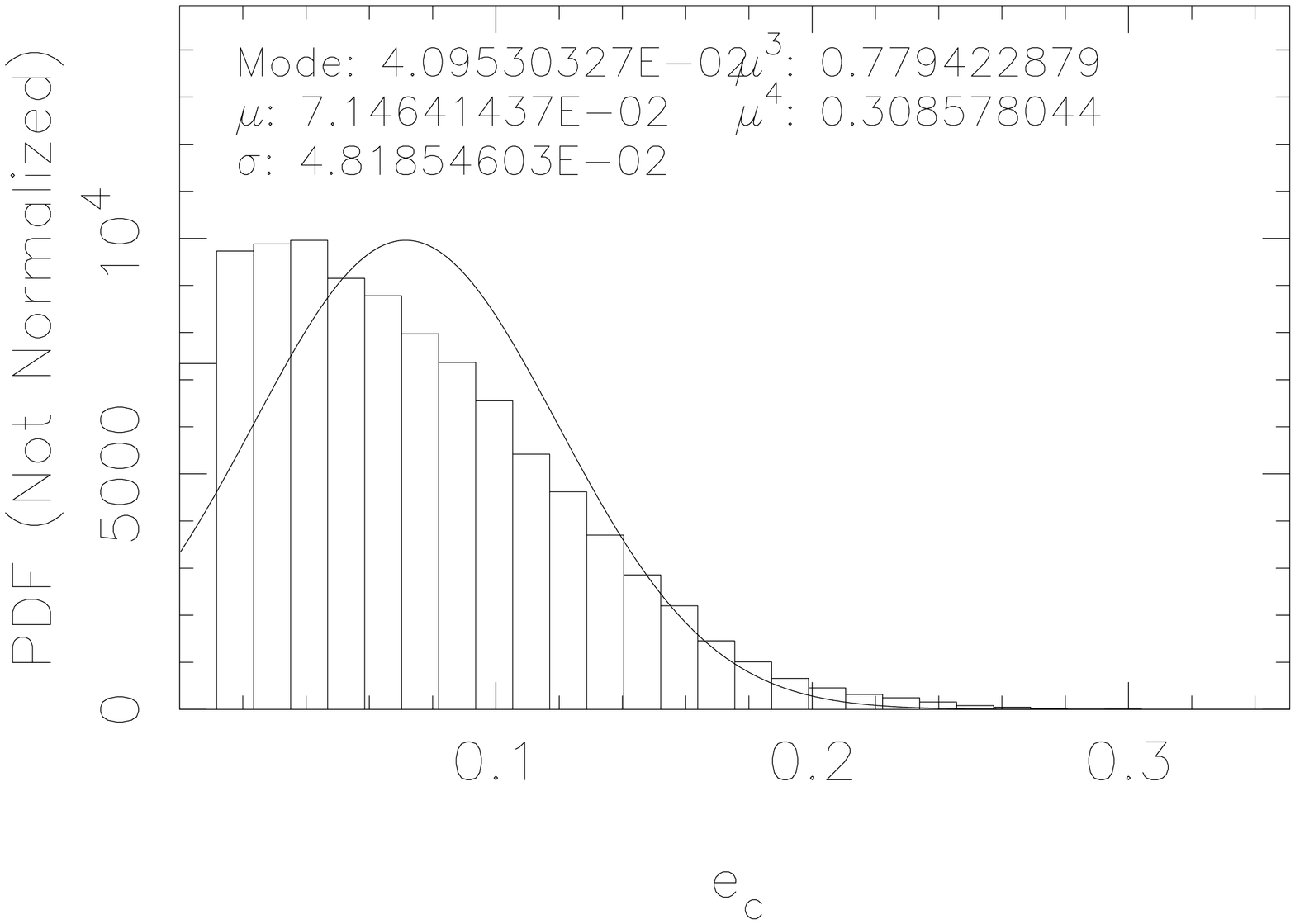}
\includegraphics[angle=270, width=0.23\textwidth,clip,trim=4cm 0 0 0]{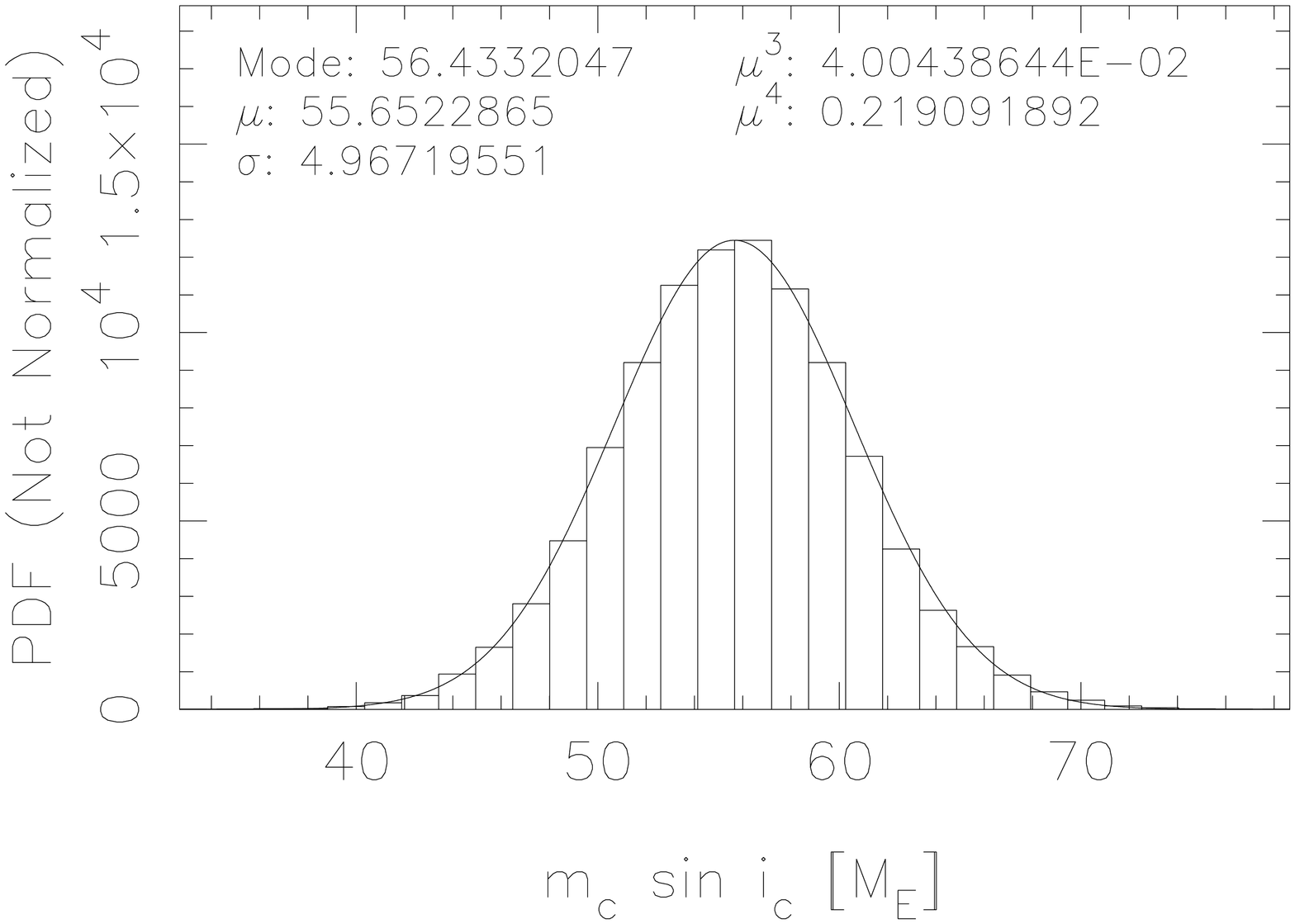}

\includegraphics[angle=270, width=0.23\textwidth,clip,trim=4cm 0 0 0]{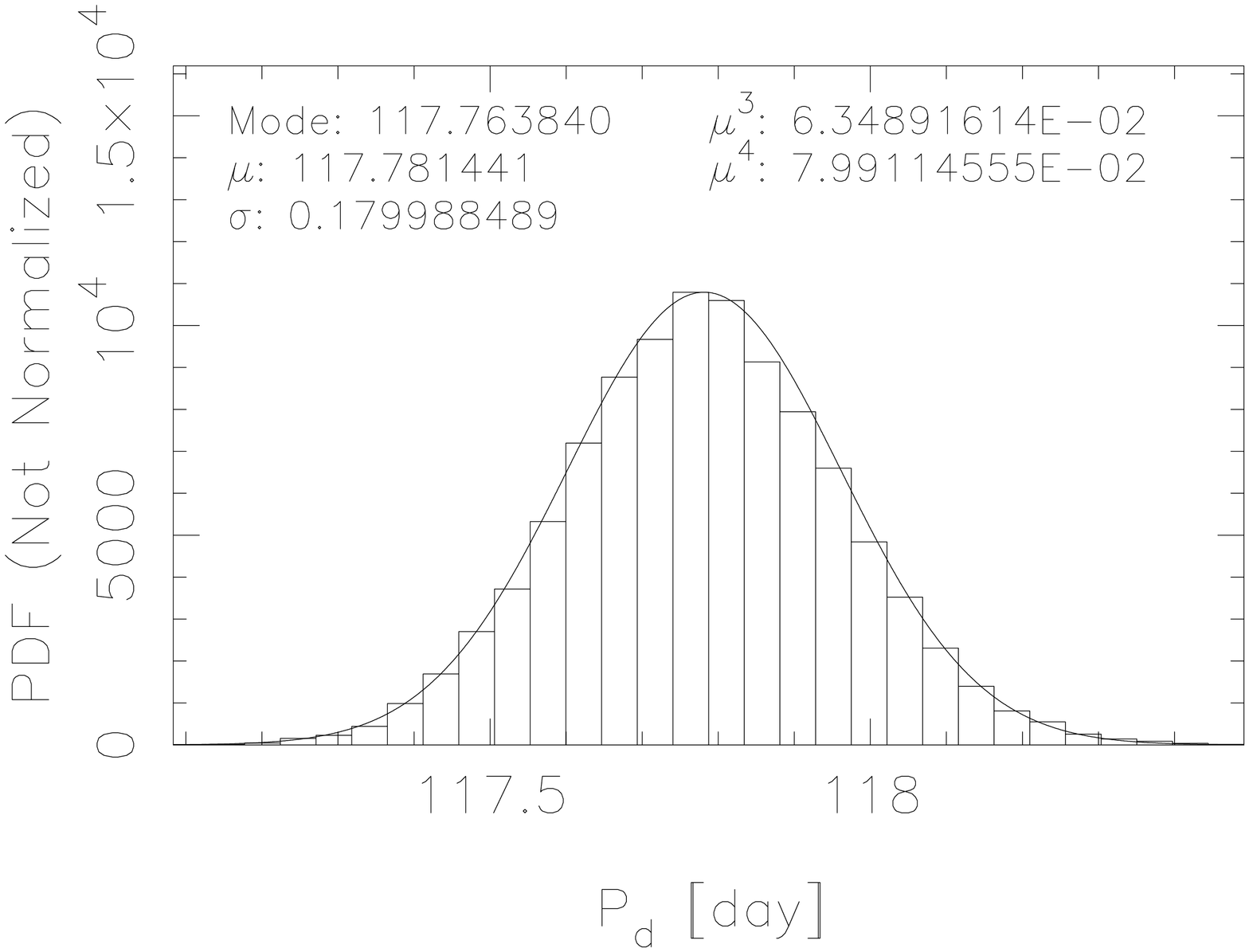}
\includegraphics[angle=270, width=0.23\textwidth,clip,trim=4cm 0 0 0]{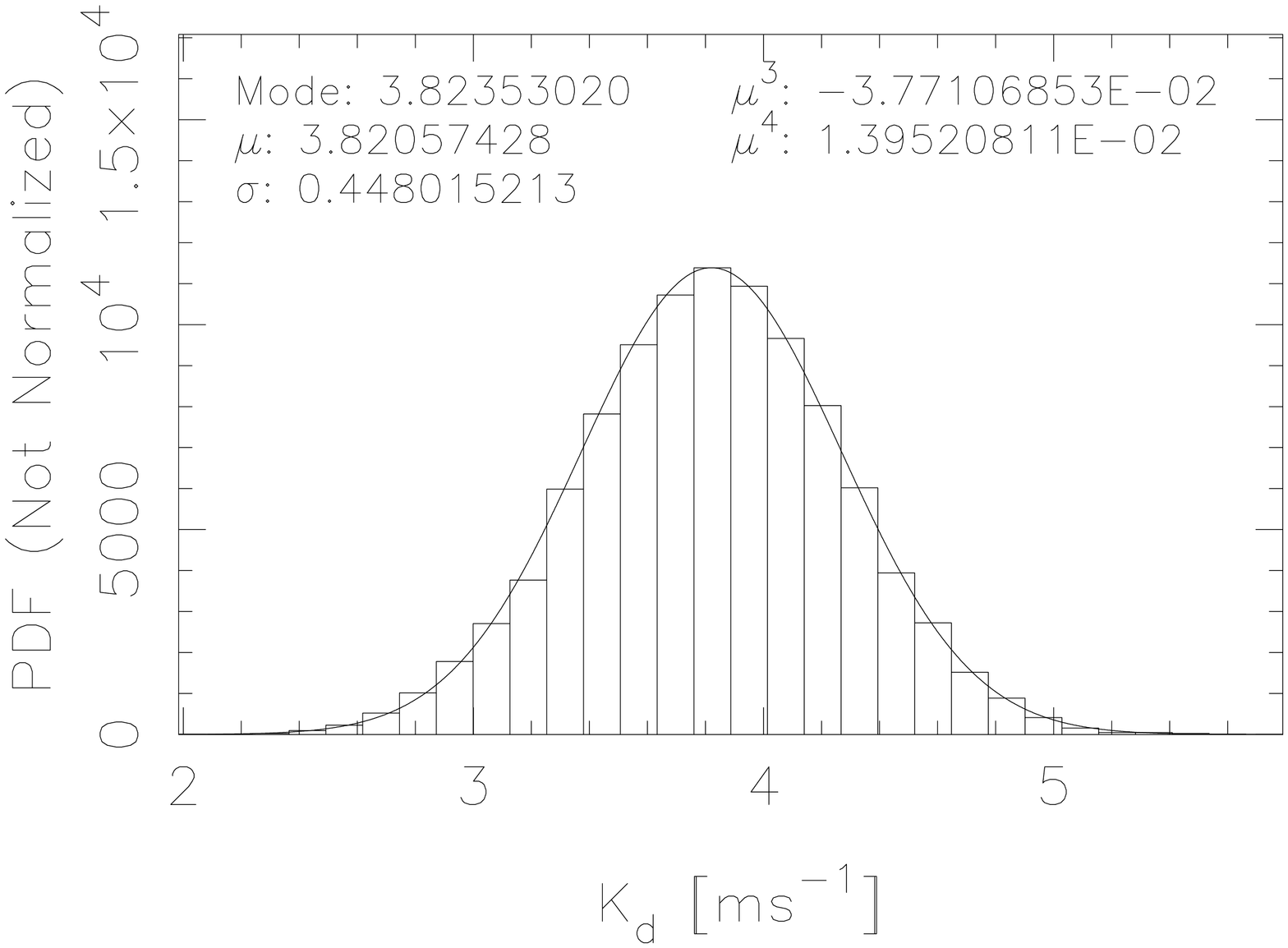}
\includegraphics[angle=270, width=0.23\textwidth,clip,trim=4cm 0 0 0]{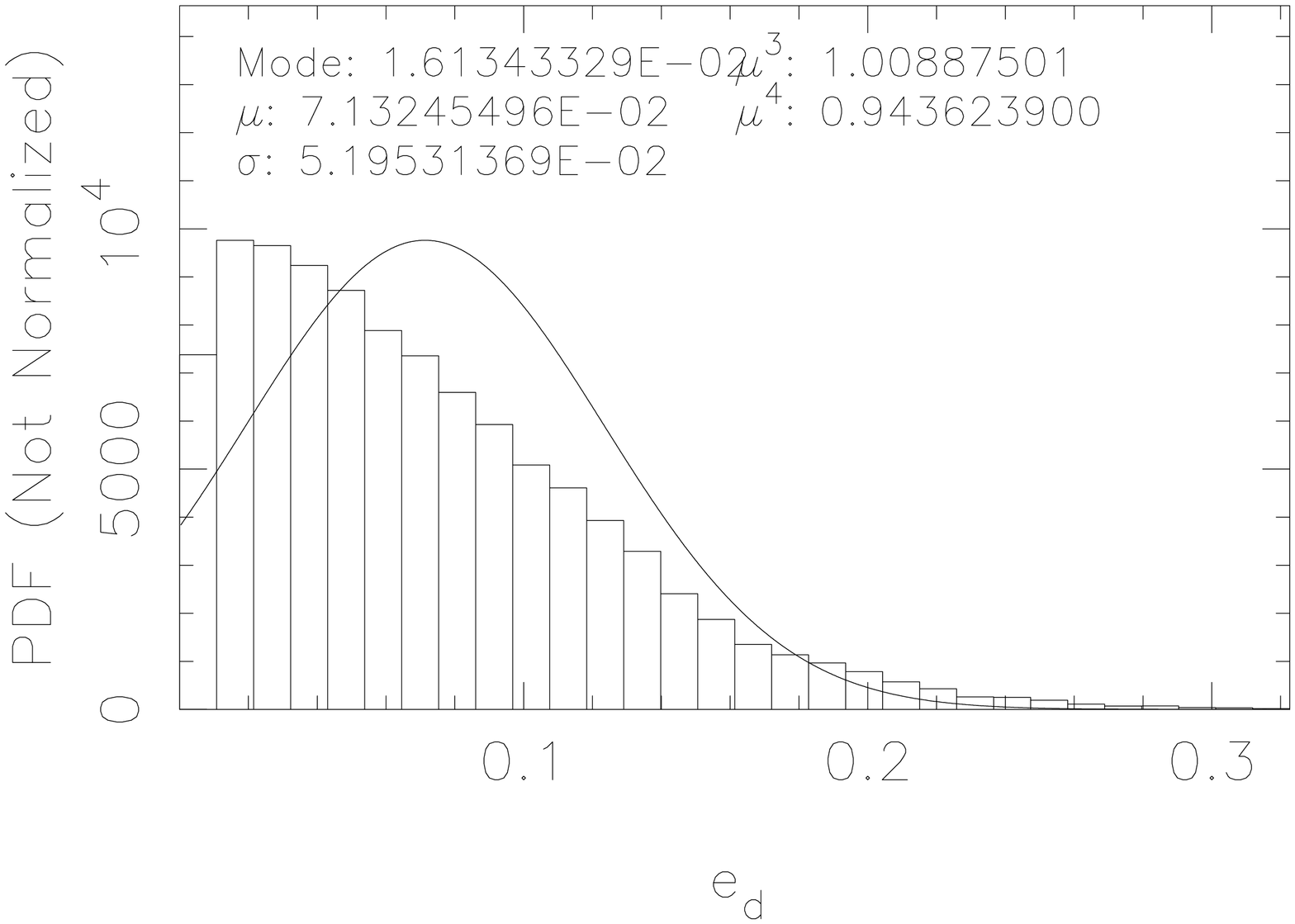}
\includegraphics[angle=270, width=0.23\textwidth,clip,trim=4cm 0 0 0]{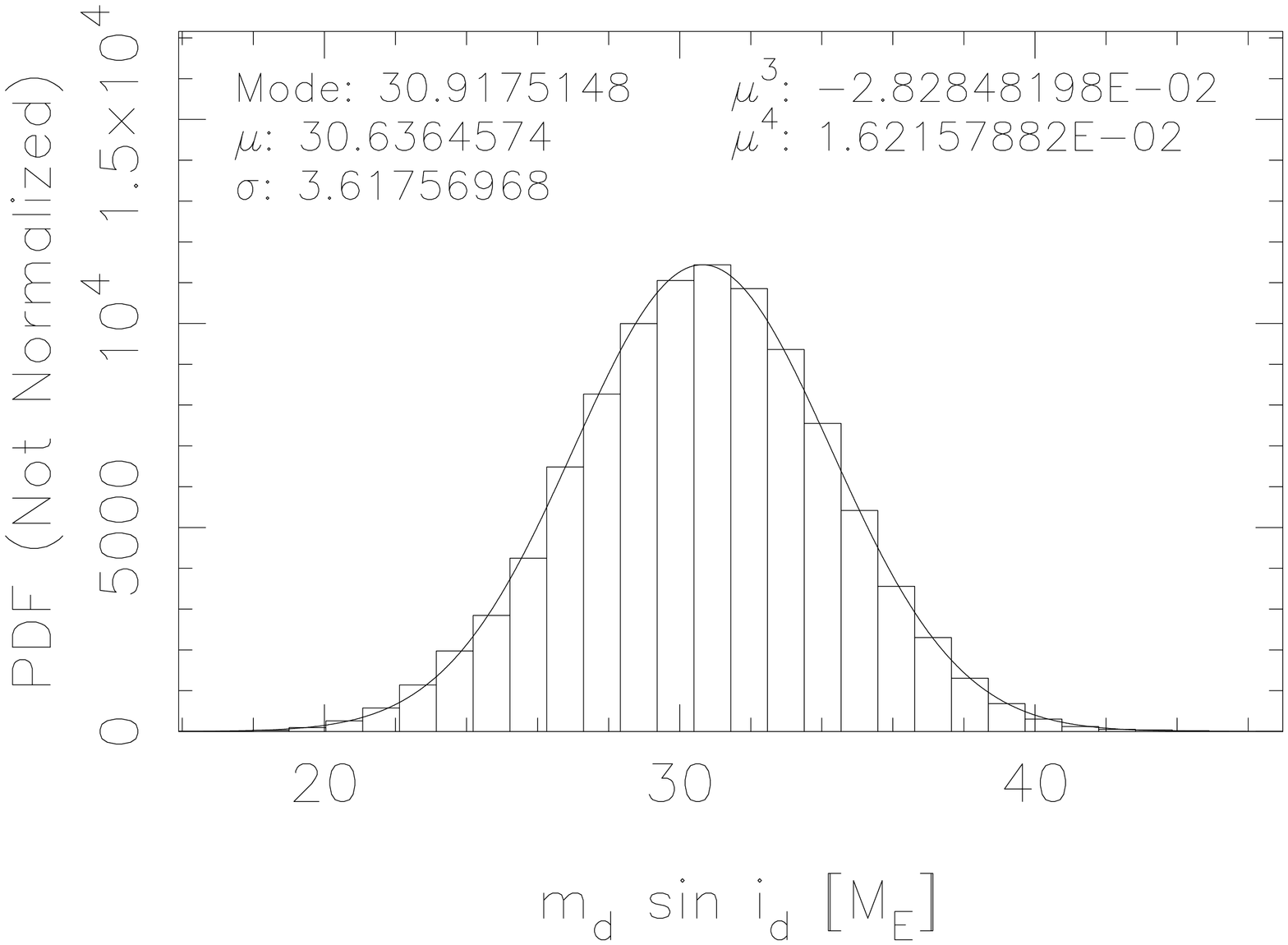}

\includegraphics[angle=270, width=0.23\textwidth,clip,trim=4cm 0 0 0]{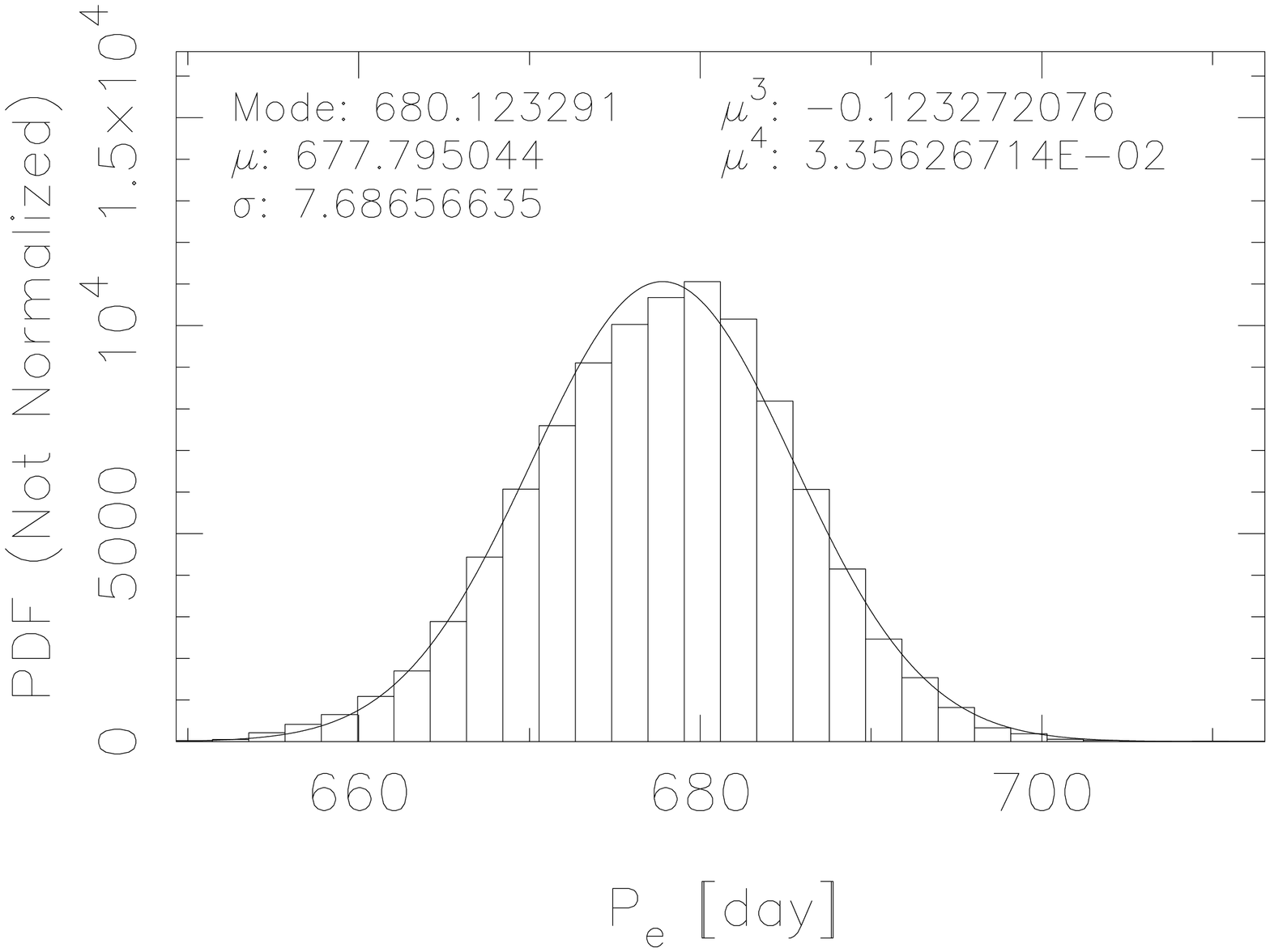}
\includegraphics[angle=270, width=0.23\textwidth,clip,trim=4cm 0 0 0]{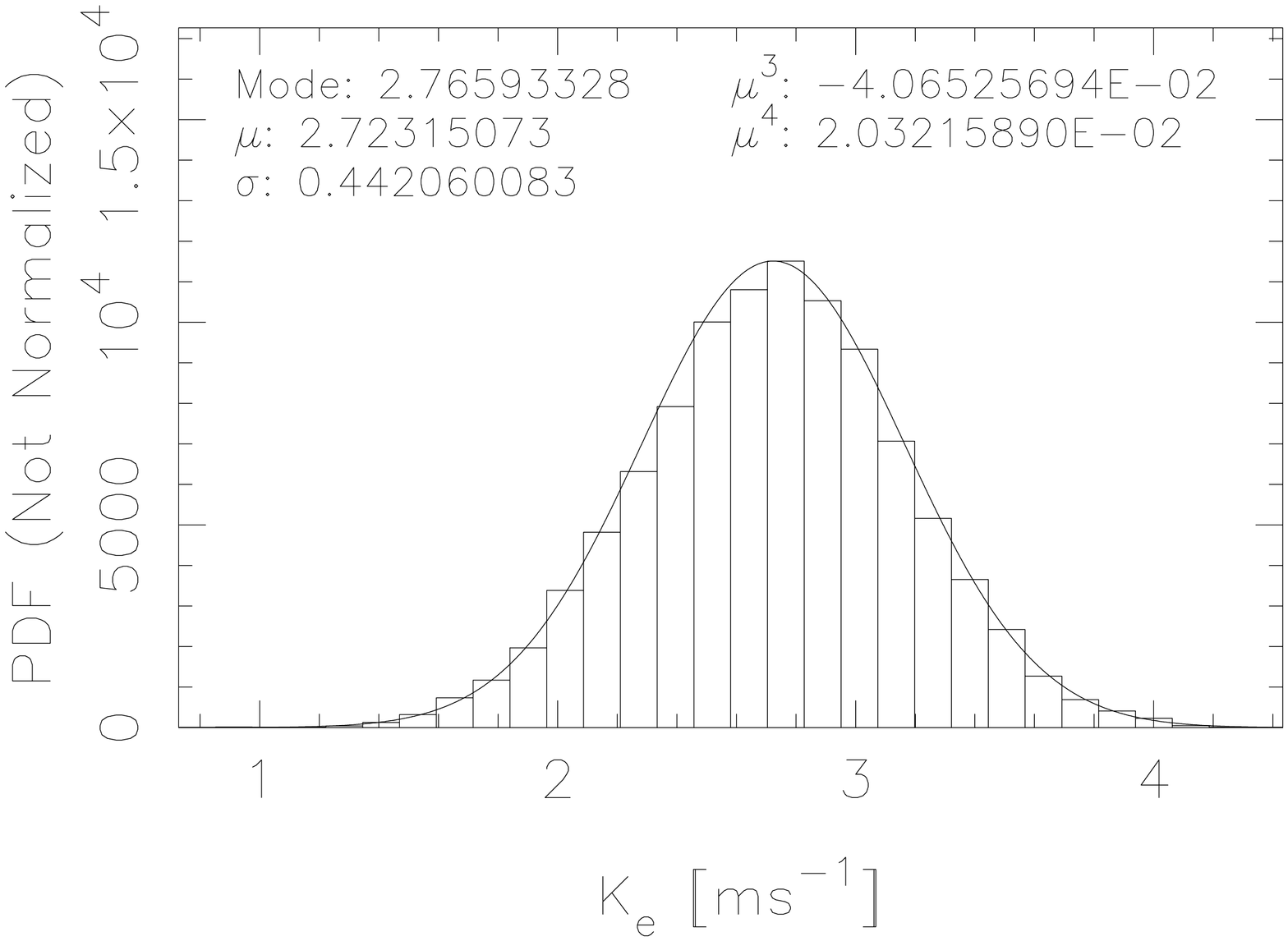}
\includegraphics[angle=270, width=0.23\textwidth,clip,trim=4cm 0 0 0]{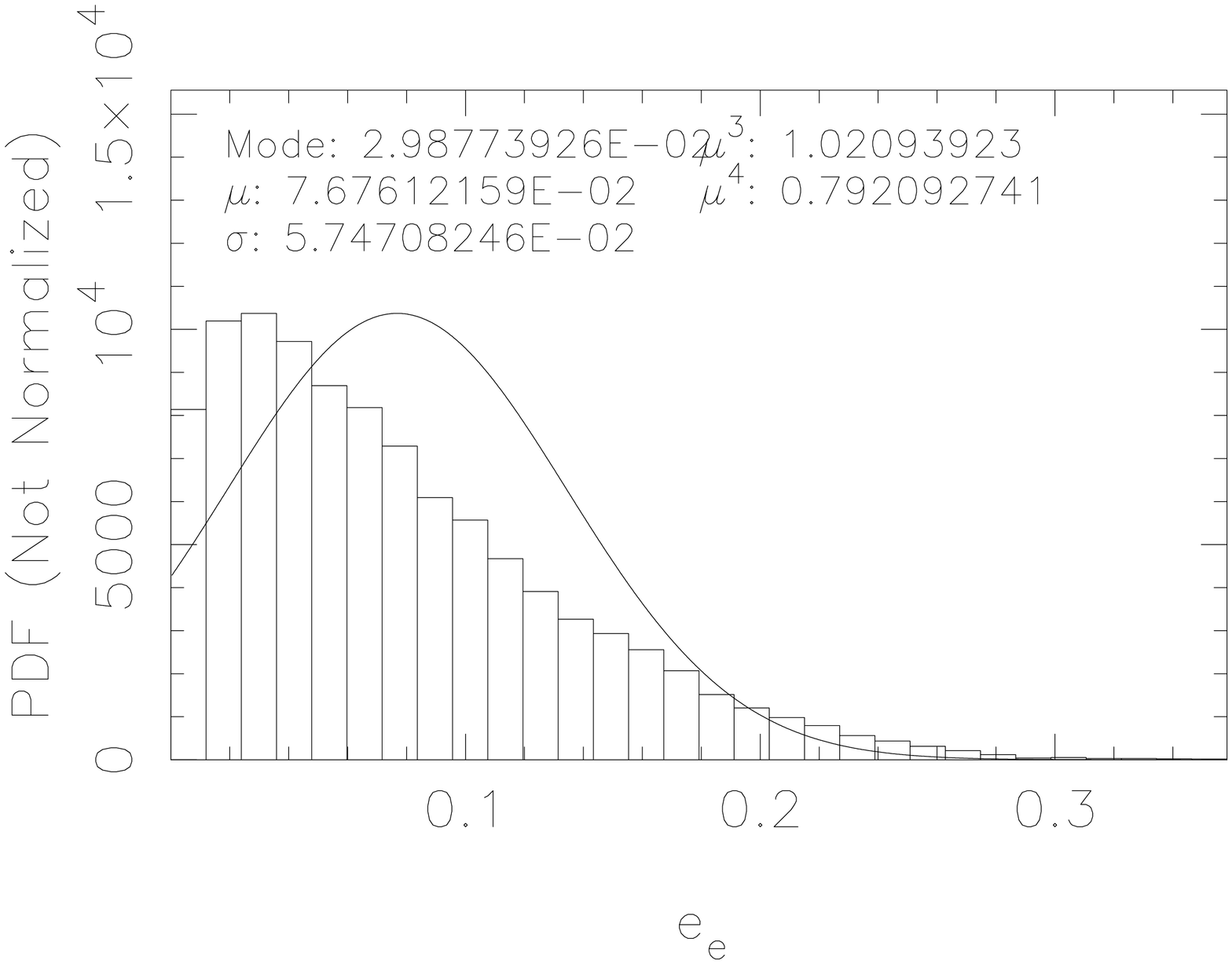}
\includegraphics[angle=270, width=0.23\textwidth,clip,trim=4cm 0 0 0]{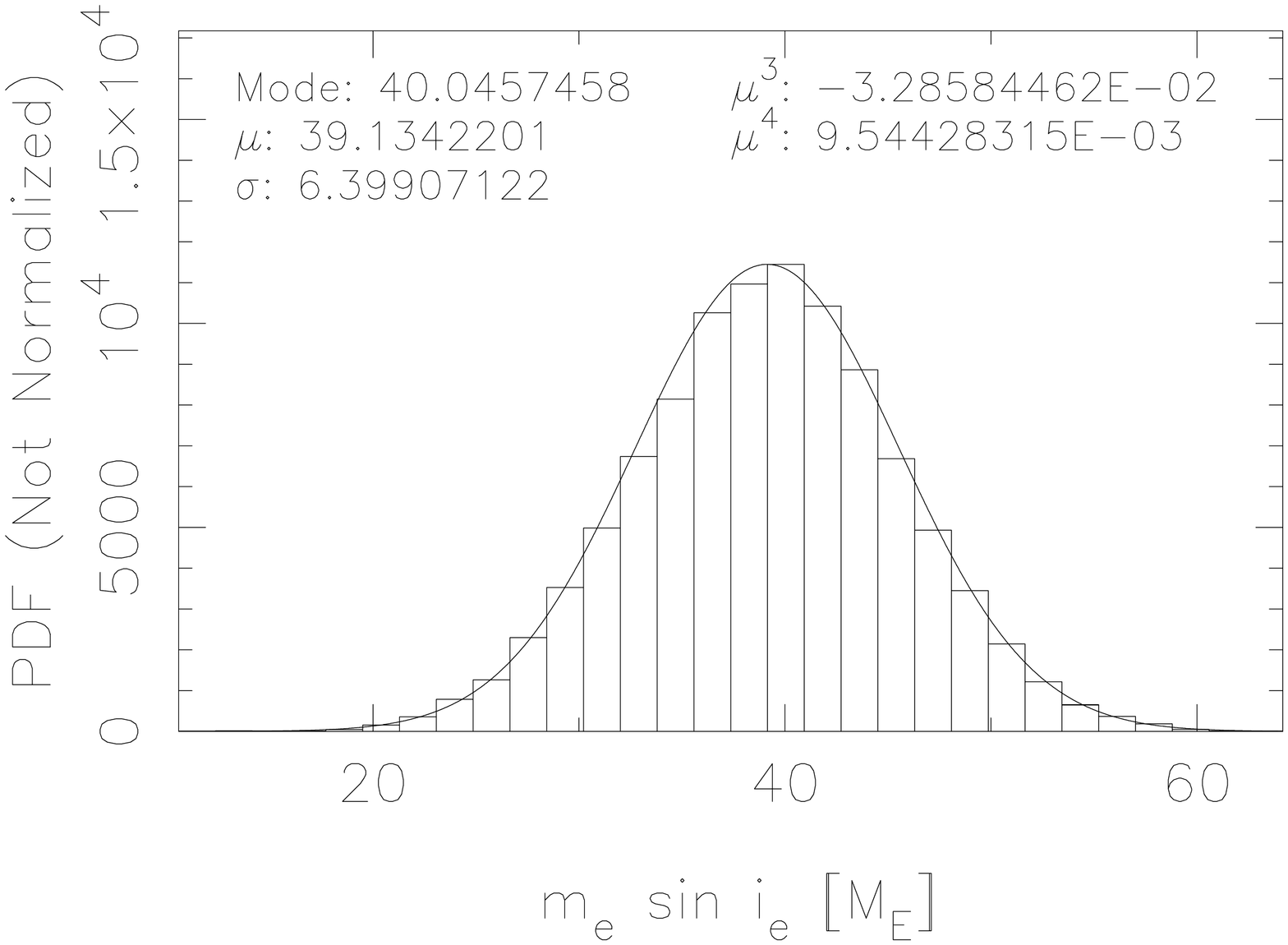}

\includegraphics[angle=270, width=0.23\textwidth,clip,trim=4cm 0 0 0]{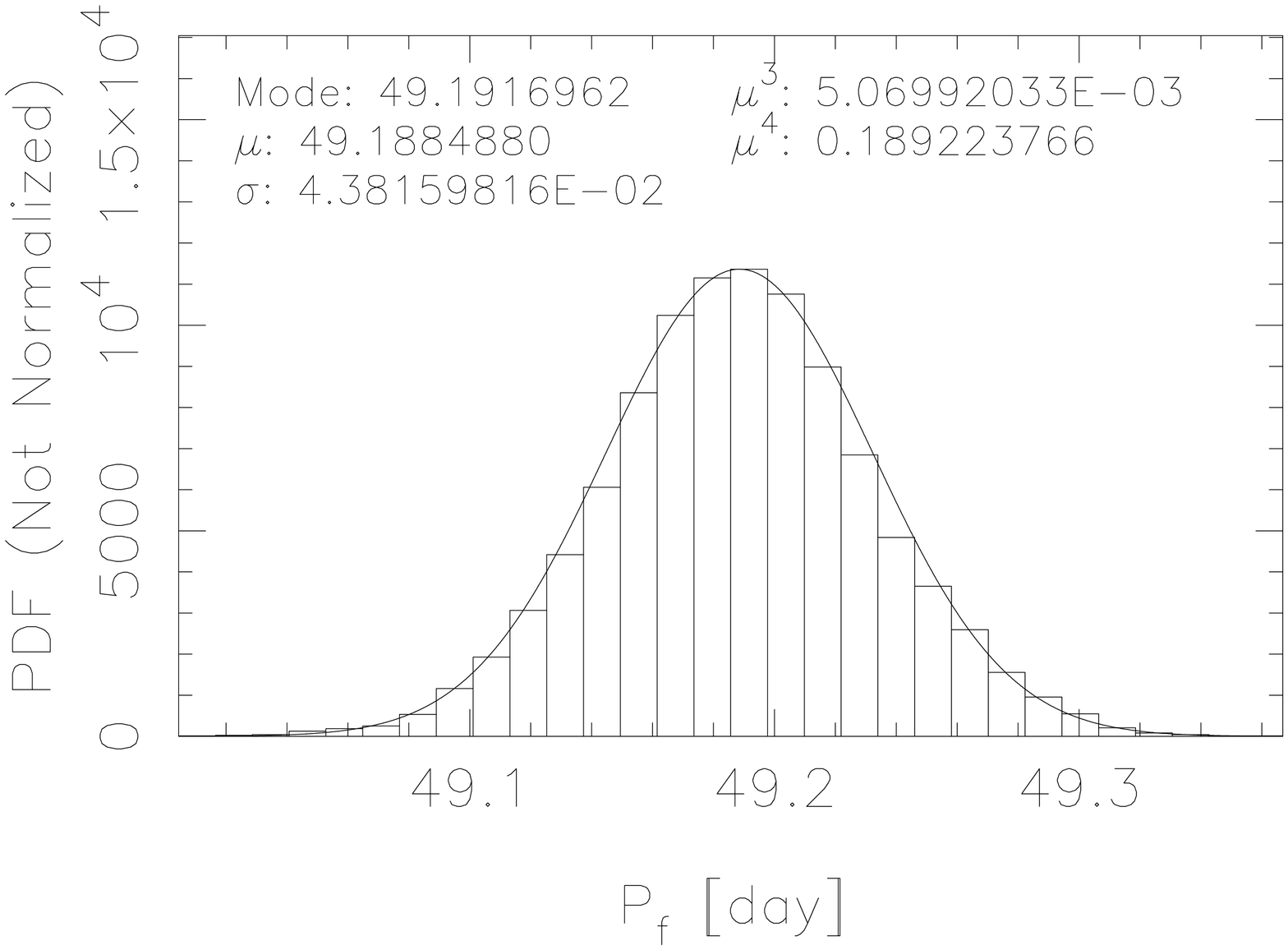}
\includegraphics[angle=270, width=0.23\textwidth,clip,trim=4cm 0 0 0]{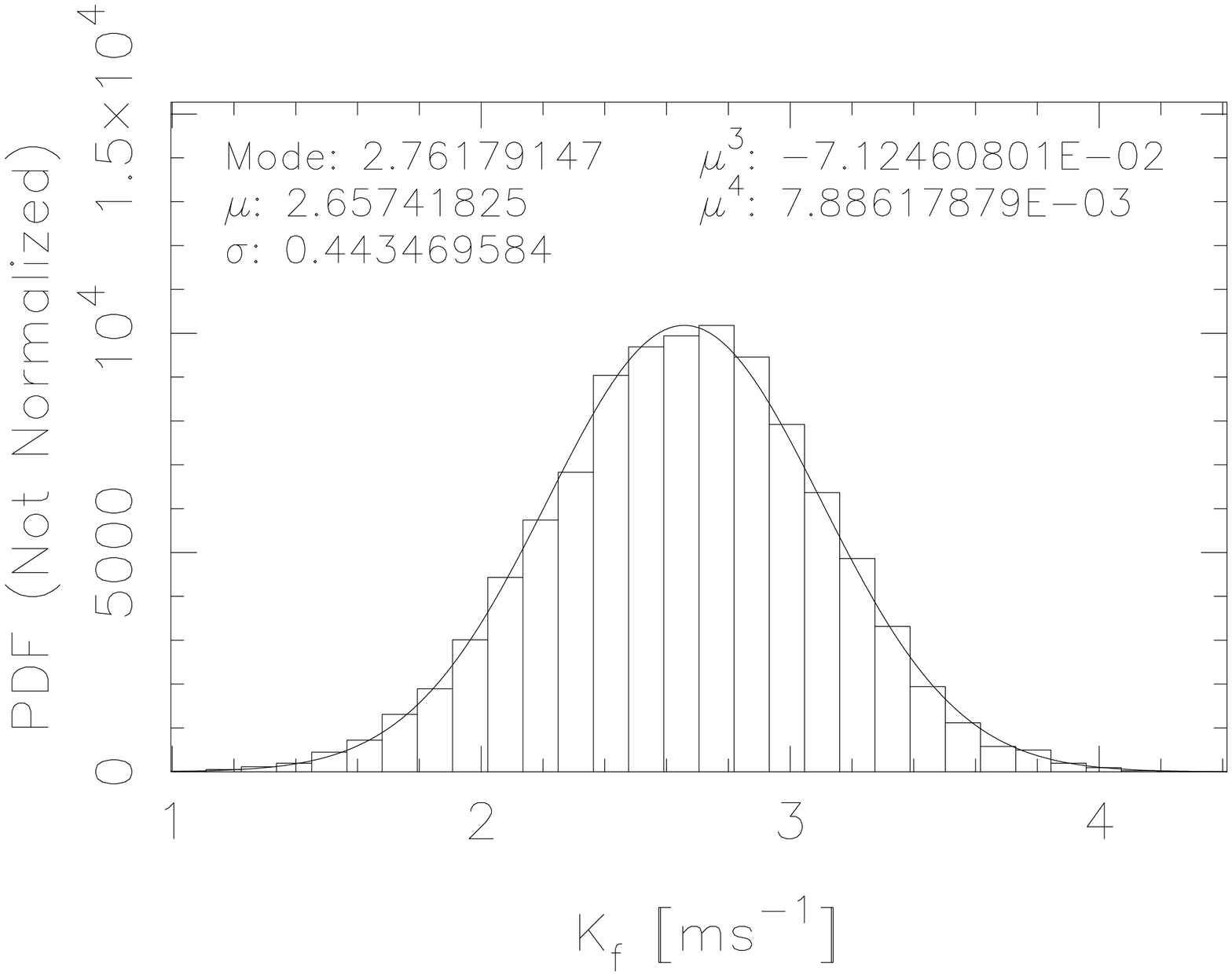}
\includegraphics[angle=270, width=0.23\textwidth,clip,trim=4cm 0 0 0]{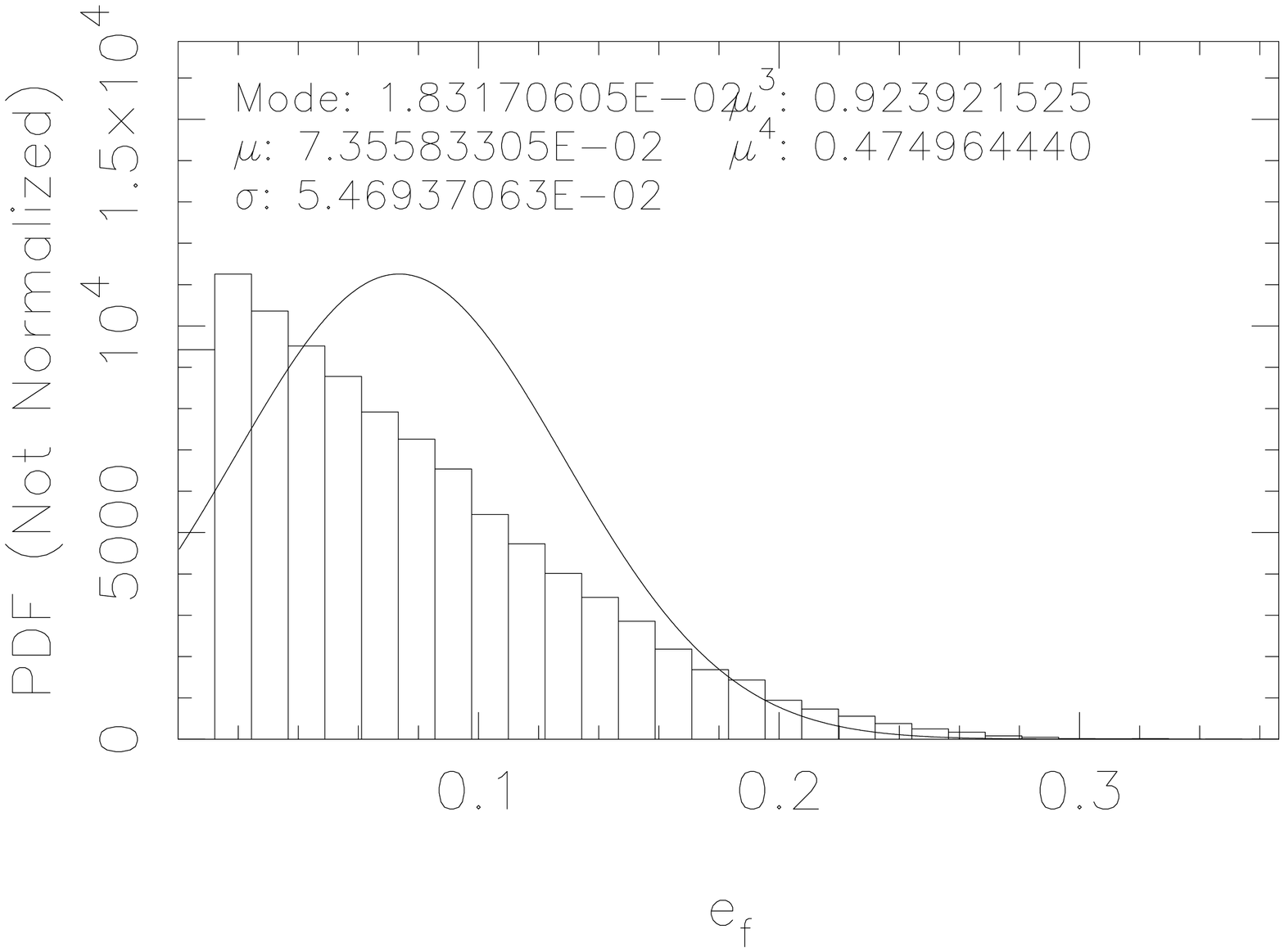}
\includegraphics[angle=270, width=0.23\textwidth,clip,trim=4cm 0 0 0]{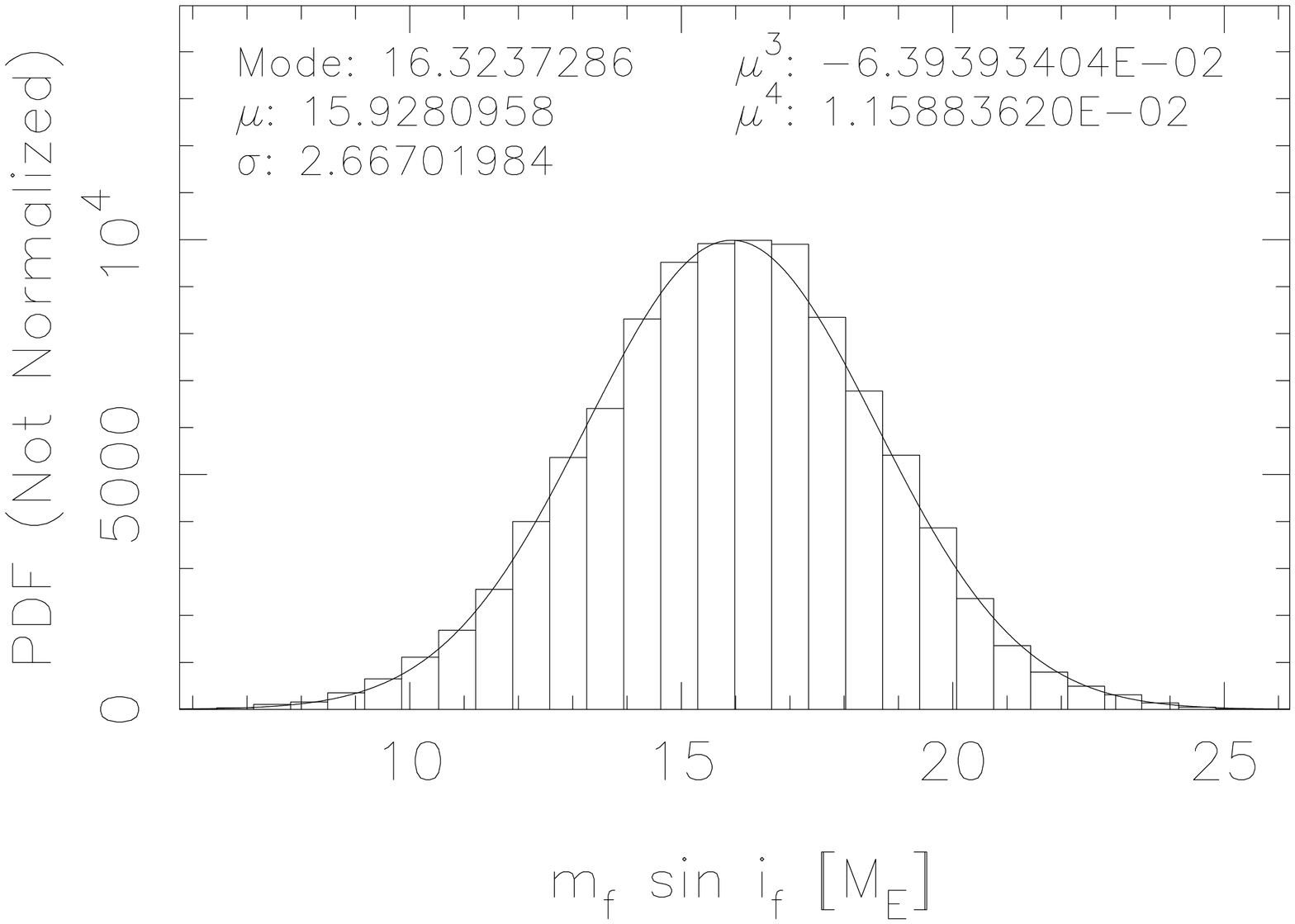}
\caption{Estimated posterior probability densities of selected Keplerian parameters and minimum masses of the planet candidates based on posterior samplings given a five-Keplerian model and a second-order polynomial acceleration. The solid curves denote Gaussian functions with the same mean ($\mu$) and variance ($\sigma^{2}$) as the estimated distributions have. The uncertainty in the stellar mass has been accounted for when obtaining estimates for the minimum masses.}\label{fig:densities}
\end{figure*}

We present statistics quantifying the significances of the signals in Table \ref{tab:likelihoods}. According to these results, we detect all five signals significantly when applying the likelihood-ratio criterion \citep[see e.g.][]{butler2017}. \cite{Howard2010} reported a value of 0.27 $\pm$0.07 for the eccentricity of the single 1049-day planet of their model. However, typically when one is modelling a superposition of several signals with a model containing only one, the model has a bias that shows up as an increased eccentricity. Although the detection threshold corresponding to a false-alarm probability (FAP) of 0.1\% is $\alpha = 20.52$ for a Keplerian model of the signals with five free parameters, we find that the eccentricities of all signals are consistent with zero (Table \ref{tab:parameters} and Fig. \ref{fig:densities}) implying that circular solutions can be used corresponding to a likelihood-ratio detection threshold of $\alpha=16.27$ (three degrees of freedom rather than five). The only signal whose inclusion in the model does not result in likelihood-ratio in excess of the former ratio is the one with a period of 680 days. However, all of the signals correspond to likelihood ratios that satisfy the latter detection threshold. Similarly, we calculated the Bayes factors quantifying evidence in favor of the signals by using the Bayesian information criterion (BIC) approximation \citep{liddle2007}. We choose this method because the BIC appears to provide a convenient and reasonably reliable compromise, avoiding false positives and negatives when a value of 150 -- corresponding to ``very strong evidence'' according to the Jeffrey's scale as presented by \citep{kass1995} -- is used as a detection threshold \citep{feng2016}. As can be seen in Table \ref{tab:likelihoods}, inclusion of all the signals in the model yields Bayes factors in excess of a value of $\ln 150 \approx 5.01$ implying they are significantly detected.

\begin{table*}
\caption{Maximum likelihoods and likelihood ratios of models with $k=0, ..., 5$ Keplerian signals. The ratios are defined such that $\Delta \ln L_{k} = \ln L_{k} - \ln L_{k-1}$. The last column shows the Bayes factors in favor of the signals.}\label{tab:likelihoods}
\begin{minipage}{\textwidth}
\begin{center}
\begin{tabular}{lcccccccc}
\hline \hline
$k$ & $\ln L_{k}$ & $\ln L_{k}$ & $\ln L_{k}$ & $\ln L_{k}$ & $\ln L_{k}$ & $\ln L_{k}$ & $\Delta \ln L_{k}$ & $\ln B_{k,k-1}$ \\
& All data & PFS & APF & HET10 & KECK & KECK10 & All data \\
\hline
0 & -1179.41 & -185.19 & -81.10 & -192.86 & -696.97 & -23.29 & & \\
1 & -1058.15 & -181.67 & -69.90 & -169.46 & -617.72 & -19.39 & 121.26 & 112.55 \\
2 & -1021.21 & -178.19 & -69.78 & -161.36 & -593.44 & -18.43 & 36.94 & 28.23 \\
3 & -991.24 & -176.78 & -64.75 & -157.80 & -573.89 & -18.02 & 29.97 & 21.26 \\
4 & -973.26 & -178.54 & -58.33 & -156.62 & -561.50 & -18.27 & 17.98 & 9.27 \\
5 & -952.25 & -175.55 & -56.52 & -157.53 & -544.79 & -17.86 & 21.01 & 12.29 \\
\hline \hline
\end{tabular}
\end{center}
\end{minipage}
\end{table*}

\begin{figure}
\center
\includegraphics[angle=270, width=0.23\textwidth,clip,clip,trim=2cm 0 0 0]{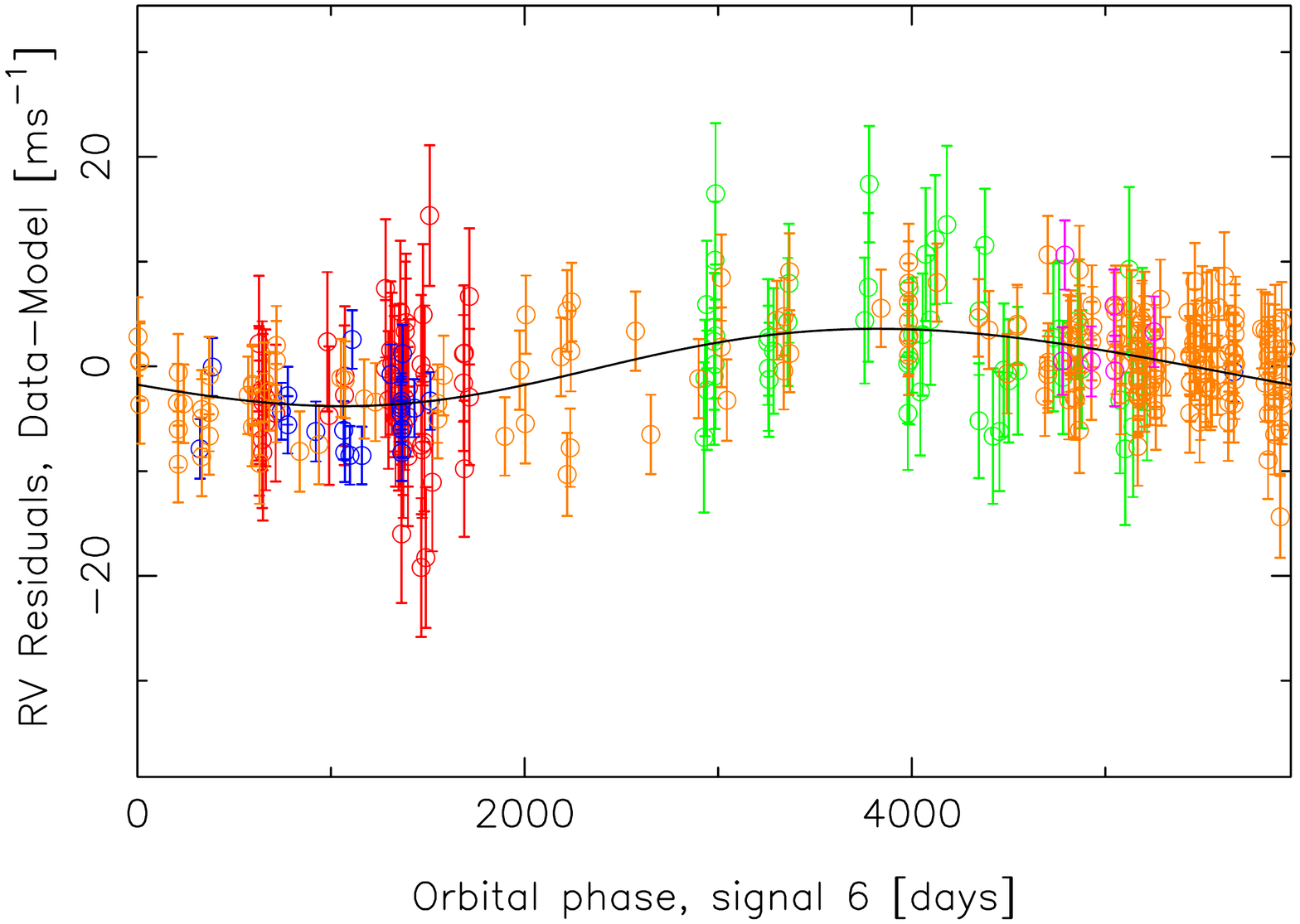}
\includegraphics[angle=270, width=0.23\textwidth,clip,clip,trim=2cm 0 0 0]{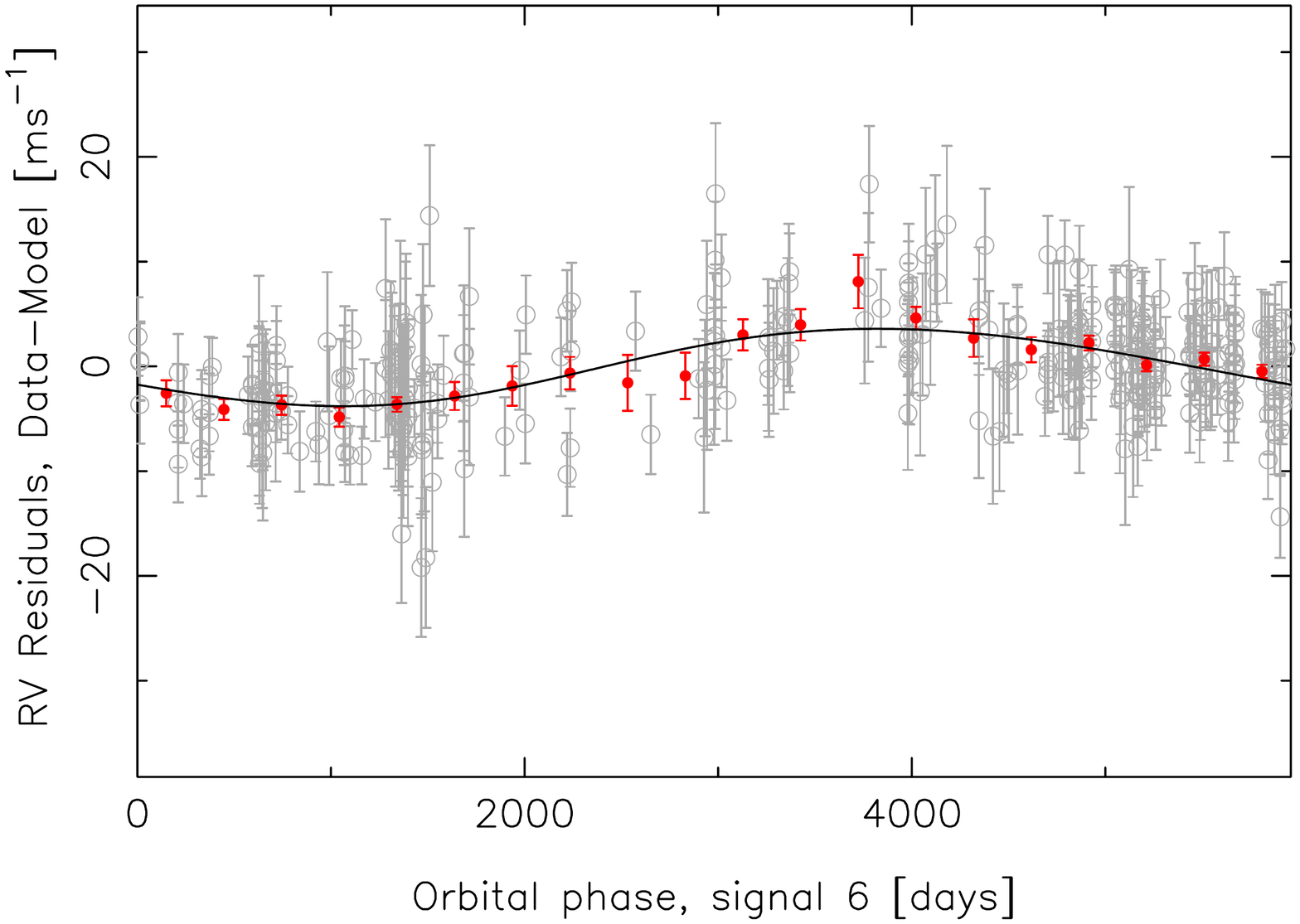}
\caption{As in Fig. \ref{fig:phased} but for the long-period signal.}\label{fig:phased_long}
\end{figure}

\begin{figure*}
\center
\includegraphics[angle=270, width=0.23\textwidth,clip,trim=4cm 0 0 0]{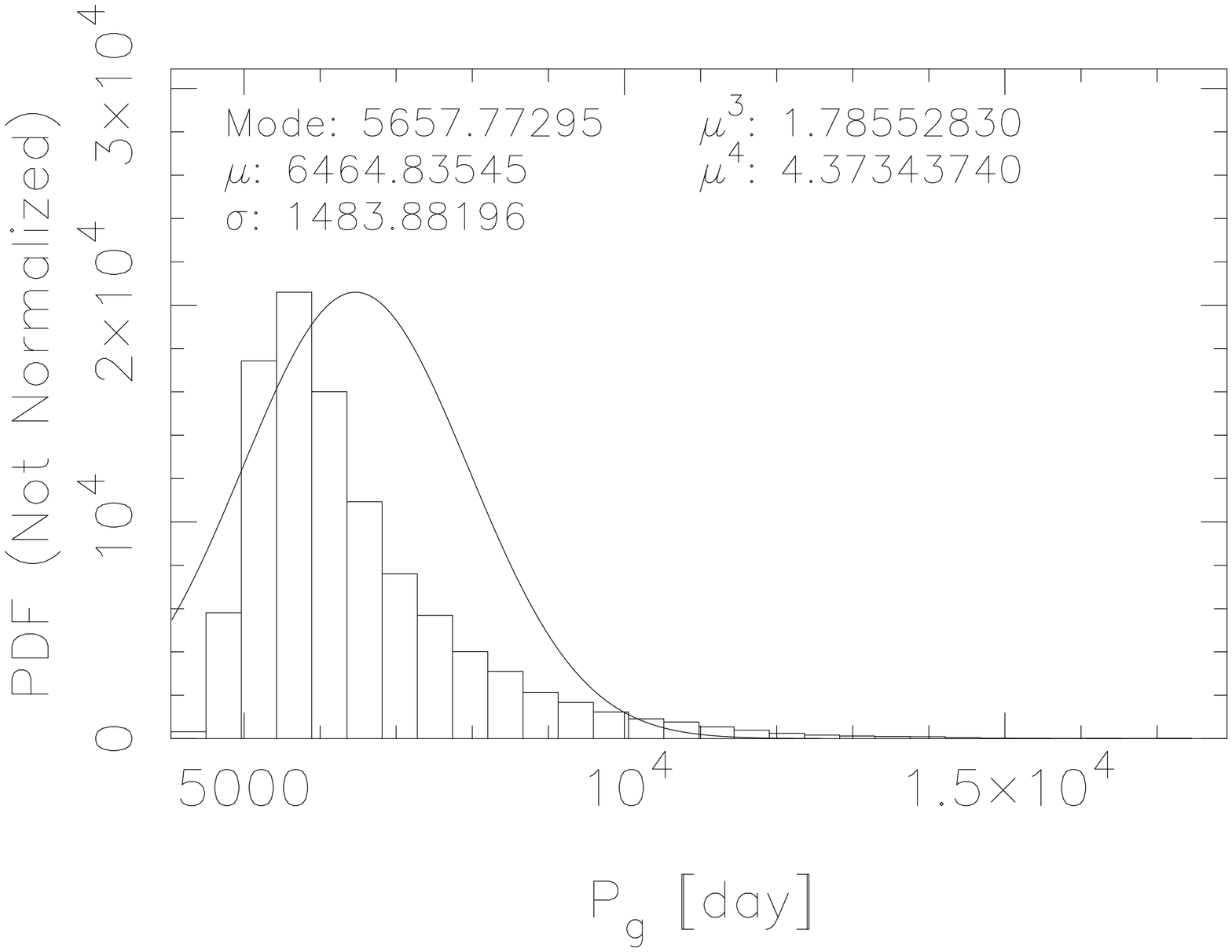}
\includegraphics[angle=270, width=0.23\textwidth,clip,trim=4cm 0 0 0]{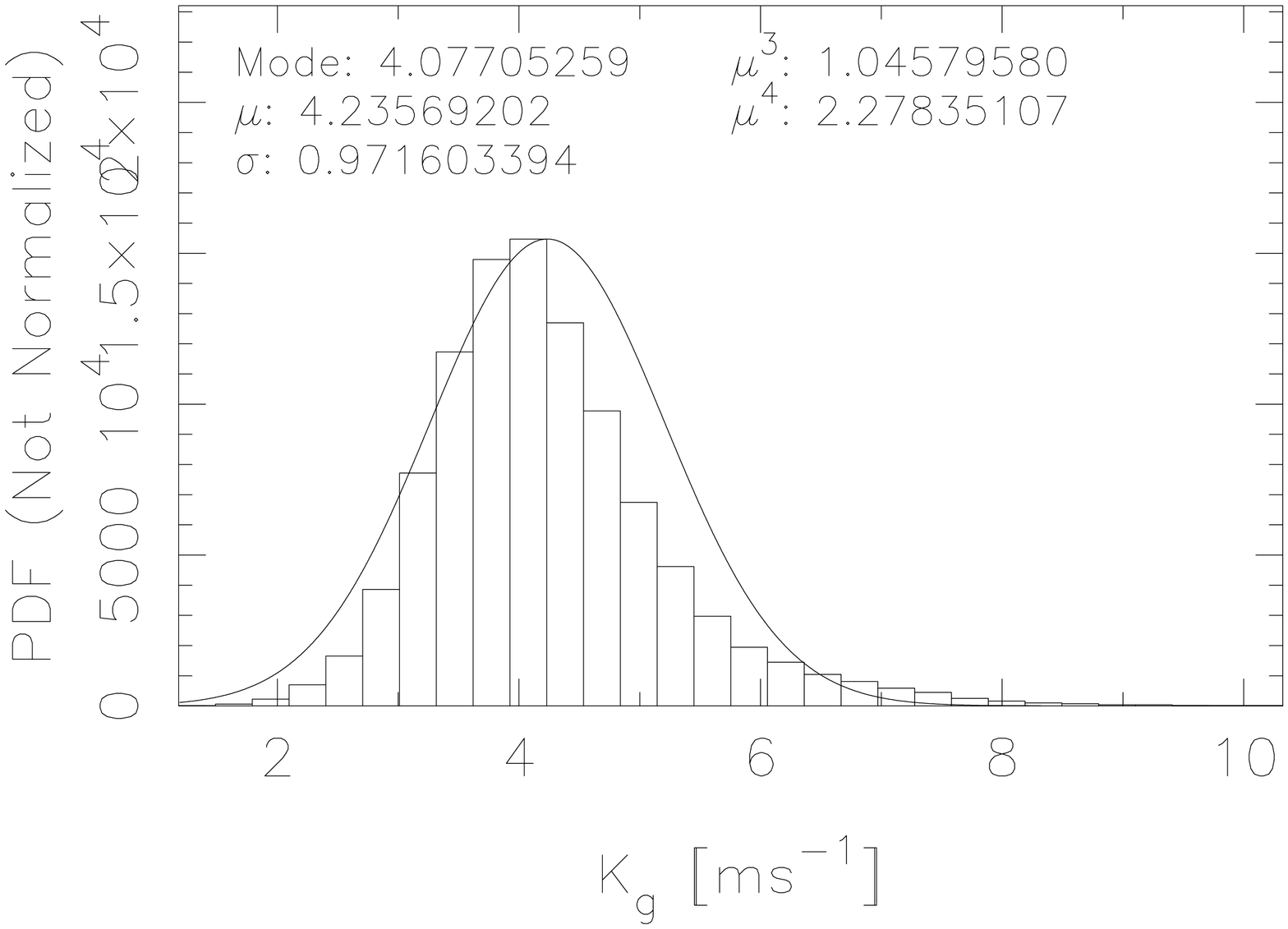}
\includegraphics[angle=270, width=0.23\textwidth,clip,trim=4cm 0 0 0]{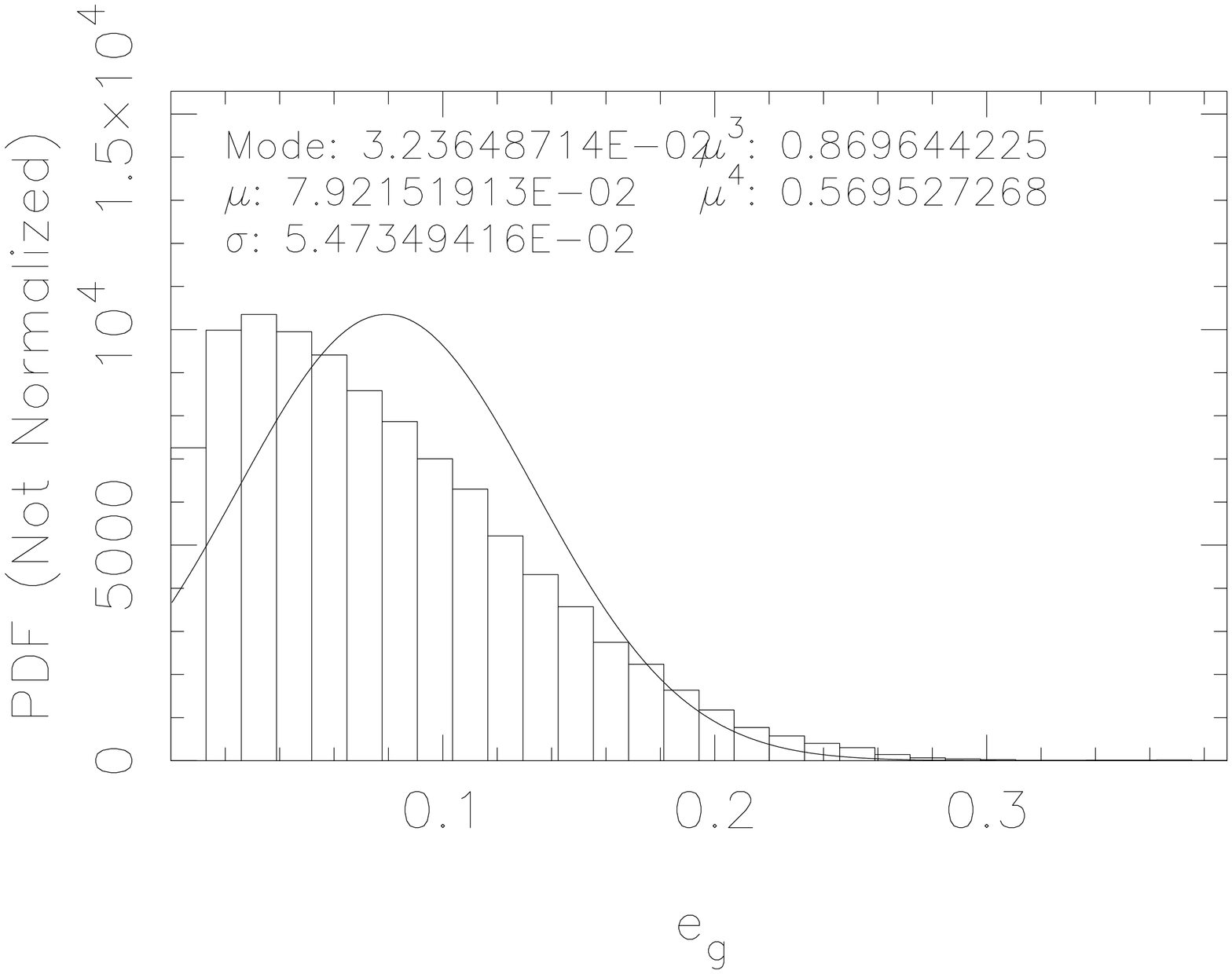}
\includegraphics[angle=270, width=0.23\textwidth,clip,trim=4cm 0 0 0]{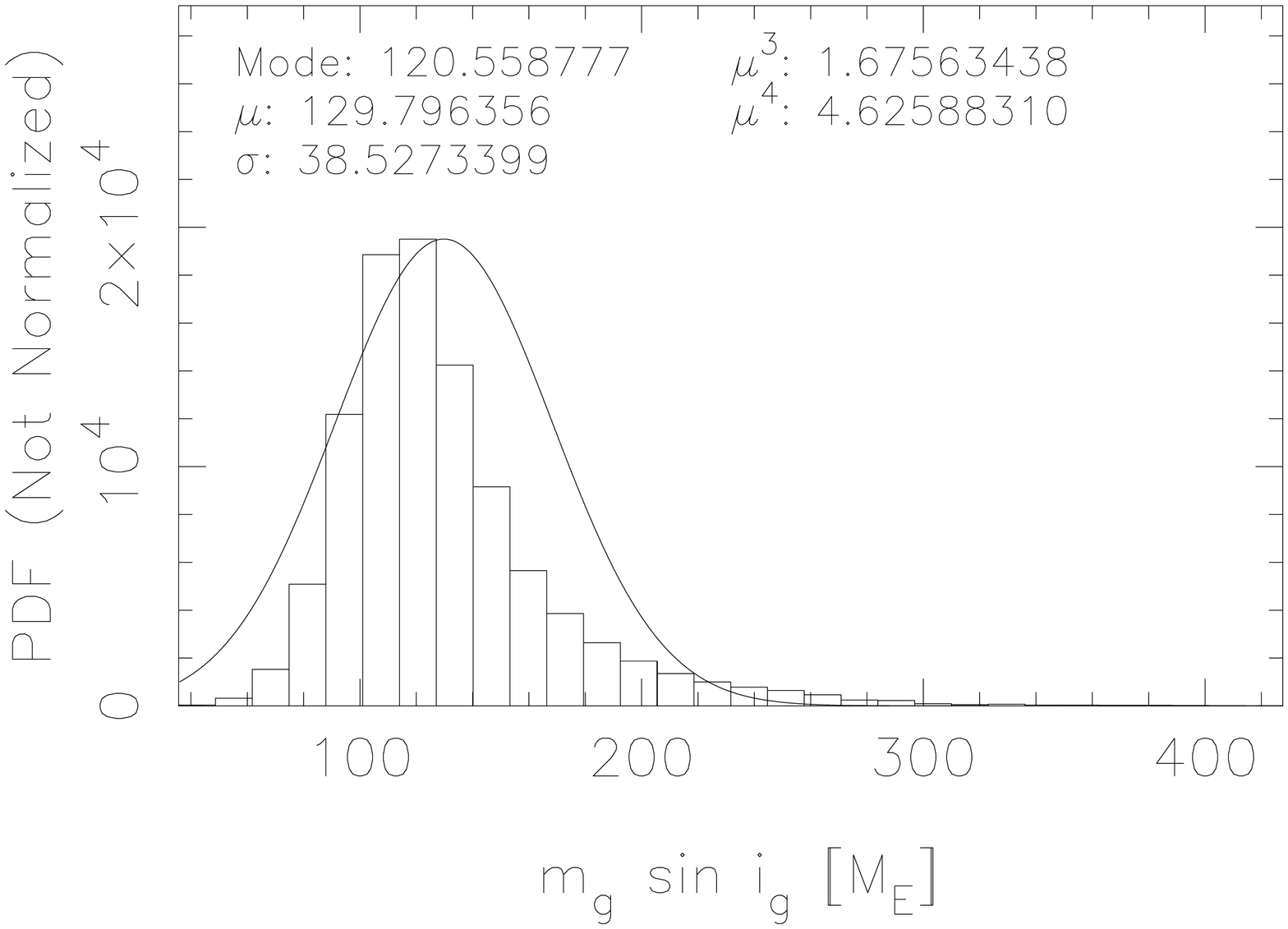}
\caption{As in Fig. \ref{fig:densities} but for the parameters of the long-period signal given a six-Keplerian model.}\label{fig:densities_long}
\end{figure*}

As mentioned above, there is evidence for acceleration in the combined data that can be described by using a second-order polynomial. This acceleration cannot be modeled with only a first-order polynomial because the second-order term $\gamma_{2}$ is significantly different from zero. We tested whether the corresponding signal could be modeled by using a Keplerian function, i.e. whether the corresponding signal satisfied the requirements that its amplitude and period were well-constrained from below and above in the parameter space \citep[according to the signal detection criteria of][]{tuomi2012}. We show the probability distributions of selected Keplerian parameters of this signal in Fig. \ref{fig:densities_long} together with the estimated distribution of minimum mass, and show the radial velocities folded on the signal phase in Fig. \ref{fig:phased_long}. According to our results, the signal is significantly present in the data and well-constrained in the parameter space. However, regardless of whether we model it with a second-order polynomial or with a Keplerian signal, we obtain a roughly equally good model according to likelihood ratios and Bayes factors. This means that there is some evidence for a sixth candidate planet in the system but continued radial velocity monitoring is required to better constrain the Keplerian parameters of this potential long-period candidate. We have tabulated our six-Keplerian solution in Table \ref{tab:parameters_six}.

\begin{table*}
\caption{Parameter estimates of the statistical model with six Keplerian signals. The uncertainties are denoted by both standard 1-$\sigma$ errors and 99\% credibility intervals. When calculating the uncertainties of the semi-major axes and minimum masses, we have accounted for the uncertainty in the stellar mass.}\label{tab:parameters_six}
\begin{minipage}{\textwidth}
\begin{center}
\begin{tabular}{lcccccc}
\hline \hline
Parameter & HD 34445 b & HD 34445 c & HD 34445 d \\
\hline
$P$ (days) & 1056.7$\pm$4.7 [1042.2, 1069.8] & 214.67$\pm$0.45 [213.40, 215.95] & 117.87$\pm$0.18 [117.30, 118.34] \\
$K$ (ms$^{-1}$) & 12.01$\pm$0.52 [10.48, 13.55] & 5.45$\pm$0.50 [4.10, 6.95] & 3.81$\pm$0.48 [2.41, 5.09] \\
$e$ & 0.014$\pm$0.035 [0, 0.117] & 0.036$\pm$0.071 [0, 0.217] & 0.027$\pm$0.051 [0, 0.234] \\
$\omega$ (rad) & 2.0$\pm$1.6 [0, 2$\pi$] & 2.6$\pm$1.3 [0, 2$\pi$] & 4.3$\pm$1.7 [0, 2$\pi$] \\
$M_{0}$ (rad) & 3.9$\pm$1.6 [0, 2$\pi$] & 4.1$\pm$1.4 [0, 2$\pi$] & 2.2$\pm$1.6 [0, 2$\pi$] \\
\hline
$a$ (AU) & 2.075$\pm$0.016 [2.034, 2.120] & 0.7181$\pm$0.0049 [0.7034, 0.7314] & 0.4817$\pm$0.0033 [0.4715, 0.4909] \\
$m_{p} \sin i$ (M$_{\oplus}$) & 200.0$\pm$9.0 [175.0, 224.9] & 53.5$\pm$5.0 [40.0, 68.4] & 30.7$\pm$3.9 [19.2, 41.2] \\
$m_{p} \sin i$ (M$_{\rm J}$) & 0.629$\pm$0.028 [0.550, 0.708] & 0.168$\pm$0.016 [0.125, 0.215] & 0.097$\pm$0.13 [0.060, 0.130] \\
\hline
Parameter & HD 34445 e & HD 34445 f & HD 34445 g \\
\hline
$P$ (days) & 49.175$\pm$0.045 [49.053, 49.311] & 676.8$\pm$7.9 [654.4, 701.8] & 5700$\pm$1500 [4000, 11400] \\
$K$ (ms$^{-1}$) & 2.75$\pm$0.47 [1.35, 4.04] & 2.74$\pm$0.46 [1.30, 3.93] & 4.08$\pm$0.98 [1.79, 7.58] \\
$e$ & 0.090$\pm$0.062 [0, 0.287] & 0.031$\pm$0.057 [0, 0.264] & 0.032$\pm$0.080 [0, 0.259] \\
$\omega$ (rad) & 5.3$\pm$2.1 [0, 2$\pi$] & 3.7$\pm$1.5 [0, 2$\pi$] & 4.1$\pm$1.6 [0, 2$\pi$] \\
$M_{0}$ (rad) & 0.7$\pm$1.8 [0, 2$\pi$] & 1.4$\pm$1.8 [0, 2$\pi$] & 3.0$\pm$1.6 [0, 2$\pi$] \\
\hline
$a$ (AU) & 0.2687$\pm$0.0019 [0.2632, 0.2741] & 1.543$\pm$0.016 [1.500, 1.590] & 6.36$\pm$1.02 [5.04, 10.29] \\
$m_{p} \sin i$ (M$_{\oplus}$) & 16.8$\pm$2.9 [8.1, 24.8] & 37.9$\pm$6.5 [18.7, 57.0] & 120.6$\pm$38.9 [48.7, 257.8] \\
$m_{p} \sin i$ (M$_{\rm J}$) & 0.0529$\pm$0.0089 [0.025, 0.078] & 0.119$\pm$0.021 [0.058, 0.180] & 0.38$\pm$0.13 [0.15, 0.82] \\
\hline \hline
\end{tabular}
\end{center}
\end{minipage}
\end{table*}

\subsection{\textbf{Correlation with Stellar Activity}}
We tested whether the radial velocities were linearly connected to the available activity indicators \citep[see e.g.][]{butler2017}. The parameters quantifying these connections are tabulated in Table \ref{tab:parameters} and demonstrate that the APF and PFS radial velocities are not connected to the corresponding S-indices statistically significantly with a 99\% credibility. However, the KECK radial velocities show a significant connection with a 99\% credibility, which is indicated by the fact that parameter $c_{\rm S,KECK}$ has a 99\% credibility interval of [21, 369] which does not include zero. Some of the variability in the KECK radial velocities is thus connected to the variations in the S-indices and originates in stellar activity.

\begin{figure}
\center
\includegraphics[angle=270, width=0.23\textwidth,clip,trim=1cm 0 0 0]{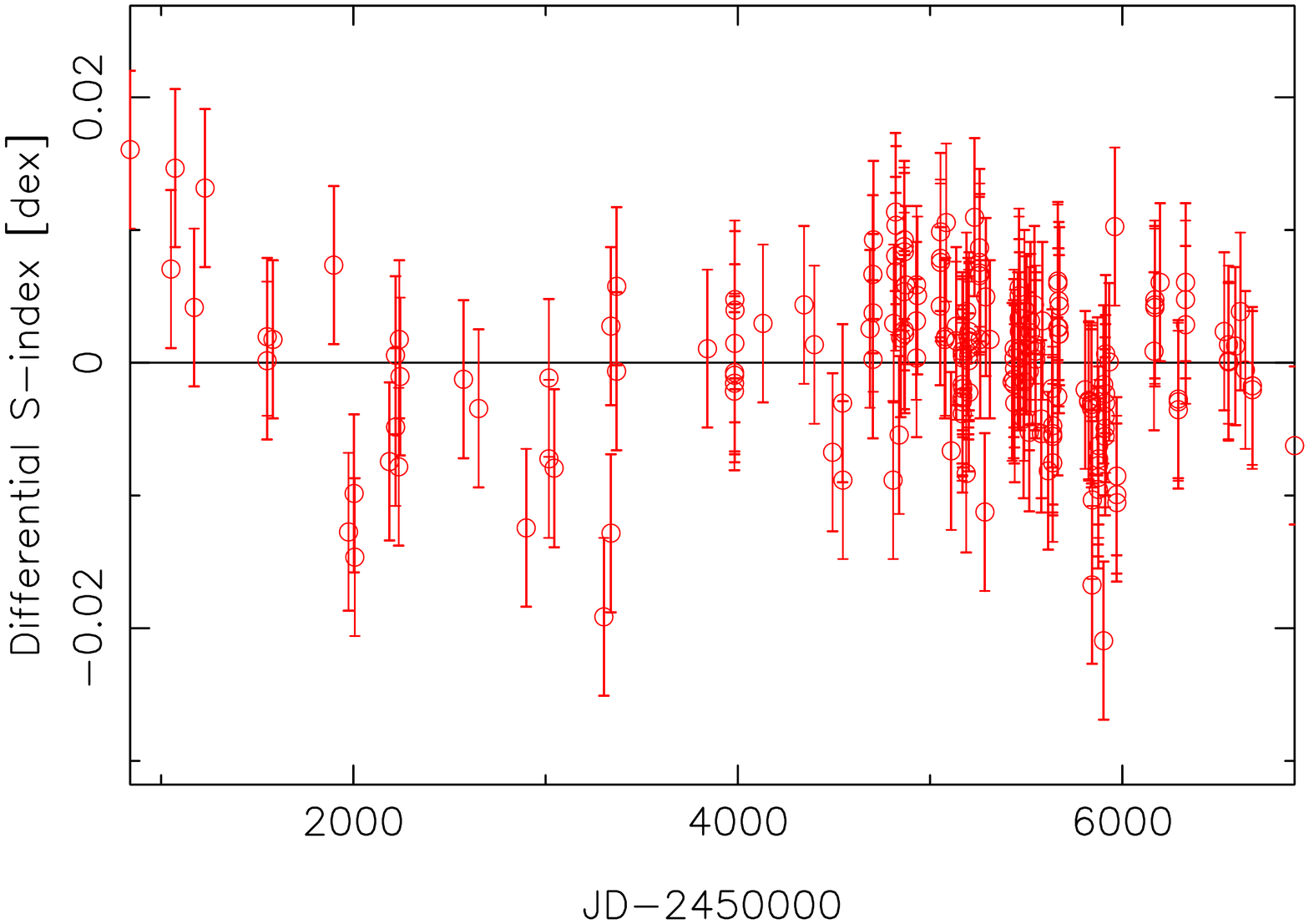}
\includegraphics[angle=270, width=0.23\textwidth,clip,trim=1cm 0 0 0]{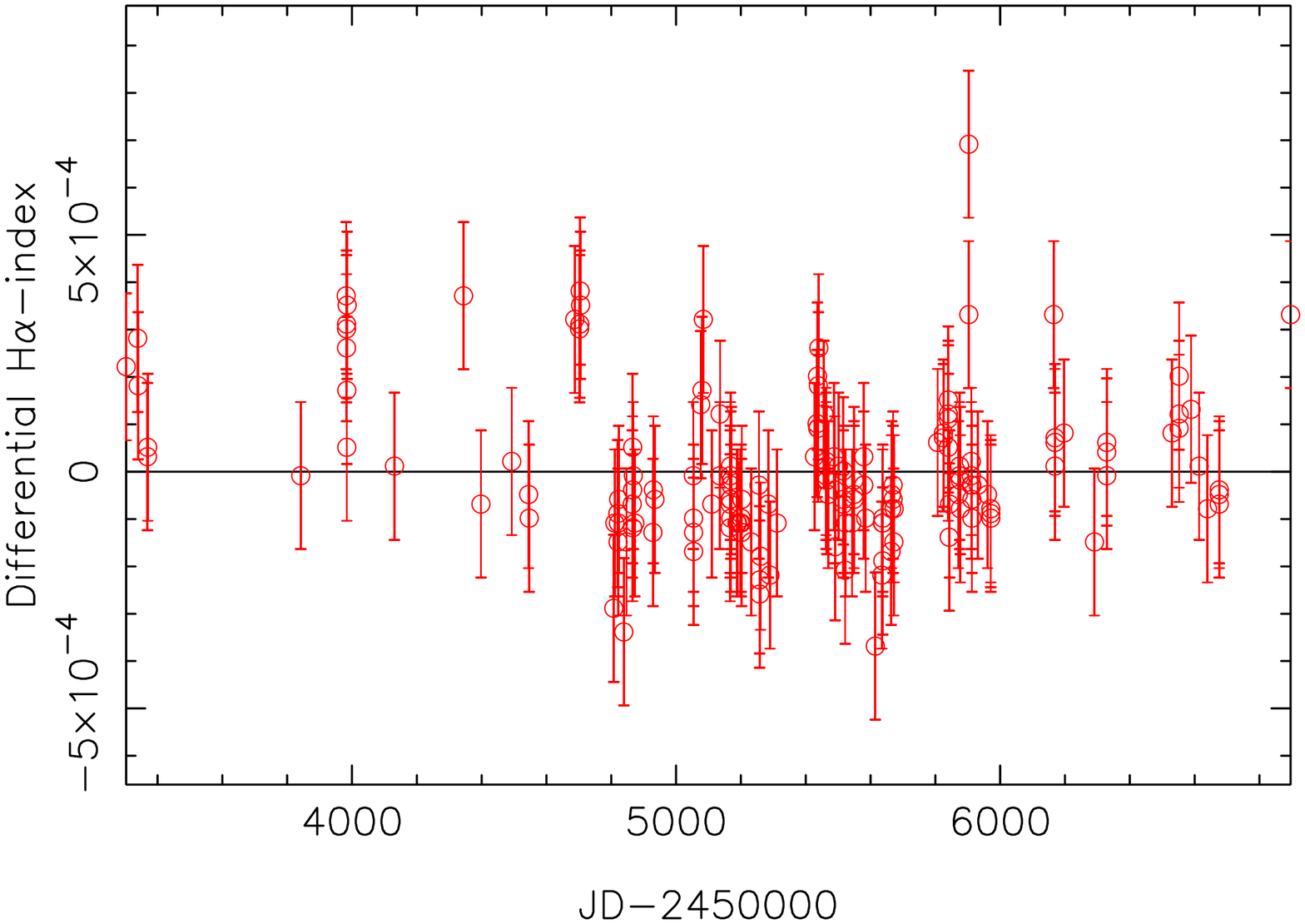}

\includegraphics[angle=270, width=0.23\textwidth,clip,trim=3cm 0 0 0]{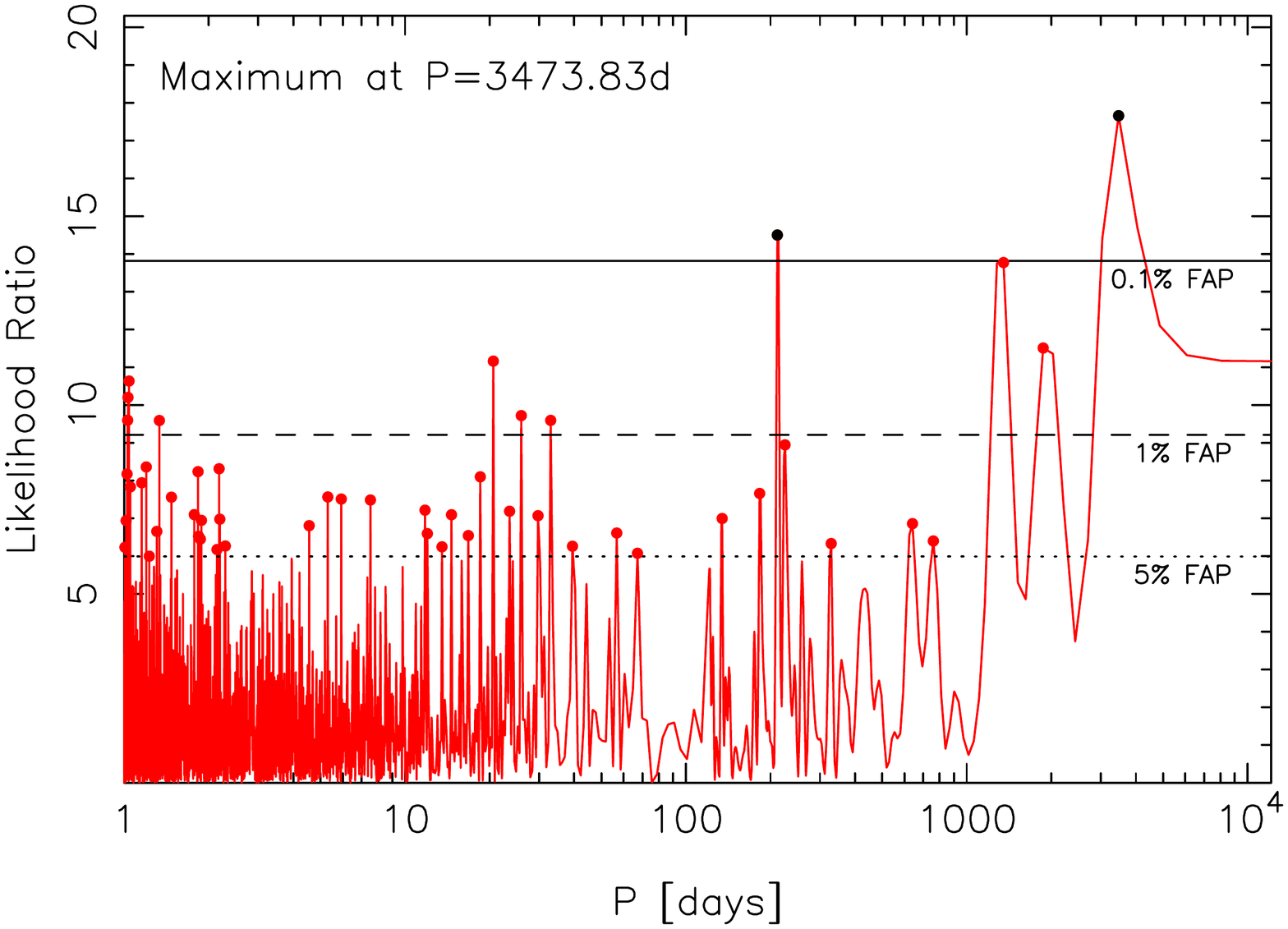}
\includegraphics[angle=270, width=0.23\textwidth,clip,trim=3cm 0 0 0]{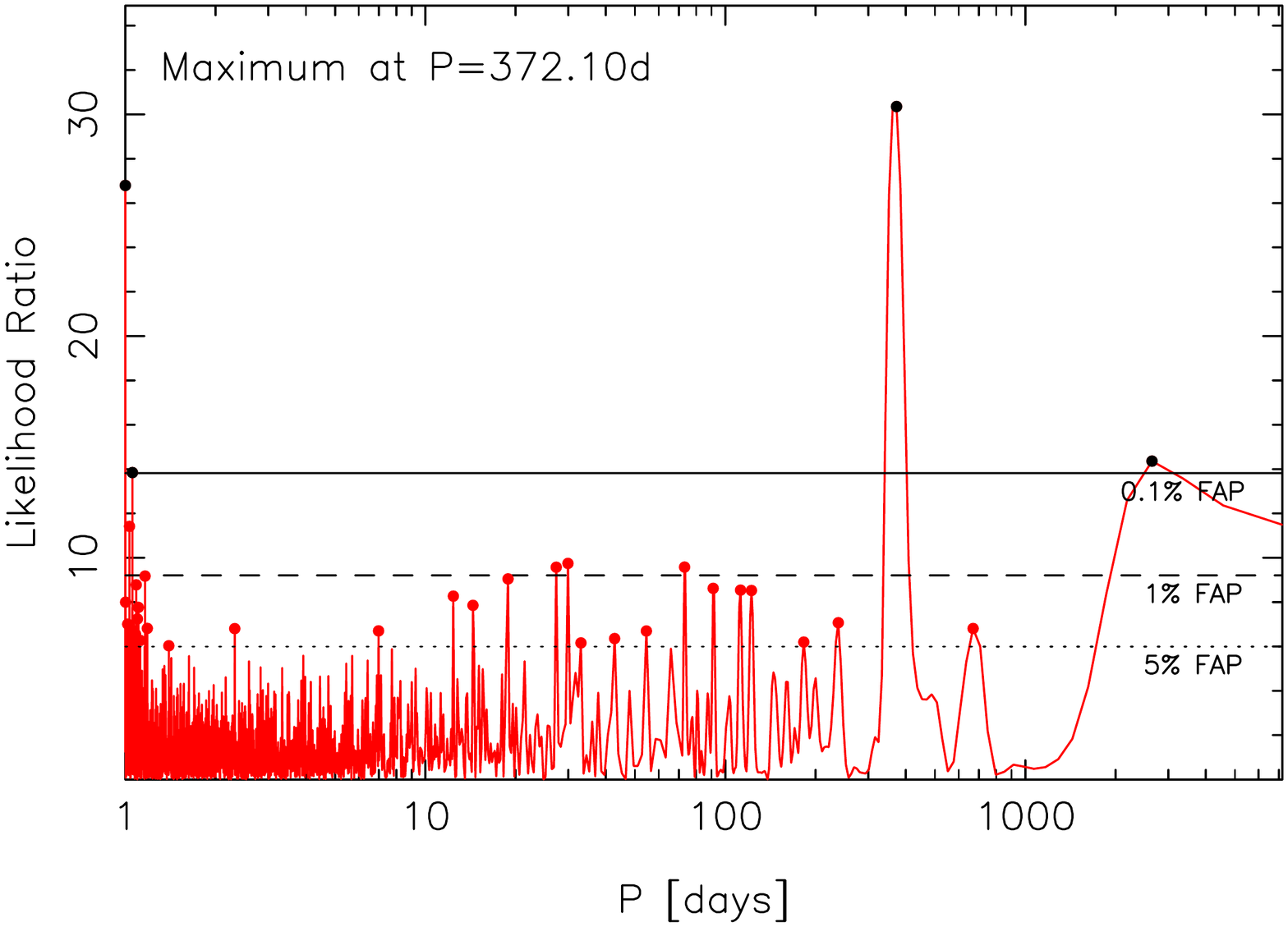}

\includegraphics[angle=270, width=0.23\textwidth,clip,trim=3cm 0 0 0]{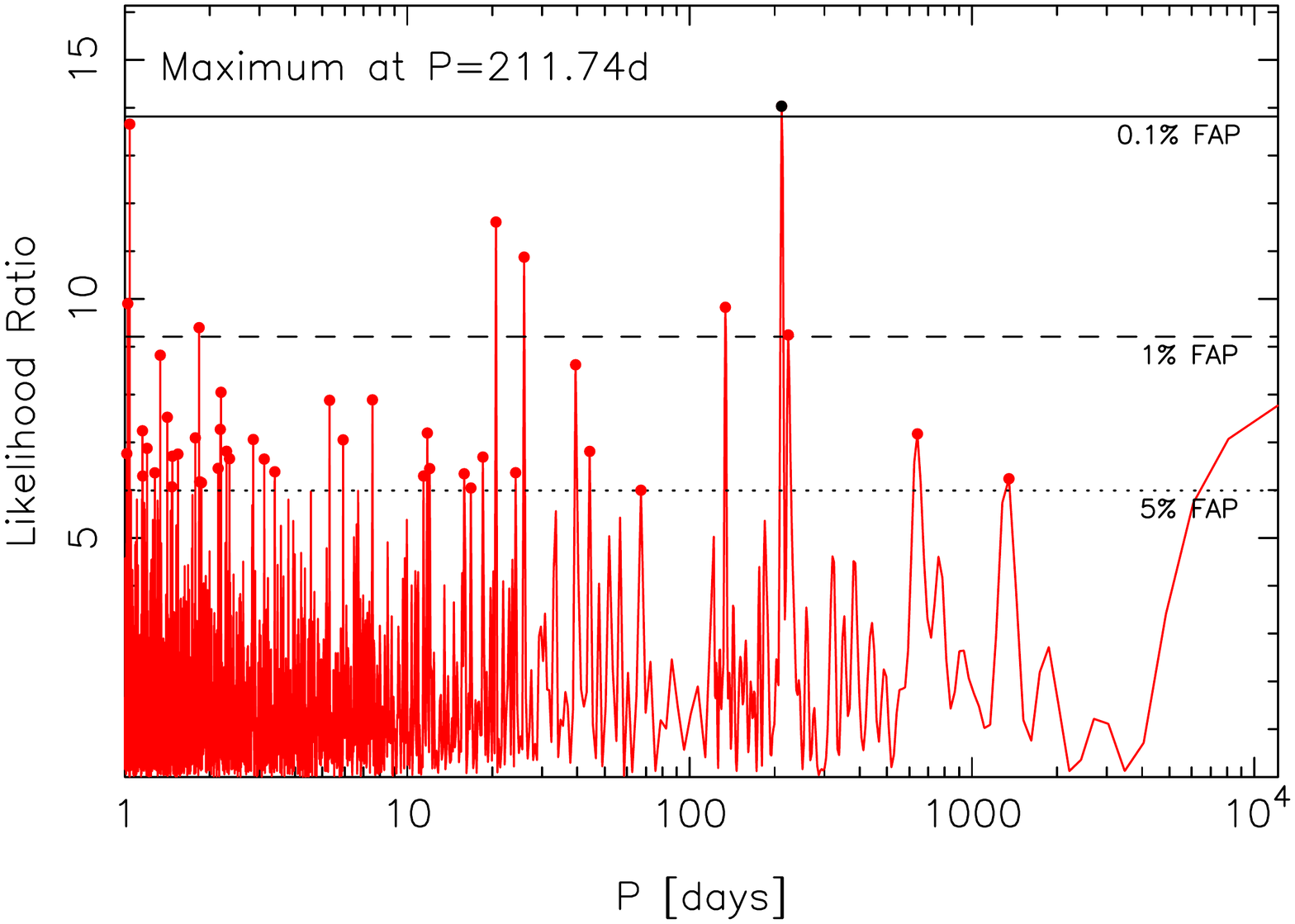}
\includegraphics[angle=270, width=0.23\textwidth,clip,trim=3cm 0 0 0]{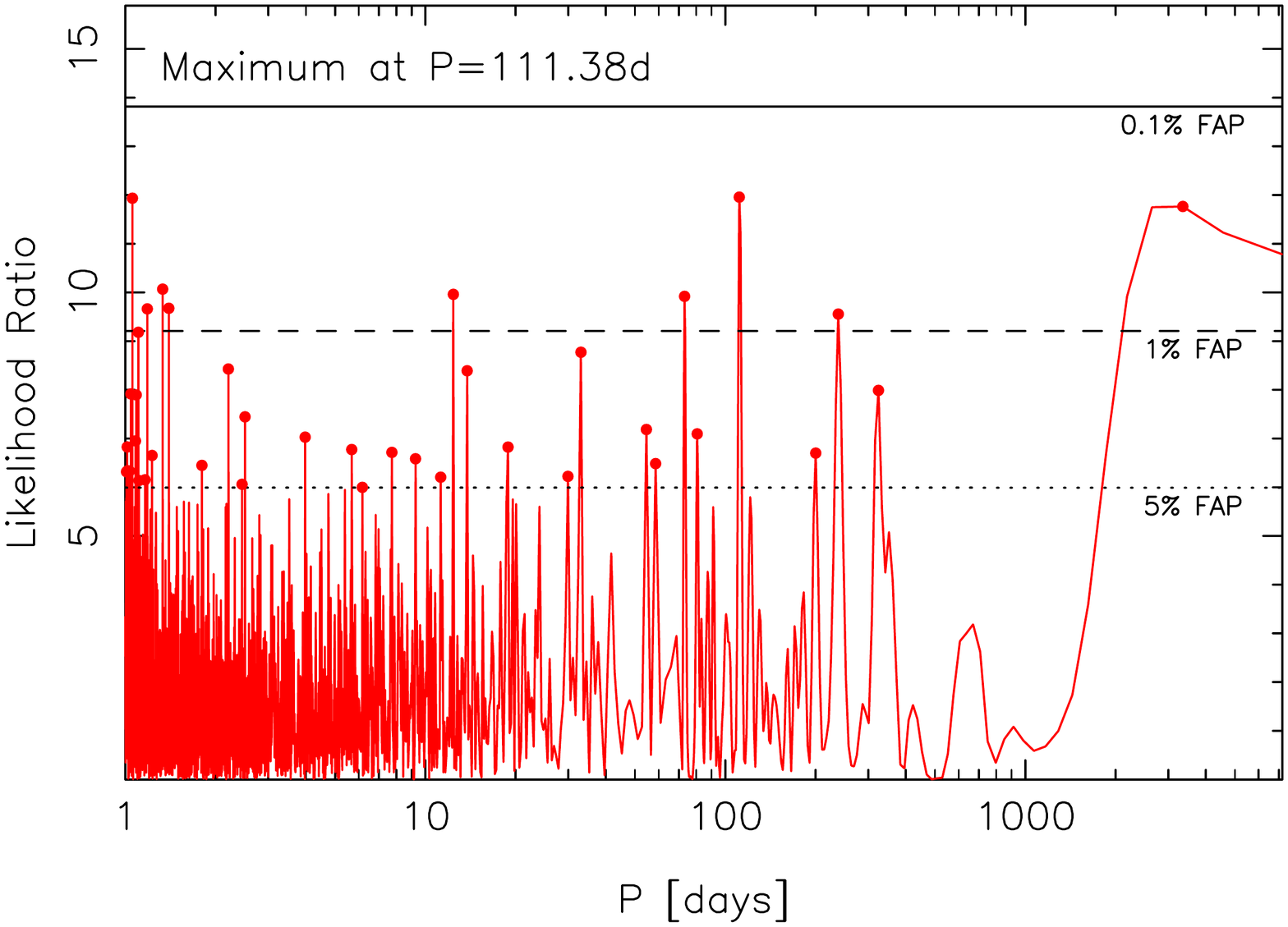}
\caption{Differential KECK S-index (left panels) and H-$\alpha$ (right panels) data of HD 34445 with respect to their mean values (top) and their likelihood-ratio periodograms (middle). The bottom panels shows the residual periodograms after subtracting the most significant signals. The black (red) dots denote likelihood ratios in excess of the 0.1\% (5\%) FAP.}\label{fig:keck_s}
\end{figure}

We subjected the KECK, APF, and PFS S-indices to searches for periodicities to see if any of the radial velocity signals had counterparts in this activity index. Although PFS and APF S-indices showed no evidence for periodicities, we observed a significant signal in the KECK S-indices at a period of 3470 days (Fig. \ref{fig:keck_s}). This signal is likely indicative of a stellar magnetic activity cycle. However, it does not coincide with the long-period radial velocity signal, which suggests that they correspond to different physical phenomena. This supports our interpretation that the long-period radial velocity signal with a period of 5700 days is likely caused by a planet orbiting the star. However, there is also another signal in the KECK S-indices at a period of 211.7 days (Fig. \ref{fig:keck_s}) that appears to be close to the radial velocity signal of 214.7 days in the period space. Yet, this activity-signal is distinct from the radial velocity signal by being statistically significantly different in period and because the parameters of the radial velocity signal are independent of the S-indices strongly suggesting planetary rather than activity-induced origin for the signal.

The H-$\alpha$ values of KECK and PFS were also subjected to a periodogram analysis. We observed a significant periodicity in the KECK H-$\alpha$ values at a period of 372 days (Fig. \ref{fig:keck_s}). This signal might be of astrophysical interest but it cannot be said to be a counterpart of any of the radial velocity signals. Moreover, it is close to a significant yearly periodicity ($P_{\rm max}=365.78\,{\rm d}$) in the periodogram of the sample times (the window function). We also looked into correlations of the PFS H-$\alpha$ values with the RV's and found no differences between results with or without a correlation term in the model. There was such a small number of PFS data that the results are not sensitive to RV variability related to H-$\alpha$  emission variations.

\begin{figure}
\center
\includegraphics[angle=270, width=0.40\textwidth,clip,trim=2cm 0 0 0]{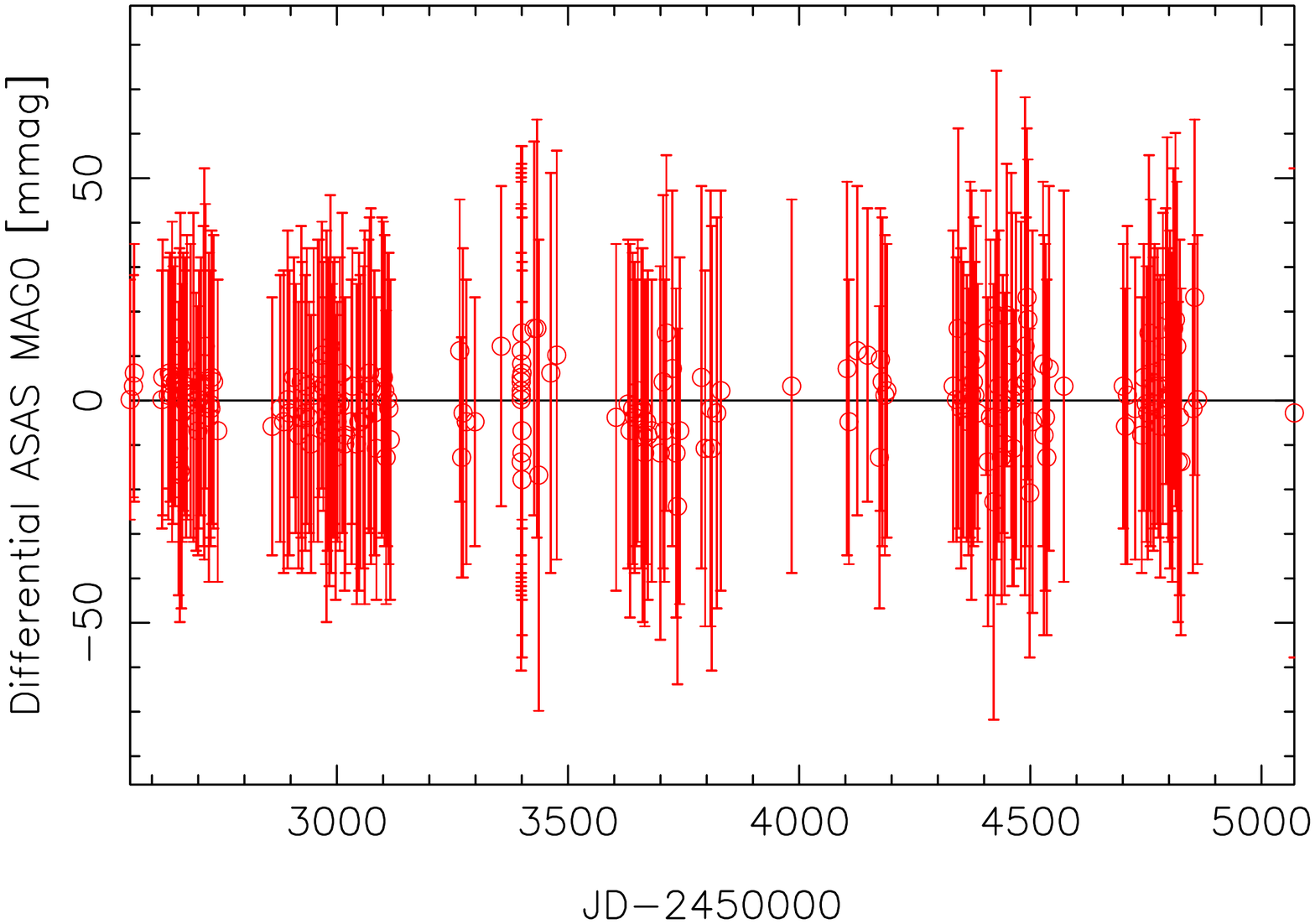}

\includegraphics[angle=270, width=0.40\textwidth,clip,trim=3cm 0 0 0]{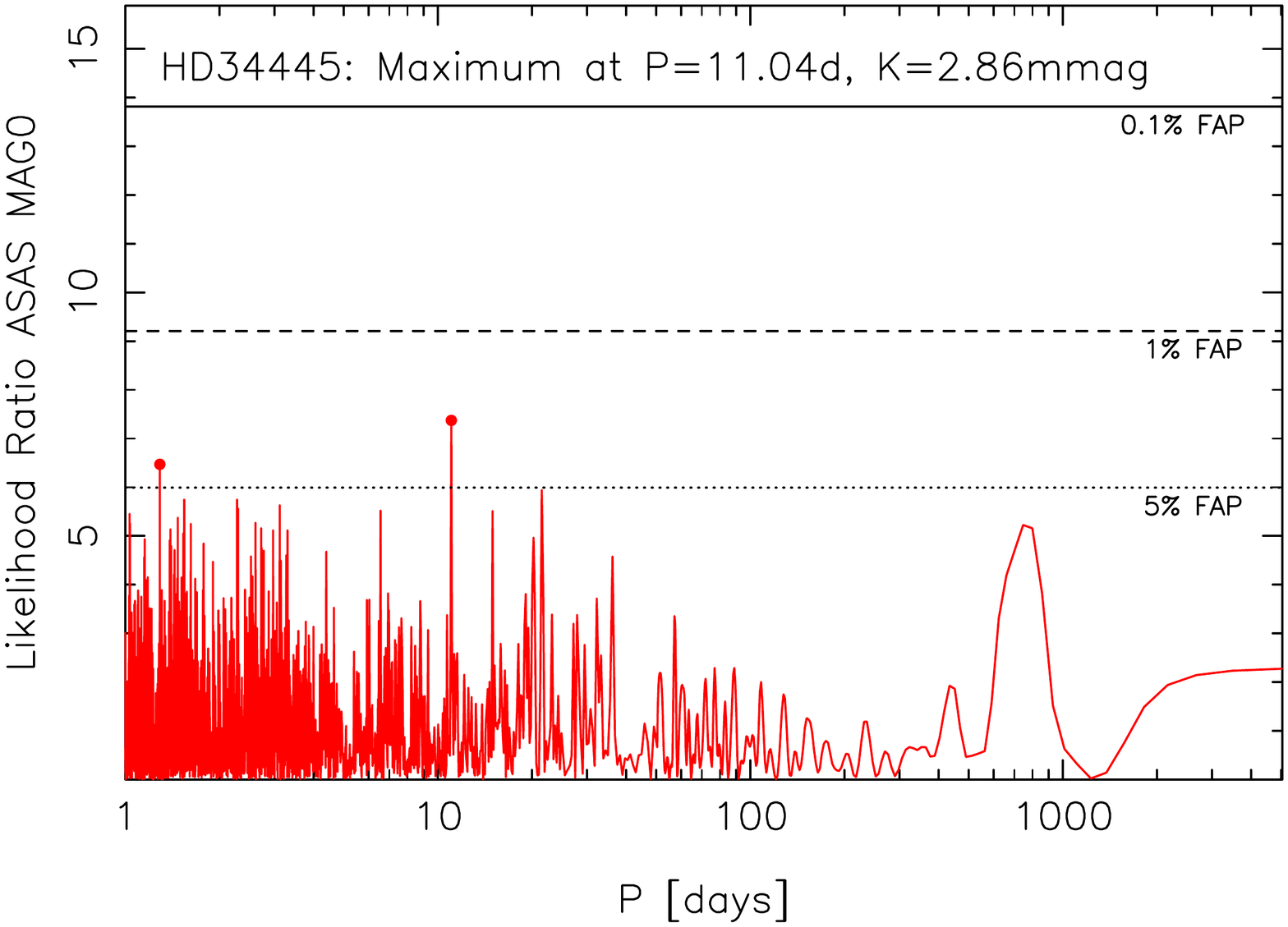}
\caption{ASAS V-band photometry data of HD 34445 (top panel) and its likelihood-ratio periodogram (bottom panel).}\label{fig:ASAS}
\end{figure}

The ASAS V-band photometry \citep{pojmanski1997} data of HD 34445 did not show evidence in favour of periodic signals. We selected the aperture with the smallest variance and grade `A' data of the ASAS measurements and removed all remaining 5-$\sigma$ outliers. This resulted in a set of 225 measurements with a mean of 7319.0$\pm$8.4 mmag. We have plotted the ASAS time-series and the corresponding likelihood-ratio periodogram in Fig. \ref{fig:ASAS}. As can be seen, there is no credible evidence in favor of periodic signals in the ASAS photometry. This might be caused by the fact that the estimated photon noise in the ASAS data appears to be overestimated and much higher in magnitude than the apparent variability (Fig. \ref{fig:ASAS}, top panel).

Apart from hints (with 5\% FAP) of a photometric periodicity of 11 days (Fig. \ref{fig:ASAS}), we did not obtain any evidence for a stellar rotation period in the ASAS photometry data or the spectral activity indicators. Specifically, we could not confirm the $\rhk$-based rotation estimate of $\sim$22 days stated in \citet{Howard2010}. As an additional point of evidence, we note that a stellar rotation period of $P\sim26$ days is implied by the $R=1.38\,R_{\odot}$ and $v\sin (i)=2.7\,{\rm km\,s^{-1}}$ measurements of \citet{Howard2010}. This period is substantially shorter than the $P=49.175\,{\rm d}$ period of the innermost candidate planet of our velocity model. In summary, our assessment of the effect of stellar activity on the derived Doppler velocities leads us to conclude that the periodicities we observe are best explained by stellar reflex motion in response to the Keplerian signals listed in Table 9.

\subsection{\textbf{Dynamical Stability}}

\begin{figure}
\center
\includegraphics[angle=0, width=0.5\textwidth,clip]{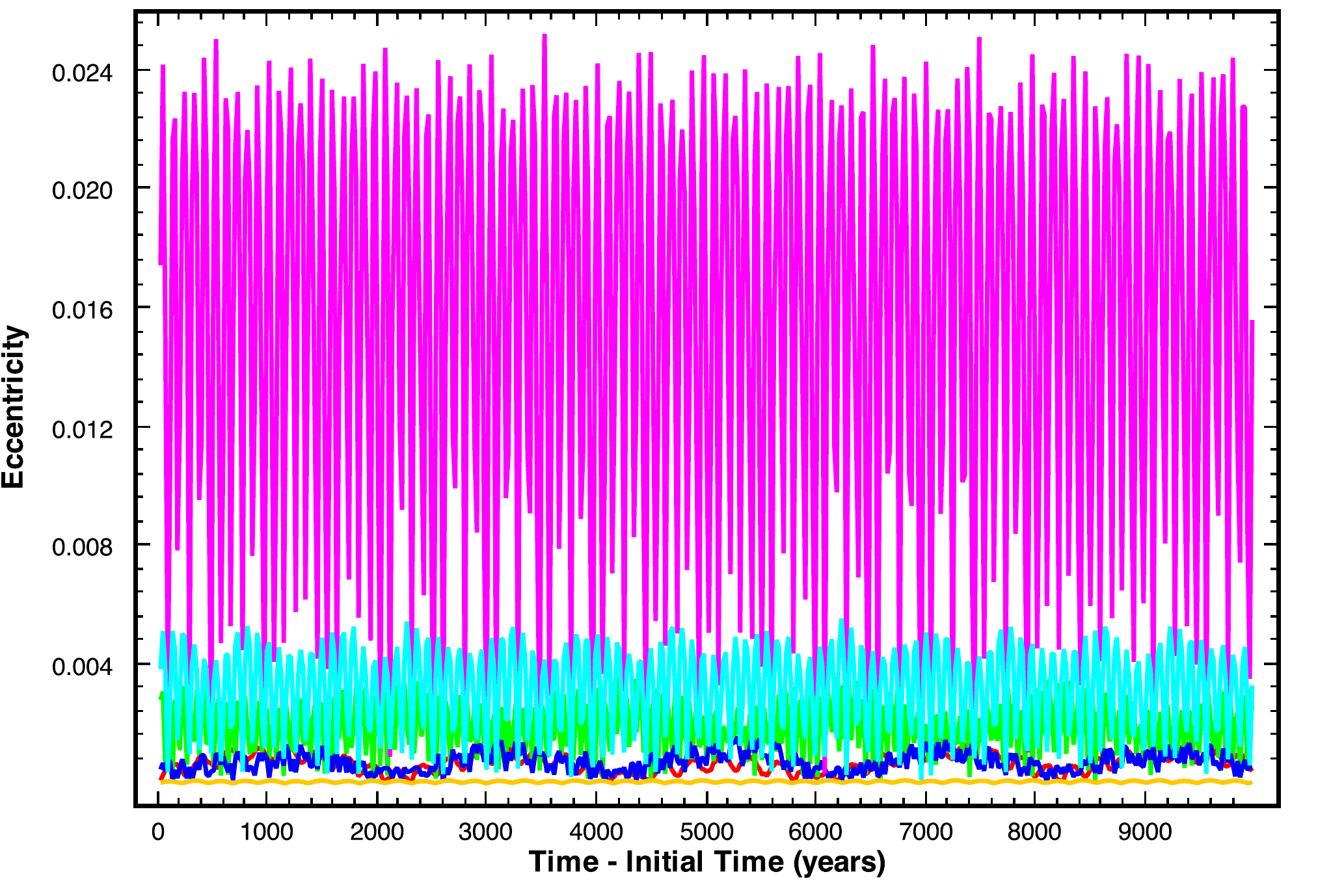}
\caption{Orbital evolution of the eccentricities of the 6-planet Keplerian model over 10,000 years.}\label{fig:stability}
\end{figure}

The 6-planet system seems to be nominally dynamically stable, especially since the eccentricities are not significantly different from zero. Figure \ref{fig:stability} shows a 10,000 year Bulirsch-Stoer integration of the 6-planet Keplerian solution of Table \ref{tab:parameters_six}. Here colors denote increasing orbital periods in the order red, green, blue, magenta, turquoise, and gold. A detailed investigation of the orbital dynamics of the system is left to other studies, but there appear to be possibilities of 5:1, 3:2, and 7:1 mean motion resonances.

\section{Discussion}

\begin{figure*}
\center
\includegraphics[angle=270, width=0.80\textwidth,clip]{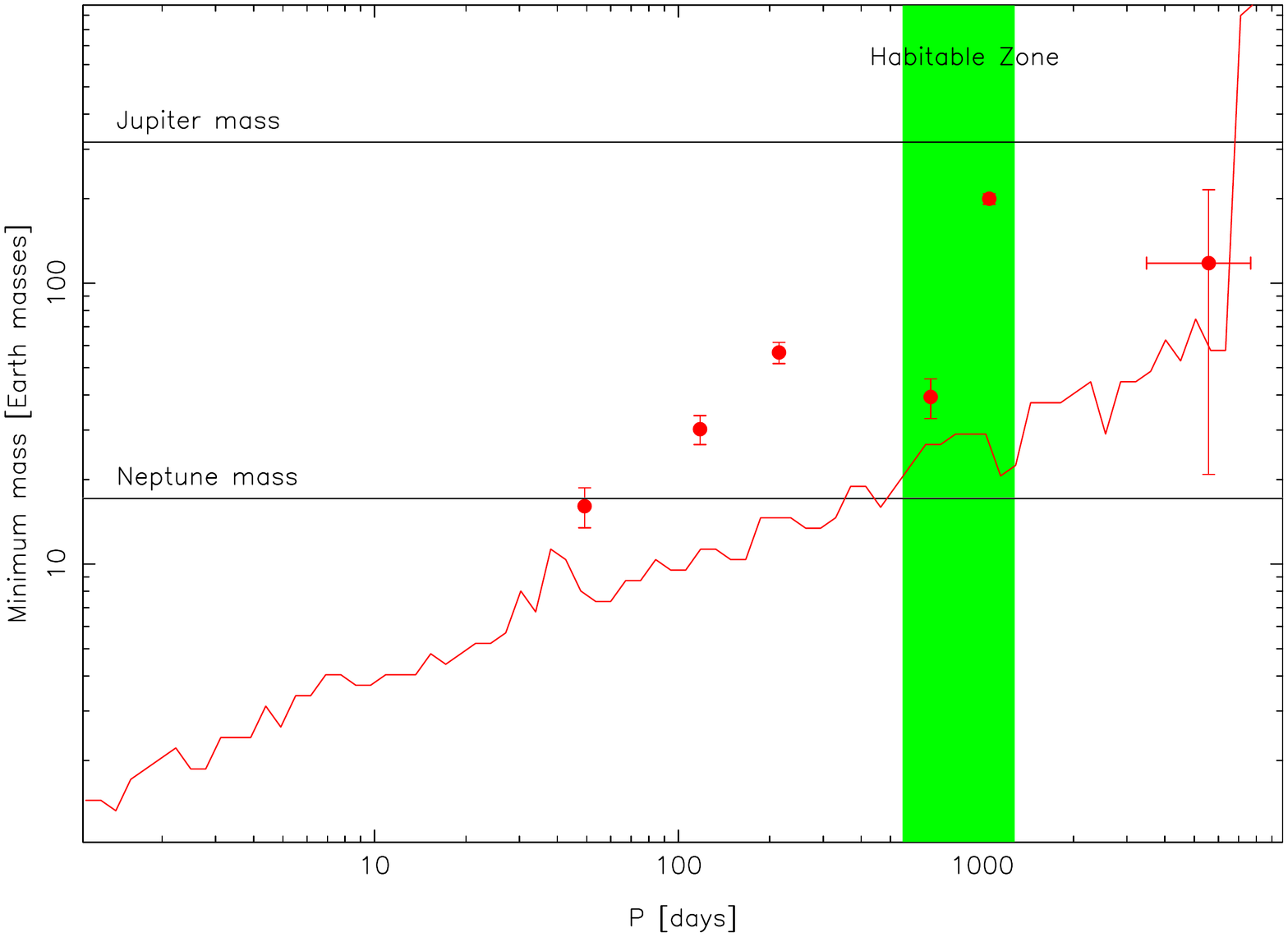}
\caption{Estimated detection threshold of additional planets orbiting HD 34445. The red dots correspond to the detected planet candidates and the green area highlights the estimated liquid-water habitable zone of the star.}\label{fig:threshold}
\end{figure*}

We plotted the detection threshold for additional planets orbiting HD 34445 in Fig.\ \ref{fig:threshold}. This threshold was calculated according to \citet{tuomi2014}. The positions of the planet candidates as functions of minimum mass and orbital period are shown for reference and indicate that five detected candidate planets are clearly above the detection threshold. We have also plotted the estimated liquid-water habitable zone of the star for comparison based on $T_{\rm eff} =$ 5836 K and $L_{\star} =$ 2.01 L$_{\odot}$ \citep{Howard2010} and estimated according to the equations of \citet{kopparapu2014}. This liquid-water habitable zone between `runaway greenhouse' and `maximum greenhouse' limits of \citet{kopparapu2014} is located between 1.34 and 2.36 AU for Earth-mass planets and indicates that there is no room for Earth-like habitable planets in the system. However, Earth-sized moons orbiting HD 34445b and HD 34445f could potentially have liquid water on their surfaces. Interestingly, even if the HD 34445 system as presented here is scaled down in mass by a factor of ten (as one might envision occurring if the primary lay near the bottom of the main sequence, and assuming a constant planetary-to-stellar mass ratio), the planets would all be substantially more massive than a terrestrial-like system of the type accompanying, e.g., TRAPPIST-1 \citep{Gillon2017, Wang2017}

\begin{figure}
\center
\includegraphics[angle=0, width=0.5\textwidth,clip]{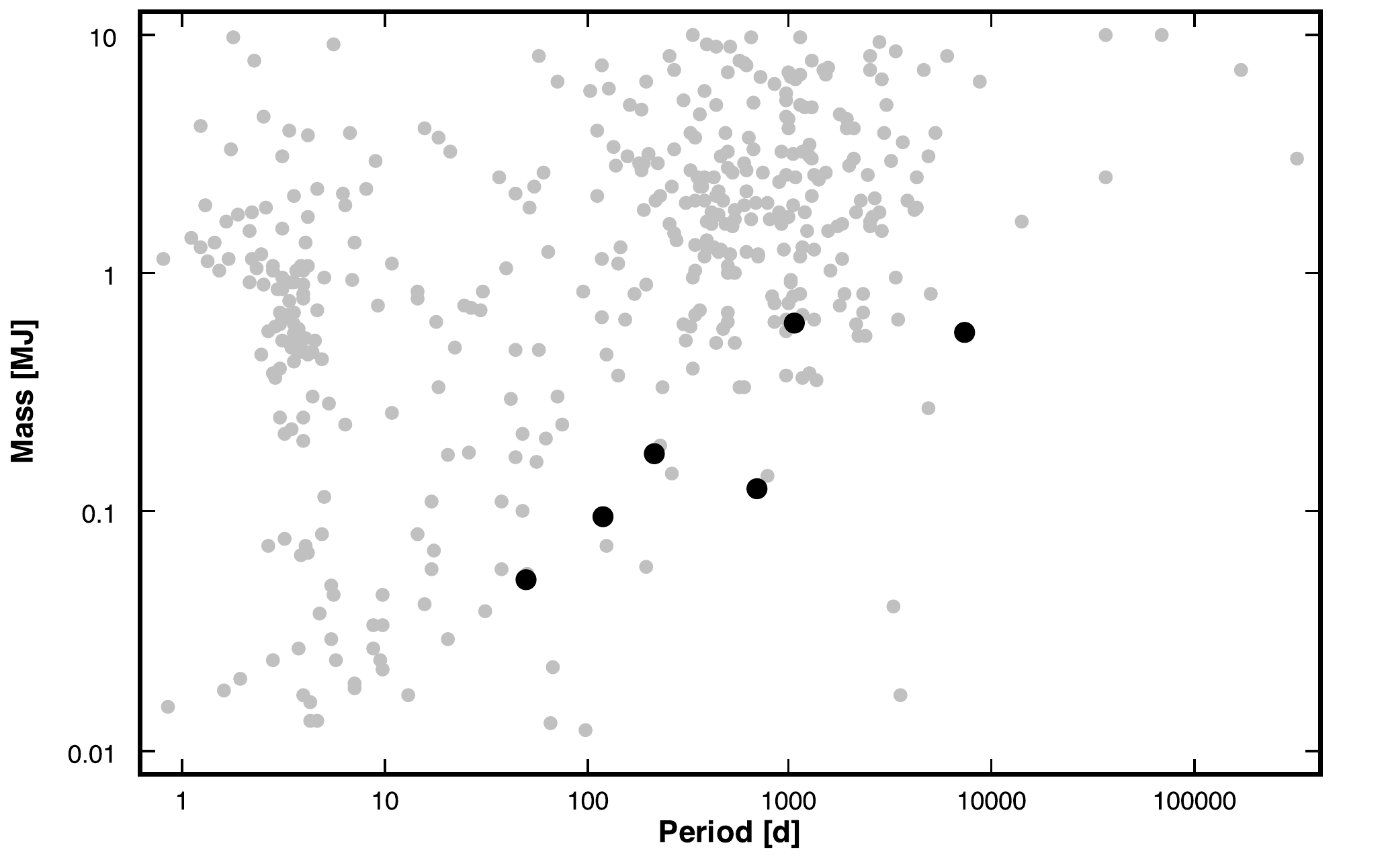}
\caption{Mass vs. period for known RV planets. The six planets of the present work are shown as bold points, and they tend to delineate the lower envelope of the distribution. In this respect (and in terms of overall multiplicity), the system resembles the set of planets orbiting HD 219134 \citep{Vogt2015, Motalebi2015}, albeit without harboring detectable bodies at the shortest orbital periods.}\label{fig:mass-period}
\end{figure}

Finally, in Figure \ref{fig:mass-period} we shows the ensemble of known RV-detected planets (gray points), together with the six planets of the present work (bold black points).

We note that HD 34445's $V=7.3$ brightness would permit high signal-to-noise measurements of the atmospheric properties of its planets in the event that they were observable in transit. The {\it a-priori} geometric probability of transit for the $P=49.175$ d innermost planet, e, is $P_e\sim2.4\%$, with the odds of planet-star occultations decreasing for the bodies orbiting further out. Planet d has a transit probability of just over 1\%, while the remaining planets in the system have probabilities that are substantially lower still.

The planetary candidates described in this paper all have radial velocity half-amplitudes in the range $2 \,{\rm m\, s^{-1}} \lesssim K \lesssim 5\,{\rm m\, s^{-1}}$, while orbiting with periods ranging from tens to thousands of days. This configuration is thus quite unlike either the Kepler multiple-transiting systems \citep{batalha13} which, with their summed planet-to-star mass ratios and system densities resemble scaled-up versions of the Jovian satellite systems, or our own solar system, in which only Jupiter and Saturn have $K\gtrsim 3\,{\rm m\,s^{-1}}$. The detection of systems resembling HD~34445 requires stable, high-precision Doppler monitoring, and critically, a very long base line of observations, a confluence that has arisen only recently.

If the catalog of bright, nearby solar-type stars is kept under surveillance, it is plausible that other multiple-planet systems similar to the one described here will be discovered.
It seems unlikely, however, that such discoveries will be commonplace. A study such as the one described here is a very costly enterprise (in terms of telescope time), and given that these planets seem clearly uninhabitable, it is difficult to sustain the required cadence of observations on a heavily competed telescope. Indeed, it is only because of the long-term Keck precision radial velocity program sanctioned by the UC TAC that we were able to use the Keck/HIRES facility to obtain few-\ms precision over the 18.7 year time base of the present investigation.

\acknowledgments
SSV gratefully acknowledges support from NSF grant AST-0307493. RPB gratefully acknowledges support from NASA OSS Grant NNX07AR40G, the NASA Keck PI program, and from the Carnegie Institution of Washington. GL acknowledges support from the NASA Astrobiology Institute under Cooperative Agreement Notice NNH13ZDA017C issued through the Science Mission Directorate. MD acknowledges the support of CONICYT-PFCHA/Doctorado Nacional-21140646, Chile.The work herein is based on observations obtained at the W. M. Keck Observatory, which is operated jointly by the University of California and the California Institute of Technology, and we thank the UC-Keck and NASA-Keck Time Assignment Committees for their support. This research has made use of the Keck Observatory Archive (KOA), which is operated by the W. M. Keck Observatory and the NASA Exoplanet Science Institute (NExScI), under contract with the National
Aeronautics and Space Administration. We also wish to extend our special thanks to those of Hawaiian ancestry on whose sacred mountain of Mauna Kea we are privileged to be guests. Without their generous hospitality, the Keck observations presented herein would not have been possible. This paper includes data gathered with the 6.5 meter Magellan Telescopes located at Las Campanas Observatory, Chile. The work herein was also based on observations obtained at the Automated Planet Finder (APF) facility and its Levy Spectrometer at Lick Observatory. This research has made use of the SIMBAD database, operated at CDS, Strasbourg, France.

{\it Facilities:} \facility{Keck (HIRES)}, \facility{Magellan (PFS)}, \facility{APF (LEVY)}.

 

\end{document}